%% file: Multiple_Exclusion_Statistics_the_kmers_problem.tex
\documentclass[12pt,a4paper,openany,oneside]{book}
\usepackage[utf8]{inputenc}
\usepackage[spanish]{babel}
\usepackage{amsmath,amsfonts,amssymb}
\usepackage{graphicx}
\usepackage{float}
\usepackage{csquotes}
\usepackage{cite} 
\usepackage{}
\usepackage[version=3]{mhchem}
\usepackage{nicefrac}

\usepackage{fancyhdr}
\begin{document}

	\begin{titlepage}
		\centering
		{\scshape\LARGE Universidad Nacional de San Luis \par}
		\vspace{1cm}
		{\scshape\Large Trabajo Final para optar por el título de Licenciado en Física\par}
		\vspace{1.5cm}
		{\huge\bfseries Mecánica Estadística de Múltiple Exclusión de Estados y su
			aplicación al problema de k-meros\par}
		\vspace{2cm}
		{\Large\itshape Julián José Riccardo\par}
		\vfill
		Director:\par
		Antonio José \textsc{Ramirez-Pastor}
		
		Co-Director:\par
		Pedro Marcelo \textsc{Pasinetti}
		
		\vfill
		
	\end{titlepage}

\newpage

\begin{center}
	{\bf Abstract}
\end{center}
A new distribution for systems of particles obeying statistical exclusion of correlated states is presented following the Haldane's state counting. It relies upon a conjecture to deal with the multiple exclusion that takes place when the states available to single particles are spatially correlated and it can be simultaneously excluded by more than one particle. The Haldane's statistics [F. D. M. Haldane, Phys. Rev. Lett. 67, 937 (1991)] and Wu's distribution [Y.-S. Wu, Phys. Rev. Lett. 52, 2103 (1984)] are recovered in the limit of non-correlated states (constant statistical exclusion) of the multiple exclusion statistics. In addition, the exclusion spectrum function $G(n)$ is introduced to account for the dependence of the statistical exclusion on the occupation-number $n$. Results of thermodynamics and state occupation are shown for ideal lattice gases of linear particles of size $k$ ($k$-mers) where multiple exclusion occurs. Remarkable agreement is found with simulations from $k=2$ to $10$ where multiple exclusion dominates as $k$ increases.


\thanks{No es una tarea sencilla la de llegar a ser Licenciado en Física, por eso creo muy
	importante agradecer a todos los que me acompañaron en este recorrido tan especial. Quiero
	agradecer a mis profesores, siempre bien predispuestos a formarnos como físicos de la mejor
	manera. A mis amigos (además compañeros de carrera) con los que pasé por incontables noches
	de estudio y risas. Con los que compartí muchísimas vivencias a lo largo de esta travesía que
	atravesamos para llegar a ser físicos. No estando ellos, sin duda, todo hubiera sido más
	complejo. A mi director, co-director y demás personas que me guiaron de forma excelente a lo
	largo de este trabajo final. A la querida UNSL, institución que me brindó formación académica
	de calidad y de la cual me siento parte. A la FCFMyN y Depto. de Física. Seguramente me olvido de varios en estos agradecimientos y la razón de esto es que siendo estudiante universitario  una gran cantidad de personas pasan por tu vida y cada una de ellas de alguna u otra manera estuvo presente aportando algo, gracias también a ustedes. Por último, pero no por ello menos importante, quiero agradecer especialmente a mi familia: padre, madre, hermanos y hermanas, sobrino, abuelos y abuelas, tíos y tías, primos y primas. Ellos fueron inspiración, sostén y principal motor de apoyo de este logro, nunca podré terminar de agradecerles.}
\tableofcontents
\listoffigures
\include{./texto/TF_Introduccion}
\include{./texto/TF_Intro_Mecanica_Estadistica}
\include{./texto/TF_Estadistica_Fraccionaria}
\include{./texto/TF_Interpretacion_Cinetica_Ads_Des}

\include{./texto/TF_Simulacion_Monte_Carlo}

\include{./texto/TF_Estadistica_de_Multiple_Exclusion_notacion_mas_reducida}

\include{./texto/TF_Determinacion_de_g}

\include{./texto/TF_Conclusiones}


\end{document}

%% file: texto/TF_Introduccion.tex
\chapter{Introducción}

La descripción de la cinética y el equilibrio termodinámico de líquidos o gases en interacción con sólidos es en la práctica de suma importancia en catálisis heterogénea, almacenamiento y separación de gases o hidrocarburos, tecnología de películas delgadas, lubricación, y un sinnúmero de procesos derivados de la interacción entre moléculas y sólidos \cite{ref1,ref2,ref3,ref4,ref5,ref6}. \\

Es relevante entender los mecanismos elementales de adsorción y difusión en la interfase gas-sólido tanto por su implicancia tecnológica y práctica como por la complejidad que surge al desarrollar las funciones termodinámico-estadísticas cuando se trata con sistemas de partículas constituidas por múltiples unidades atómicas (ej. \ce{O2}, \ce{N2}, \ce{CO2}) o grupos funcionales como en el etano (\ce{C2H6}), propano (\ce{C3H8}).\\ 

Los mayores esfuerzos en desarrollos teóricos en sistemas modelos se han basado en considerar que las partículas interactuantes son geométricamente simples (usualmente esféricas o partículas puntuales que pueden ocupar sitios sobre una red geométrica de sitios que representan muy idealmente el conjunto de mínimos locales del potencial de interacción gas-sólido. Éste sistema de partículas y red de sitios es lo que se denomina ``gas de red'' y representa una forma relativamente sencilla y manejable para desarrollar analíticamente las funciones termodinámicas del sistema de estudio. Decimos que es un forma relativamente sencilla porque solo en casos especiales es posible calcular exactamente las funciones termodinámico estadísticas de un gas de red en una dimensión (\textbf{1D}), o cuando las partículas no interaccionan con otras vecinas (interacción lateral) y ocupan solo un sitio de la red en dos dimensiones (\textbf{2D}) y tres dimensiones (\textbf{3D})\cite{ref7} , y en ciertos casos especiales de partículas con interacciones laterales en \textbf{2D} y \textbf{3D}\cite{ref9,ref10,ref11}.\\

Sin embargo el problema se vuelve mas complejo cuando las partículas interactuantes están compuestas de múltiples unidades y ocupan más de un sitio de la red: Como lo describen algunos autores , éste es el prototipo de problema de partículas sobre una red\cite{ref12}.\\

Se han hecho numerosas contribuciones a este complejo problema de la física estadística, desde la consideración de mezclas binarias (polímeros en solvente de monómeros)que es formalmente equivalente  la adsorción de cadenas lineales sobre una red regular de sitos \cite{ref4,ref13,ref14,ref15,ref16,ref17,ref18,ref18p,ref18pp} hasta las más recientes considerando los efectos de multiple ocupación y heterogeneidad energética en la interacción de las partículas con la red \cite{ref19,ref20}. \\

En el campo experimental, el desarrollo de diversas técnicas que permiten observar o inferir la configuración espacial de partículas en interacción con la superficie de un sólido o su migración sobre ella,como Scanning Tunneling Electron Microscopy \textbf{STEM}), Scanning Tunneling Microscopy(\textbf{STM}), Field Emmision Microscopy(\textbf{FEM}))  \cite{ref21,ref22,ref23,ref24}, la síntesis de materiales de baja dimensionalidad como nanotubos \cite{ref25,ref26,ref27}, zeolitas unidimensionales \ce{AlPO4-5} \cite{ref28,ref28p} y la formación de fases moleculares en su interior \cite{CUNHA2018312} o el estudio de la migración superficial de especies moleculares o clusters sobre fases cristalinas de metales (ej. dímeros sobre metales \cite{ref29}, \ce{H2O}/\ce{Ni(110)} \cite{ref30}, difusión de \ce{Pt2}, \ce{Pt3}, \ce{Pt4} sobre tungsteno \cite{ref31},\ce{CO}/\ce{Ni(110} \cite{ref32}, n-alkanos sobre superficies regulares  \cite{Ishida}), representan un estímulo complementario y evidencias para avanzar en la comprensión de la mecánica estadística de moléculas poliatómicas con interacción. En lo que sigue muchas veces usaremos el término partícula poliatómica o partícula multisitio para referirnos a un modelo de especie molecular poliatómica.\\

En este trabajo nos proponemos comprender mejor y desarrollar el marco teórico basado en la extensión del formalismo de Estadística Cuántica Fraccionaria de Haldane\cite{ref33} propuesto en  la ref.\cite{ref34} para describir el fenómeno complejo de adsorción de moléculas poliatómicas o en general para sistemas de partículas con estructura geométrica (cadenas o k-meros lineales, k-meros flexibles, k-meros no-lineales, etc.)a través de una aproximación formalmente simple y manejable tanto analíticamente como en su aplicación práctica denominada Teoría Estadística Cuántica Fraccionaria para Adsorción (\textbf{FSTA} por su acronismo en inglés\cite{ref34}) \\

La idea básica detrás de esta aproximación es que todo sistema de partículas confinado en un volumen $V$ determinado del espacio tiene un espectro de estados accesibles con un número total $G$ de ellos y que cada partícula que se agrega en el volumen $V$ excluye $g$ estados del total. El parámetro $g$ denominado \textbf{parámetro de exclusión estadística} toma valores en el rango $[0,1]$ en el formalismo de Haldane\cite{ref33}, siendo $g=0$ y $g=1$ los casos conocidos de bosones y fermiones, respectivamente, y $0\leq g \leq 1$ corresponde a cuasi-partículas con parámetro de exclusión fraccionaria \cite{ref34p,ref34p1,ref34p2,ref34p3,ref34p4}. Toda la termodinámica de partículas con $g\in[0,1]$ se puede desarrollar de forma sencilla del formalismo de Haldane cuando los estados a los que pueden acceder las partículas son independientes. \\


En este trabajo abordamos el problema de la mecánica estadística de partículas de tamaño $k$ (compuestas $k$ unidades idénticas) sobre una red de sitios extendiendo el formalismo de Haldane para partículas ``clasicas'' con $g>1$ \cite{ref34}. Se obtiene una estadística generalizada donde el valor del parametro $g$ se relaciona con el tamaño y la forma de la partícula en el estado adsorbido sobre la red. de esta forma $g$ adquiere una significación física relevante y su valor accesible a partir de experimentos termodinámicos.\\

De esta forma se puede desarrollar una descripción simple y compacta de las funciones termodinámicas que comprehende  un amplio espectro de sistemas gas-sólido y además obtener en principio no solo información de las interacciones gas-sólido y gas-gas sino además sobre la configuración espacial con que las partículas ocupan la red y la estructura de las fases ordenadas que pueden desarrollarse por efecto de las interacciones laterales. \\

Una parte importante de este trabajo esta dedicada a la simulación de la adsorción de k-meros lineales con y sin interaccion lateral, con $k=2$ a $k=7$, y al cálculo tanto de las isotermas de adsorción, como a funciones estadísticas que nos permitan: a) comprender el efecto de exclusión estadística de estados en un gas de red, b) estudiar la forma de ocupación del conjunto de estados accesibles, c) contrastar el modelo de estadística fraccionaria para adsorción, d) proponer una nueva estadística para el fenómeno de multiple exclusión, d) relacionar directamente la exclusión estadística $g$ con los observables y proponer la forma para determinarla a partir de las isotermas de adsorción.     

realizamos simulaciones de diversos sistemas de k-meros con interacciones repulsivas que resultan  interesantes por el desarrollo de fases ordenadas y los reinterpretamos en términos del concepto de exclusión estadística.\\

En la formulación \textbf{Estadística Cuántica Fraccionaria} de Haldane \cite{ref33}
la exclusión efectiva de estados por cada partícula caracterizada por el parámetro $g$ es constante e independiente del número de partículas presentes en el volúmen $V$. Esto es equivalente a decir que los estados accesibles a cada partícula son independientes entre sí en tanto los estados \textbf{excluidos }por una partícula cuando se incorpora a $V$ no son  excluidos a la vez por otra partícula diferente. Esto parece aceptable en sistemas cuánticos donde $0\leq{g}\leq{1}$ y la mayor exclusión de un estado es la ocupación plena por un fermión $g=1$. Sin embargo en nuestra representación el volumen $V$ esta definido por la red geométrica de sitios que introduce una correlación espacial entre los sitios (estados) y es ocupado por partículas en el rango de interés $g>1$ que se comportan como ``superfermiones'' ya que ocupan mas de un sitio de la red. En consecuencia los estados que ocupa y excluye cada partícula \textbf{no son independientes entre sí} y ocurren configuraciones con \textbf{múltiple exclusión de estados}, dando lugar a un conjunto mucho mas complejo de estados accesibles. En general para un k-mero aislado sobre la red, formalmente el cubrimiento $\theta\to 0$ , el número de estados excluidos es $g\propto{k}$ , sin embargo en la saturación, $\theta\to 1$ y $g\to 1$. En conclusión el de problema múltiple exclusión de estados es intrínseco al de partículas que ocupan múltiples estados sobre una red como consecuencia de las correlaciones espaciales de entre los sitios. Cuanto mayor el tamaño mas importante es este efecto. \\

Para resolver este problema en este trabajo presentamos las bases una nueva estadística para sistemas de partículas con \textbf{Múltiple Exclusión de Estados}, obtenemos las funciones termodinámicas, comparamos  con la formulación estadística fraccionaria de Haldane e independientemente con simulaciones de adsorción en equilibrio de k-meros lineales sobre red cuadrada. La Estadística Fraccionaria de Haldane resulta un caso límite de la Estadística de Múltiple Exclusión. Los resultados muestran que la múltiple exclusión de estados es relevante cuanto más grande es $k$ y las consecuencias de esta nueva estadística propuesta son estimulantes para profundizar este estudio en el futuro.\\

Por último desarrollamos una función simple y robusta para determinar el parámetro de exclusión estadística $<g>$ a partir de medidas termodinámicas como la isoterma de adsorción, su dependencia con la densidad o cubrimiento y relacionarlo con la configuración espacial de la partícula en estado adsorbido. Esta metodología nos permite entender con claridad el comportamiento de k-meros con interacciones repulsivas en términos del concepto de exclusión estadística y el desarrollo de distintas fases adsorbidas ordenadas en función del cubrimiento cuando el sistema se encuentra por debajo de la temperatura crítica.       

En nuestro mejor conocimiento, de los modelos teóricos y soluciones analíticas disponibles, la Estadística Fraccionaria  y la Estadística de Múltiple Exclusión presentada en este trabajo aparecen como los mas elaborados disponibles para  describir en forma generalizada el complejo problema estadístico del gas de red de k-meros asi como un amplio espectro de sistemas experimentales que van desde gases poliatómicos simples como \ce{O2}, \ce{N2}, \ce{CO} , hasta moléculas mas complejas como alcanos, hidrocarburos o proteínas adsorbidas sobre superficies sólidas.   

%% file: texto/TF_Intro_Mecanica_Estadistica.tex
\chapter{Introducción a la Mecánica Estadística}

\section{Conjunto Microcanónico: Principios}
El estudio de este trabajo lo realizamos dentro del marco de la Teoría Mecánica Estadística. Ésta tiene por objeto describir el comportamiento macroscópico de un sistema de muchas partículas a partir de las propiedades microscópicas de sus componentes incluidas las interacciones entre ellos \cite{Giorgio}. \\

\noindent En un sistema aislado de $N$ partículas en un volúmen $V$ y energía constante $U$ \textbf{todos los estados} microscópicos(cuánticos) accesibles \textbf{son igualmente probables} y por lo tanto el valor promedio de cualquier observable físico $A$ se expresa como 

\begin{equation}\label{eq.1.1}
<A>=\int A \: P\left( A\right)  dA
\end{equation}

\noindent 
donde $P(A) dA$ es la probabilidad de encontrar el observable $A$ con valores entre $A$ y $A+dA$. \\

\noindent Para calcular este promedio Gibbs \cite{Gibbs} introdujo la noción de \textbf{Conjunto Estadístico}, denominado Microcanónico cuando $N,V,U=\textrm{ctes}$, como un numero muy grande (estadísticamente relevante) de réplicas del sistema en el cual aparecen representados los distintos estados microscópicos \textbf{con la misma probabilidad}, de forma que la probabilidad $P(A) dA$ se puede calcular sobre este conjunto como : 

\begin{equation}\label{eq.1.2}
P\left( A\right)  dA=\lim_{\mathcal{N_{\mathcal{O}}}\to\infty} \dfrac{\mathcal{N}\left( dA\right) }{\mathcal{N_{\mathcal{O}}}}
\end{equation}
\noindent donde $\mathcal{N}(dA)$ es el numero de réplicas del conjunto estadístico donde el obsevable $A\in(A,A+dA)$ y $\mathcal{N_{\mathcal{O}}}$ es el número total de sistemas del conjunto. Esta propiedad de equiprobabilidad para los estados de un sistema microcanónico es uno de los postulados fundamentales de la teoría Mecánica Estadística\\

\noindent La relación del promedio  $<A>$ sobre el conjunto estadístico y el promedio temporal $\overline{A(t)}$ que se observa en un experimento de medida de $A$ está basado en el \textbf{Teorema Ergódico} demostrado por Birkhoff \cite{TeoremaErgodico}

\begin{equation}\label{eq.1.3}
<A>= \lim_{t\to\infty}  \dfrac{1}{t} \int_{0}^{t} A\left( \mathbf{r}\left( t\right) ,\mathbf{p}\left( t\right) \right)  dt
\end{equation}
	
\noindent  donde el promedio temporal se hace sobre una trayectoria del sistema compatible con $N,V,U=\textrm{ctes}$ y $\mathbf{r}(t),\mathbf{p}(t)$ su posición en el espacio de fases al tiempo $t$ sobre la superficie $U=cte$.\\

\noindent Todo sistema que cumple con la condición\eqref{eq.1.3} decimos que es \textbf{ergódico}. Sin embargo,
no todos los sistemas cumplen con la condición \eqref{eq.1.3}. A éstos los denominamos \textbf{no ergódicos} (ej.: si existe otra cantidad conservada ademas de $U$ como en los superconductores, o vidrios donde el sistemas pasa mucho tiempo en alguno de sus estados comparado con otros).\\

\noindent La relación fundamental del Conjunto Microcanónico con la termodinámica se completa a través de la relación de Boltzmann entre el número de estados microscópicos accesibles $ \Omega\equiv\Omega(N;V;U)$ y la entropía $S\equiv S(N;V;U)$\cite{Boltzmann}

\begin{equation}\label{eq.1.4}
S=k_{B} \ln \varOmega    
\end{equation} 

\noindent con $ k_{B}=1.3807 \textrm{x}10^{-23} \textrm{J}/\textrm{K}$ para que sea compatible con la escala de temperatura de Kelvin ${1}/{T}=\partial S/\partial U$. El Postulado de Boltzmann expresado en la ec. \eqref{eq.1.4} cumple con el segundo y tercer Principio de la Termodinámica y es el segundo postulado fundamental de la Mecánica Estadística. \\

\section{Conjunto Canónico}

\noindent Si en vez de considerar un sistema aislado consideramos al sistema $\mathbb{S}$ en contacto con un baño térmico $\mathbb{S}^{'}$ en el cual $N,T$ y $V$ son constantes entonces el postulado de equiprobabilidad de estados accesibles del conjunto microcanónico aplicado ahora al sistema aislado $\mathbb{S}^{*}=\mathbb{S}+\mathbb{S}^{'}$ conduce a la definición de la Función de Partición Canónica 

\begin{equation}\label{eq.1.5}
Z=\sum_{j}\exp{-\beta E_{j}}
\end{equation}

  \noindent donde $\beta=1/k_{B}T$ y la suma se realiza sobre todos estados de energía del sistema. \\
  
  \noindent Para $N,V,T$ constantes la Energía Libre de Helmholtz $F$ es la función termodinámica adecuada porque se relaciona directamente la función de partición $Z$. Así
  
  \begin{equation}\label{eq.1.6}
F=U-TS  
  \end{equation}
  \begin{equation}\label{eq.1.7}
  F=-\frac{1}{\beta} \ln Z
  \end{equation}
  
  \noindent y la probabilidad $f_{j}$ de que $\mathbb{S}$ se encuentre en un estado de energía $E_{j}$ es 
  
  \begin{equation}\label{eq.1.8}
  f_{j}=\frac{e^{-\beta E_{j}}}{Z}  
  \end{equation}  

\noindent Podemos decir que la forma exponencial de $f_{j}$ es una consecuencia directa de la relación fundamental de Boltzmann (logarítmica) entre la entropía y el número total de configuraciones del sistema aislado (eq. \eqref{eq.1.4}),dado que la entropía es aditiva mientras que el número de configuraciones es multiplicativo. La definición de la función de partición $Z$ (ec. \eqref{eq.1.5}) y su relación con $F$ (ec. \eqref{eq.1.7}) surge de la condición de normalización de las probabilidades $f_{j}$, $\sum_{j} f_{j}=1$. \\

  \noindent Con $f_{j}$ de \eqref{eq.1.8}  los promedios $<A>$ de los observables físicos se pueden calcular de forma directa como 
  
  \begin{equation}\label{eq.1.9}
  <A>=\sum_{j} A_{j}f_{j}
  \end{equation}
 
  \noindent para el caso de la energía del sistema, $<A>=U$ y  
  
  \begin{equation}\label{eq.1.10}
  U=\sum_{j} E_{j}f_{j}= -\frac{\partial\ln Z}{\partial \beta}=\frac{ \partial\left(  \beta F\right) }{\partial \beta}
  \end{equation}
 
 \noindent Por simplicidad prescindimos de escribir explicítamente los argumentos de las funciones $Z$, $F$ y $U$ que como todas en este conjunto son funciones de $(N,T,V)$.
  
  \noindent En un sistema aislado (Microcanónico) la evolución hacia el equilibrio esta gobernada por $\Delta S \geq 0$ y el equilibrio determinado por la condición $S=\textrm{máximo}$. \\
  
  \noindent  En el Conjunto Canónico en cambio ($N,V,T$ constantes) es simple demostrar que evoluciona hacia el equilibrio con la condición $\Delta F \leq 0$ y $\Delta F=\textrm{mínimo}$ es la condición de equilibrio. Si además ocurre que la energía de cada estado $E_{j}$ se puede expresar como la suma de la energía de diferentes grados de libertad independientes (o débilmente acoplados) entonces $Z$ se puede factorizar. De esta forma, si 
  
  \begin{equation}\label{eq.1.11}
  E_{j}= \sum_{j} \epsilon_{ij}
  \end{equation}
   
  \noindent  ,donde $\epsilon_{ij}$ es la contribución del grado de libertad $i$ al estado de energía $E_{j}$, entonces 
  
  \begin{equation}\label{eq.1.12}
  Z=\prod_{i} z_{i} \qquad \qquad \textrm{con}\qquad z_{i}=\sum_{j} e^{-\beta E_{j}}
  \end{equation}
 
  \noindent y 
  
  \begin{equation}\label{eq.1.12.a}
  F=\frac{1}{\beta} \sum_{i} \ln z_{i}
  \end{equation} 
  
  \noindent donde $z_{i}$ es la función de partición del grado de libertad $i$ y las \eqref{eq.1.12} y \eqref{eq.1.12.a} expresan la \textbf{propiedad de factorización} de $Z$ que resulta muy útil en la práctica. \\ 
  
  \section{Conjunto Macrocanónico}
 
 Si consideramos un sistema $\mathbb{S}$ con volúmen $V$ en contacto con un baño térmico a $T=\textrm{cte}$ que pueden intercambiar no solo energía sino también partículas entre ambos, el postulado fundamental de equiprobabilidad de los estados accesibles aplicado al sistemas completo aislado $\mathbb{S}+\mathbb{S}^{'}=\mathbb{S}^{*}$ nos conduce a definir una nueva función $\varPsi$ denominada Gran Potencial Termodinámico 
 \begin{equation}\label{eq.1.13}
 \varPsi= U-\mu N -T S
 \end{equation} 
  
  \noindent donde $\varPsi \equiv \varPsi(\mu,T,V)$ y el potencial químico $\mu$ es la fuerza generalizada externa correspondiente a la variable $N$ (número de partículas en $\mathbb{S}$). De la ec. \eqref{eq.1.13} surge que \\
  
  \begin{equation}\label{eq.1.14}
  N=-\left( \frac{\partial \varPsi}{\partial \mu}\right) 
  \end{equation}
   
  \noindent y $\varPsi$ se puede relacionar con la Gran Función de Partición de Partición del conjunto macrocanónico, $\varXi\equiv \varXi(\mu,T,V)$
  
  \begin{equation}\label{eq.1.15.a}
  \varXi=\sum_{j} e^{-\beta(E_{j}-\mu N_{j})}
4  \end{equation}
  
  \begin{equation}\label{eq.1.15.b}
  \varPsi=-\frac{1}{\beta} \ln \varXi
  \end{equation}
  
  \noindent
  
  \begin{equation}\label{eq.1.15.c}
  f_{j}=\frac{e^{-\beta(E_{j}-\mu N_{j})}}{\varXi}
  \end{equation}
 
  \noindent Análogamente a la ec. \eqref{eq.1.10}, se puede escribir la identidad 
  
  \begin{equation}\label{eq.1.16}
   U=-\left( \frac{\partial\ln \varXi}{\partial \beta}\right) =\frac{\partial \left( \beta \varPsi\right) }{\partial \beta}
  \end{equation}  
 
  \noindent y la condición de equilibrio en el conjunto macrocanónico es $\varPsi=\textrm{mínimo}$ para $\mu,V,T \quad\textrm{contantes}$. \\
  
  Finalmente, mencionaremos por completitud, que un sistema de partículas bajo condiciones  $N,p,T \quad\textrm{contantes}$ se describe adecuadamente con el Potencial de Gibbs (también denominado Energía Libre de Gibbs), $G\equiv G(N,p,T)$,
  
  \begin{equation}\label{eq.1.17}
  G=U-TS+pV 
  \end{equation}
  
  \noindent y la Función de Partición de Gibbs, $\varGamma\equiv\varGamma(N,p,T)$  
  \begin{equation}\label{eq.1.17.a}
  \varGamma=\sum_{j} e^{-\beta H_{j}}
  \end{equation}
  
  \noindent donde $H_{j}=E_{j} + p V_{j} $ es la entalpía del estado o configuración $j$, 
  
  \begin{equation}\label{eq.1.18}
  G=-\frac{1}{\beta} \ln \varGamma
  \end{equation} 
  
  \begin{equation}\label{eq.1.18.a}
  f_{j}=\frac{e^{-\beta H_{j}}}{\varGamma}
  \end{equation}
  
  \begin{equation}\label{eq.1.18.c}
   U=-\left( \frac{\partial\ln \varGamma}{\partial \beta}\right) =\frac{ \partial\left(  \beta G\right) }{\partial \beta}
  \end{equation}
  
  \noindent siendo $\Delta G\leq 0$ y $G=\textrm{mínimo}$ las condiciones de aproximación al equilibrio y equilibrio, respectivamente. \\
  
  \section{Estadísticas Cuánticas: Bose-Einstein y Fermi-Dirac}
  
  Si consideramos un sistema de $N$ partículas idénticas en un volúmen $V$ a temperatura $T$ y no interaccionan entre sí,   $E(N)=\sum_{i=1}^{N} \epsilon_{i}$ es la energía total,  $z_{i}$ es la función de partición de la partícula $i$, entonces
  
  \begin{equation}\label{eq.1.19}
  Z(N,V,T)\propto z_{i}^{N}
  \end{equation} 
  
   \noindent Si asumiéramos sin más que esto es una igualdad, $Z=z_{i}^{N}$, y considerando que $z_{i}$ para cada partícula en el volúmen $V$ es proporcional a $V$,$z_{i}=c V$ con $c= \textrm{constante}$, entonces la energía libre de Helmholz $F$ 
   
   \begin{equation}\label{eq.1.20}
   F(N,V,T)=-\frac{1}{\beta} \ln Z=-\frac{1}{\beta} \ln z_{i}^{N}=-\frac{1}{\beta} N \ln c V
   \end{equation}
   \noindent \textbf{no sería una cantidad extensiva} como debe ser ya que si en la ec. \eqref{eq.1.20} hacemos el cambio de escala $N\rightarrow2N$ y $V\rightarrow2V$ (con $N/V \quad\textrm{cte}$)

  \begin{equation}\label{eq.1.21}
  F(2N,2V,T)\neq 2 F(N,V,T)
  \end{equation} 
  
  \noindent Hace falta dividir $z_{i}^{N}$ por $N!$ porque las $N$ partículas ``clásicas'' son indistinguibles entre sí y cualquier permutación entre ellas no resulta en un nuevo estado del sistema. De aquí que, 
\begin{equation}\label{key}
F(N,V,T)=-\frac{1}{\beta} \ln   \frac{\left( c V\right)^{N} }{ N! }
\end{equation}  
  \begin{equation}\label{eq.1.22.a}
  F(2N,2V,T)=-\frac{1}{\beta} \ln \left(  \frac{\left( c 2V\right)^{2N} }{\left( 2 N!\right) }\right)  = -\frac{1}{\beta} \left(  2 N\ln 2 c V -\ln 2 N!\right) 
  \end{equation}
   \noindent y aproximando $\ln\approx N \ln N -N\qquad \textrm{para} N\to \infty$ 
   
   \begin{equation}\label{eq.1.22.b}
   F(2N,2V,T)=-\frac{1}{\beta}\left(2N \ln 2cV -2N \ln 2N + 2N\right) 
   \end{equation}
   
   \begin{equation}\label{eq.1.22}
   F(2N,2V,T)=2 \quad F(N,V,T) \qquad \qquad \textrm{Extensiva ¡¡}
   \end{equation}
   
   \noindent  Tuvimos que introducir $N!$ para lograr que la termodinámica estadística de partículas ``clásicas'' (ideal) sea consistente.
  
    Este problema desaparece cuando consideramos que las partículas de un sistema son siempre de naturaleza intrínseca cuántica; son fermiones o bosones y eso determina su propiedades termodinámicas. No puede haber más de un fermión en cada estado cuántico como consecuencia de que la función de onda de $N$ fermiones debe ser antisimétrica ante intercambio de dos fermiones $i,j$ cualesquiera
   
   \begin{equation}\label{eq.antisim}
   \psi\left(q_{1},..,q_{i},...,q_{j},..,q_{N}\right)=-\psi\left(q_{1},..,q_{j},...,q_{i},..,q_{N}\right)
   \end{equation} 
   
   \noindent donde $q_{i},q_{j}$ son coordenadas generalizadas que definen el estado de los fermiones $i,j$. \\
     
   \noindent Si $i,j$ estan en el mismo estado $\psi=0$.\\
   
  sin embargo para bosones dado que ya que la función de onda en este caso es simétrica
  
  \begin{equation}\label{e.q.sim}
  \psi\left(q_{1},..,q_{i},...,q_{j},..,q_{N}\right)=\psi\left(q_{1},..,q_{j},...,q_{i},..,q_{N}\right)
  \end{equation} 
   
  \noindent y nada impide que haya cualquier número de partículas en el mismo estado.\\
  
   \noindent Consideramos ahora un gas ideal de partículas cuánticas en el Conjunto Macrocanónico $(\mu,T,V)$. Si el gas se encuentra en un estado de energía total $E_{j}$ con $N_{j}$ tal que $n_{1}$ partículas están en el estado $\epsilon_{1}$, $n_{2}$ en $\epsilon_{2}$, etc. , con $N_{j}=\sum_{i} n_{i}$ y $E_{j}=\sum_{i} n_{i}\epsilon_{i}$, luego
   
   \begin{equation}\label{e.q.1.24}
   \varXi=\sum_{j} e^{-\beta(E_{j}-\mu N_{j})}=\sum_{j} e^{-\beta\sum_{i}\left( \epsilon_{i}-\mu\right) n_{i}}
   \end{equation}
   
   \noindent entonces $\varXi$ se factoriza en término de cada los estados individuales de las partículas
    
   \begin{equation}\label{e.q.1.25}
   \varXi(\mu,T,V)=\prod_{i} \vartheta_{i}   \qquad\qquad \textrm{con} \qquad \vartheta_{i}=\sum_{n} e^{-\beta\left( \epsilon_{i}-\mu\right) n}
   \end{equation}

\noindent Para fermiones el número de ocupación de cada estado puede ser $n_{i}=0,1$ y conduce a 
\begin{equation}\label{e.q.1.26}
\vartheta_{i}(\mu,T,V)=1+ e^{-\beta\left( \epsilon{i}-\mu\right)} 	
\end{equation}

\noindent El número medio de ocupación del estado con energía $\epsilon_{i}$, $\overline{n_{i}}$  (equivalente al cubrimiento medio de un sitio en un gas de red de monómeros ideales) es 
\begin{equation}\label{e.q.1.27}
\overline{n_{i}}=\frac{\sum_{{0,1}} n_{i} \ e^{-\beta\left( \epsilon{i}-\mu\right) n_{i}}}{\vartheta_{i}}=\frac{e^{-\beta\left( \epsilon{i}-\mu\right)}}{1+ e^{-\beta\left( \epsilon{i}-\mu\right)}}
\end{equation}
\noindent denominada distribución de \textbf{Fermi-Dirac} \cite{Fermi,Dirac}.

\begin{figure}[H]
	\centering
	\includegraphics[width=1\linewidth]{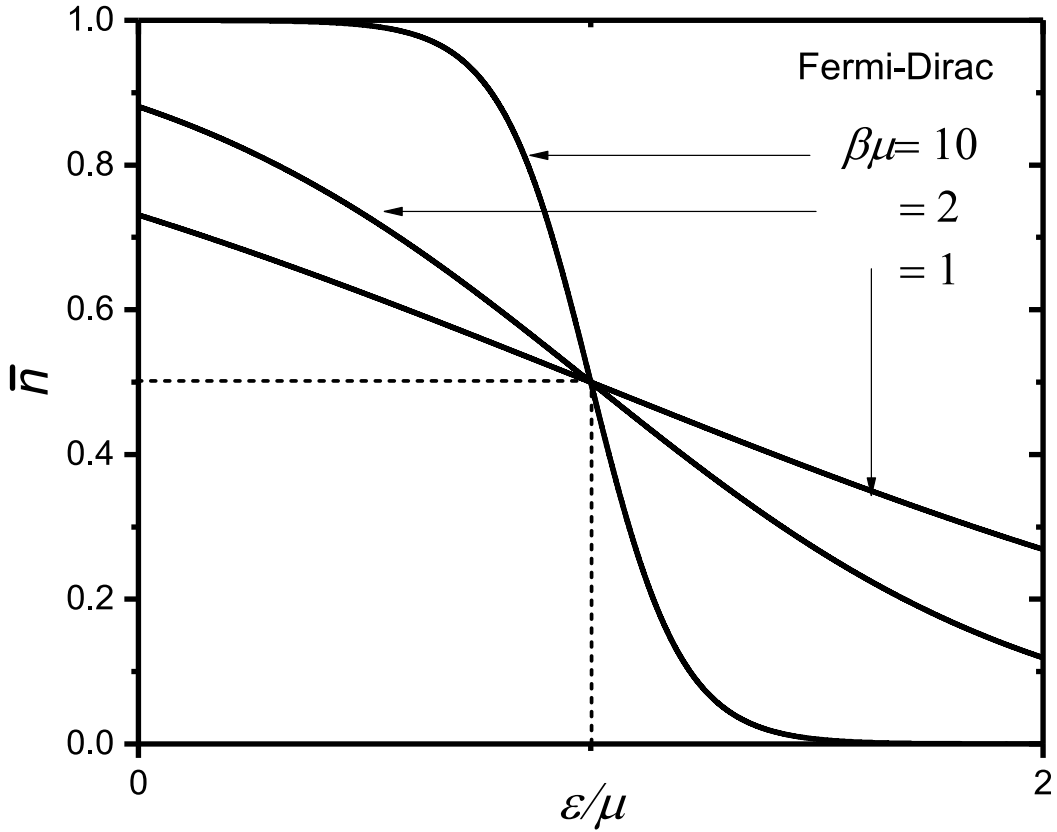}
	\caption{Distribución de Fermi\textrm{-}Dirac. Ocupación media de estados $\overline{n}(\epsilon)$ versus $\epsilon/\mu$ para distintas temperaturas, $\beta \mu=1,2 \ \textrm{y} \ 10$. $\overline{n}\left( \epsilon=\mu\right) =\nicefrac{1}{2} $. Es implícito que $\mu, \epsilon, 1/\beta$ están expresados en la misma unidad (energía)}
	\label{fig.1}
\end{figure}

\noindent $\overline{n}\left( {\epsilon}=\mu\right) =\nicefrac{1}{2}$ $\forall \quad T$ y es simétrica respecto de $\nicefrac{\epsilon}{\mu}=1$ y $\overline{n}_{\epsilon}=0.5$. En el límite $(T\to 0)$ todos los estados con $\epsilon<\mu$ están ocupados y los de $\epsilon>\mu$ vacíos; $ \epsilon_{F}=\mu$ es el Nivel de Fermi. 

Para bosones la ocupación de estados $n_{i}$ puede tomar valores $n_{i}=0,1,2,...$, ya que no hay límite para el número de partículas en un dado estado $i$. De aquí que 

\begin{equation}\label{e.q.1.28}
\vartheta_{i}(\mu,V,T)=\sum_{n=0}^{\infty} e^{-\beta \left( \epsilon_{i} n-\mu n\right)}=\sum_{n=0}^{\infty} \left[ e^{-\beta \left(\epsilon_{i} -\mu \right)}\right]^n 
\end{equation} 

\noindent y sumando la serie $\sum_{n=0}^{\infty} x^n=\nicefrac{1}{(1-x)}$
\noindent

\begin{equation}\label{e.q.1.29}
\vartheta_{i}(\mu,V,T)=\frac{1}{1- e^{-\beta\left( \epsilon{i}-\mu\right)}}
\end{equation}

\noindent y 
\begin{equation}\label{e.q.1.30}
\overline{n_{i}}=\frac{1}{e^{\beta\left( \epsilon{i}-\mu\right)}-1}  \qquad \qquad \textrm{ver fig. \ref{fig.2} }
\end{equation}

\noindent denominada distribución de Bose-Einstein \cite{Bose.2,Einstein.1,Einstein.2,Einstein.3}.

\begin{figure}[H]
	\centering
	\includegraphics[width=1\linewidth]{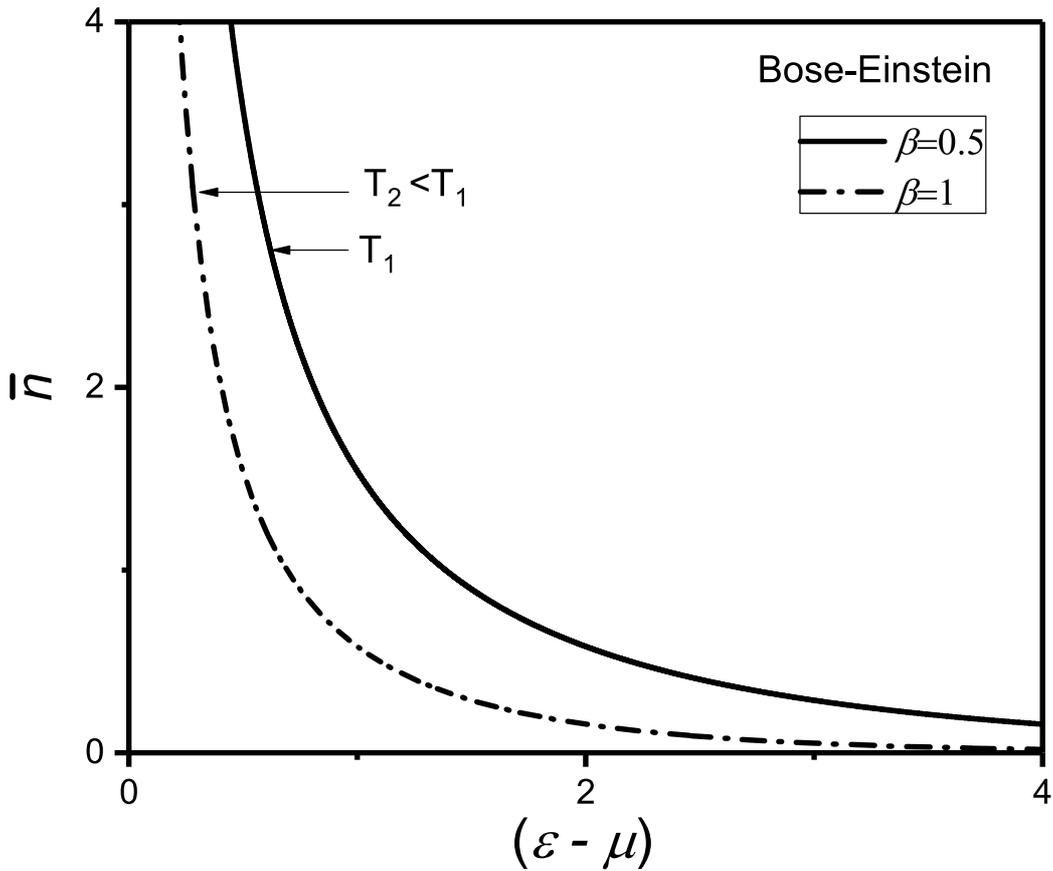}
	\caption{Distribución de Bose\textrm{-}Einstein. Ocupación media de estados $\overline{n}(\epsilon)$ versus $\left(\epsilon-\mu\right)$ para distintas temperaturas, $\beta=0.5 \ \textrm{y} \ 1$. Se asume que$\beta$, $\mu$ y $\epsilon$ están expresados en las mismas unidades}
	\label{Fig.2}
\end{figure}

\noindent Finalmente 

\begin{equation}\label{e.q.1.31}
\varXi=\prod_{i} \left[ 1\pm e^{-\beta\left( \epsilon{i}-\mu\right)}\right]^{\pm 1} 
\end{equation}

\begin{equation}\label{e.q.1.32}
\overline{n_{i}}=\frac{1}{e^{\beta\left( \epsilon_{i}-\mu\right)}\pm 1}
\end{equation}

\noindent donde $(+) $ corresponde a la distribución de \textbf{Fermi-Dirac (FD)} y $(-)$ a la de \textbf{Bose-Einstein (BE)}. 

\begin{figure}[H]
	\centering
	\includegraphics[width=1\linewidth]{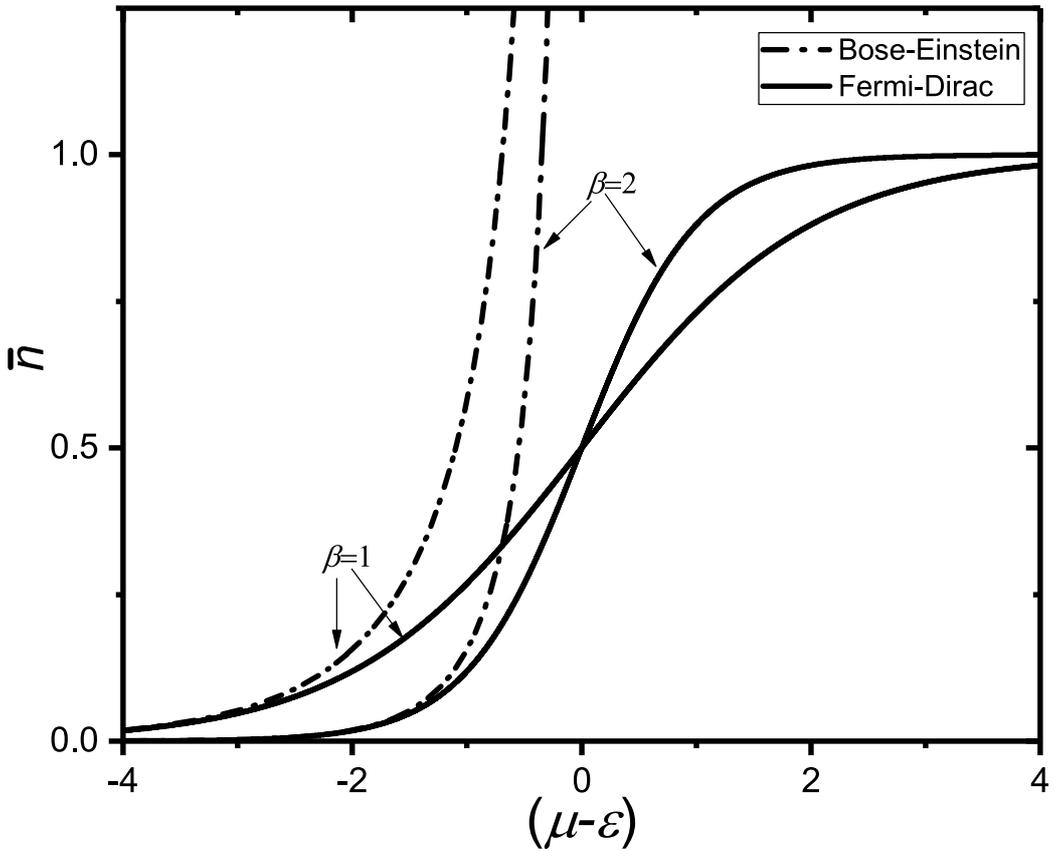}
	\caption{Comparación de las distribuciones de Fermi\textrm{-}Dirac y Bose\textrm{-}Einstein; $\overline{n}(\epsilon)$ versus $\left(\mu-\epsilon\right)$ para distintas temperaturas, $\beta=1 \ \textrm{y} \ 2$. Se asume que $\beta$, $\mu$ y $\epsilon$ están expresados en las mismas unidades}
\label{Fig.FDBE}
\end{figure}

\noindent En la fig. \ref{Fig.FDBE} comparamos el comportamiento de estas dos distribuciones en función del potencial químico y la temperatura. 
   
La ec. \eqref{e.q.1.32} se puede escribir de la forma

\begin{equation}\label{e.q.1.32.a}
\overline{n_{i}}=\frac{\lambda e^{-\beta \epsilon_{i}}}{1\pm \lambda e^{-\beta \epsilon_{i}}}
\end{equation}

\noindent donde $ \lambda=e^{\beta \mu}\equiv \textrm{fugacidad del gas}$. Cuando $\lambda\equiv e^{\beta \mu} \ll 1$, es el límite clásico de las estadísticas de \textbf{FD} y \textbf{BE} (densidad muy baja o temperatura muy alta), y 
\begin{equation}\label{e.q.1.32.b}
\overline{n_{i}}=\lambda e^{-\beta \epsilon_{i}}
\end{equation}

\noindent que es la distribución de Boltzmann  

\begin{equation}\label{e.q.1.32.c}
\varPsi(\mu,T,V)=-\frac{1}{\beta} \left( \pm\right)  \sum_{i} \pm \lambda e^{-\beta \epsilon_{i}}=\lambda \sum_{i} e^{-\beta \epsilon_{i}}=\lambda z
\end{equation}

\noindent luego $\lambda z$ es igual a $\ln \varXi $ y

\begin{equation}\label{e.q.1.32.d}
\varXi=e^{\lambda z}= \sum_{N=0}^{\infty} \frac{\left( \lambda z\right) ^{N}}{N!}= \sum_{N=0}^{\infty} Z_{N} \: e^{\beta \mu N}
\end{equation}

\noindent con $Z_{N}= z^{N}/N!$. Esto muestra que $N!$ aparece de forma natural en el límite clásico de las estadísticas cuánticas.\\

El problema de ocupación de estados por fermiones bajo condiciones $\mu,T,V$ es isomorfo al de un gas de red con partículas que ocupan un solo sitio de la red. Las partículas se asemejan a fermiones ocupando los sitios (estados posibles).  En este caso, dado que las partículas no tienen interacción, los sitios son independientes entre sí. El número medio de ocupación de estados $\overline{n}$ de la ec. \eqref{e.q.1.27} corresponde al cubrimiento medio de sitios (fracción ocupada) $\theta$ a potencial químico $\mu$ y temperatura $T$
\begin{equation}\label{e.q.1.32.e}
\overline{n}=\theta(\mu,T)= \frac{1}{e^{\beta\left( \epsilon_{i}-\mu\right)}+ 1}=\frac{e^{-\beta\left( \epsilon_{i}-\mu\right)}}{1+e^{-\beta\left( \epsilon_{i}-\mu\right)}}
\end{equation}
  
\noindent llamada isoterma de Langmuir \cite{ref36}, que ha sido muy útil para comprender el fenómeno de adsorción de moléculas sobre superficies. 

%% file: texto/TF_Estadistica_Fraccionaria.tex
\chapter{Estadística Fraccionaria para Adsorción}
El problema de muchas partículas que interaccionan es el problema central de la mecánica estadística que ha atraído la atención por décadas ya que son muy pocos los casos de sistemas que tienen solución exacta basadas en modelos muy simplificados de las partículas y sus interacciones . En contraste,la inmensa mayoría de los sistemas reales de muchas partículas que se interpretan con la teoría mecánica estadística contienen partículas que están compuestas por dos o más átomos o grupos químicos y se alejan del caso ideal de partículas que pueden ser consideradas como puntuales o esféricamente simétricas.

\noindent En este trabajo modelaremos el fenómeno de adsorción de moléculas poliatómicas mediante un gas de red. Los fenómenos de adsorción, difusión, agregación de moléculas o reacción de especies químicas son fundamentales en muchas aplicaciones tecnológicas como la separación de gases, catálisis heterogénea, crecimiento de películas, lubricación, etc. La mayoría de los gases o especies en los fenómenos reales de interacción gas-sólido son poliatómicos $(\ce{O2},\ce{N2},\ce{CO},\ce{CO2},\ce{CH4},\ce{C2H6},\ce{C3H8})$. Sin embargo los modelos analíticos e interpretación de experimentos se basan en su mayoría en modelos de partículas ideales con simetría esférica. 

\noindent  Por otra parte, cuando tratamos con partículas o moléculas más complejas como dímeros $(\ce{O2},\ce{N2})$, cadenas lineales, hidrocarburos o proteínas, su estructura espacial no esférica nos enfrenta al problema entrópico. En definitiva se trata primero de calcular el número de configuraciones posibles $\varOmega$ de muchas partículas con estructura geométrica no esférica y $S= k_{B} \ln\varOmega$. 

\noindent Lo que hace a este problema desafiante e interesante es la complejidad de aproximar o estimar adecuadamente la función $S$. Además no solo estamos interesados en desarrollar una solución para un modelo particular de adsorción de partículas con tamaño y estructura geométrica sino que nuestro objetivo es definir un modelo para este fenómeno suficientemente general y matemáticamente sencillo para poder interpretar un amplio conjunto de sistemas gas-sólido reales. 

\noindent Existen evidencias experimentales sobre equilibrio y cinética de adsorción de alcanos que muestran que es necesario tomar en cuenta la estuctura espacial de las especies adsorbidas y los efectos entrópicos\cite{ref40}.

\section{Estadística Fraccionaria de Exclusión de Haldane}{\label{Estadistica Fraccionaria}}

En 1991 Haldane \cite{haldane1983,ref33} propuso una generalización de las estadísticas cuánticas de \textbf{FD} y \textbf{BE} denominada Estadística Fraccionaria motivado por describir cuasi partículas en sistemas con efecto Hall cuántico fraccionario  y espinones en cadenas de espines antiferromagnéticos \cite{ref37,ref38,EHCFNob}. Existen otras formulaciones estadísticas fraccionarias como la de Gentile \cite{ref34ppp} y Poychronakos \cite{ref34pp} que comentaremos al final del capítulo a modo de referencia. 

El formalismo de Haldane esta basado en una generalización del Principio de Exclusión de Pauli. En un sistema de partículas con $G$ estados disponibles para cada partícula aislada en un volúmen $V$ del espacio, el número de estados accesibles para la $N\textrm{-ésima}$ partícula luego de que $(N-1)$ partículas han ocupado estados del total $G$ en $V$, es 

\begin{equation}\label{e.q.2.1}
d\left( N\right) =G-\sum_{N=1}^{N-1} g(N)=G-G_{0}\left( N\right) 
\end{equation} 

\noindent donde $g(N)$ es el número de estados que excluye la partícula $N$ y se denomina \textbf{parámetro de exclusión estadística} definido en la ec. \eqref{e.q.2.1} por la relación

\begin{equation}\label{eq.2.2}
g(N)=- \frac{\Delta d(N)}{\Delta N}  
\end{equation} 

\noindent que representa una generalización del Principio de Exclusión de Pauli, donde $d(N)$ es formalmente la dimensión del espacio de Hilbert de una partícula (número de estados accesibles) cuando ya existen $(N-1)$ partículas en el sistemas y $0\leq g(N)\leq1$ para sistemas con efectos quánticos fraccionarios e interpola entre bosones y fermiones. Haldane  en su trabajo seminal \cite{ref33} consideró $g(N)=g=$constante. A las partículas o excitaciones con $0<g<1$ se las suele  denominar genéricamente \textbf{g-ons} o \textbf{anyons}.

\noindent El número de configuraciones para $N$ partículas en $d(N)\equiv d_{N}$ estados es \cite{ref33}

\begin{equation}\label{eq.2.3}
W(N)= \frac{\left( d_{N}+N-1\right) !}{N! \left( d_{N}-1\right) !}
\end{equation}

\noindent Si $g(N)=1$ entonces la relación \eqref{eq.2.3} se reduce al factor configuracional de $N$ fermiones en $G$ estados 

\begin{equation}\label{e.q.2.3.a}
W_{F}(N)= \frac{G!}{N! \:(G-N)!}
\end{equation}

\noindent Análogamente si $g(N)=0$, recuperamos el factor configuracional de bosones

\begin{equation}\label{e.q.2.3.b}
W_{B}(N)= \frac{(G+N-1)!}{N!\: (G-1)!}
\end{equation}

\noindent La función de partición canónica de $N$ partículas , cada una con energía $\epsilon$  y parámetro de exclusión $g(N)$ es (ec. \eqref{eq.1.5}) 

\begin{equation}\label{eq.2.4}
Z(N,T,G)= \sum_{j} e^{-\beta E_{j}} q^{N}=\sum_{j} e^{-\beta N\epsilon} q^{N}=  e^{-\beta N\epsilon} q^{N} \sum_{j} 1= W(N) e^{-\beta N \epsilon} q^{N}
\end{equation}

\noindent donde el índice $j$ corre sobre todas las posibles $W(N)$ configuraciones, $q$ es la función de partición interna de las partículas, a la que le asignamos $q=1$ por simplicidad,  y el volúmen V esta parametrizado por el numero total de estados $G$.  La Energía Libre de Helmholtz es

\begin{equation}\label{e.q.2.5}
\beta F(N,T,G)=- \ln Z(N,T,G)= \beta \left( N\epsilon- \frac{1}{\beta} \ln W(N)\right) 
\end{equation}

\noindent Si asumimos que $g(N)=g= \textrm{constante}$, el número medio de partículas por estado en función del potencial químico $\mu$ y temperatura $T$ surge del formalismo canónico 

\begin{equation}\label{eq.2.6}
\overline{n}(\mu,T)= \frac{1}{\omega + g}
\end{equation}

\noindent donde $\omega$ satisface la ecuación trascendente 

\begin{equation}\label{eq.2.7}
\omega^{g} \left[1+\omega\right]^{1-g}=e^{\beta \left( \epsilon-\mu\right) }\equiv \xi  
\end{equation}

\noindent Por claridad, prescindimos de aquí en más de la notación $\overline{n}(\mu,T)$ para el número medio de ocupación de estado y designamos a ésta cantidad simplemente $n(\mu,T)$ como en la ec. \eqref{eq.2.6}.

\noindent La función \eqref{eq.2.6} es conocida como distribución de Wu \cite{ref39}, quien la derivó de esta forma a partir de la generalización de Haldane \cite{ref33}.  Para $g=0 \quad \omega(\xi)=\xi-1$ y $g=1 \quad \omega(\xi)=\xi$ y la ec. \eqref{eq.2.6} conduce a las distribuciones de \textbf{BE}  y \textbf{FD} (ecs. \eqref{e.q.1.31}).

\noindent Cuando $\xi\gg 0$  $\omega(\xi)\approx\xi$ y con $g\ll \omega(\xi)$ la ec. \eqref{eq.2.6} reproduce la distribución de Boltzmann $n_{B}(\mu,T)$ 

\begin{equation}\label{eq.2.8}
n_{B}(\mu,T)= e^{-\beta\left( \epsilon-\mu\right) }=\xi^{-1}
\end{equation}

\noindent Además como $\xi\gg0$ y $\omega\gg0$ también resulta 

\begin{equation}\label{eq.2.8.a}
n \le 1/g
\end{equation}

\noindent  En el límite clásico $T\to 0$
\begin{equation}\label{eq.2.7.n}
\xi= \begin{cases} 0 & \mbox{si } \epsilon<\mu  \\ +\infty & \mbox{si } \epsilon>\mu \end{cases} \implies \begin{cases} \omega=0  \\ \omega \to \infty\end{cases}
\end{equation}

\noindent En conclusión $n \ \textrm{a} \ T=0 \ \textrm{y} \ g\neq 0$ (tipo fermiones) tiene forma de escalón (análoga a la distribución de Fermi a $T=0$ )

\begin{equation}\label{eq.2.7.nb}
n= \begin{cases} 0 & \mbox{si } \epsilon>\epsilon_{F}  \\ 1/g & \mbox{si } \epsilon<\epsilon_{F} \end{cases} 
\end{equation}

\noindent donde la energía de Fermi $\epsilon_{F}$ esta determinada por la condición $G=gN$. En la ec. \eqref{eq.2.7}, $\epsilon_{F}=\mu$.

Es útil ver como se comporta el número medio de partículas $n$ para diferentes valores de exclusión $g$. Para esto tomamos las ecs. \eqref{eq.2.6} y \eqref{eq.2.7} y resolvemos $\beta\mu$ en función de $n$

\begin{equation}\label{eq.2.8.n}
e^{\beta(\mu-\epsilon)}=\frac{n \left[ 1-n\left( g-1\right) \right]^{g-1}}{\left[ 1-n g\right]^{g}}
\end{equation}

\noindent o  

\begin{equation}\label{e.q.2.9}
\beta \mu= \ln n + (g-1) \ln \left[ 1- n (g-1)\right]- g \ln \left[ 1-n g\right]+\beta \epsilon  
\end{equation}

\begin{figure}[H]
	\centering
	\includegraphics[width=1\linewidth]{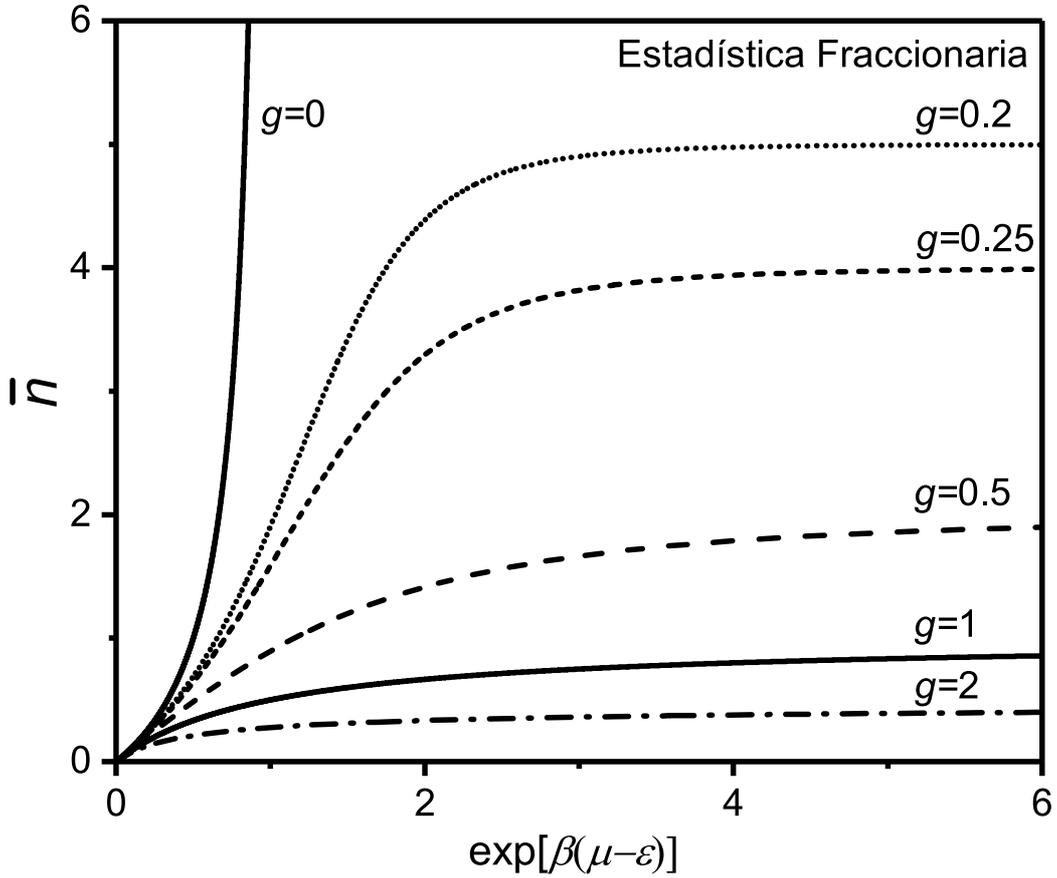}
	\caption{Estadistica Fraccionaria de Haldane. Ocupación media de estado $n$ versus $\beta(\mu-\epsilon)$ para $g=0,0.2,0.25,0.5,1$  a partir de la ec. \eqref{eq.2.8.n}. Como referencia, las líneas sólidas corresponden a los casos $g=0$ (BE) y $g=1$ (FD). Se muestra también el caso ``no cuantico'' $g=2$.}
	\label{Fig.Estadistica_Fraccionaria}
\end{figure}
\noindent La fig. \ref{Fig.Estadistica_Fraccionaria} nos muestra el comportamiento de la ocupación media de estados versus $\beta(\mu-\epsilon)$. La función varía desde un comportamiento tipo fermión, para $\nicefrac{1}{2}<g<1$, a uno tipo bosón para  $0<g<\nicefrac{1}{2}$.  

\begin{figure}[H]
	\centering
	\includegraphics[width=1\linewidth]{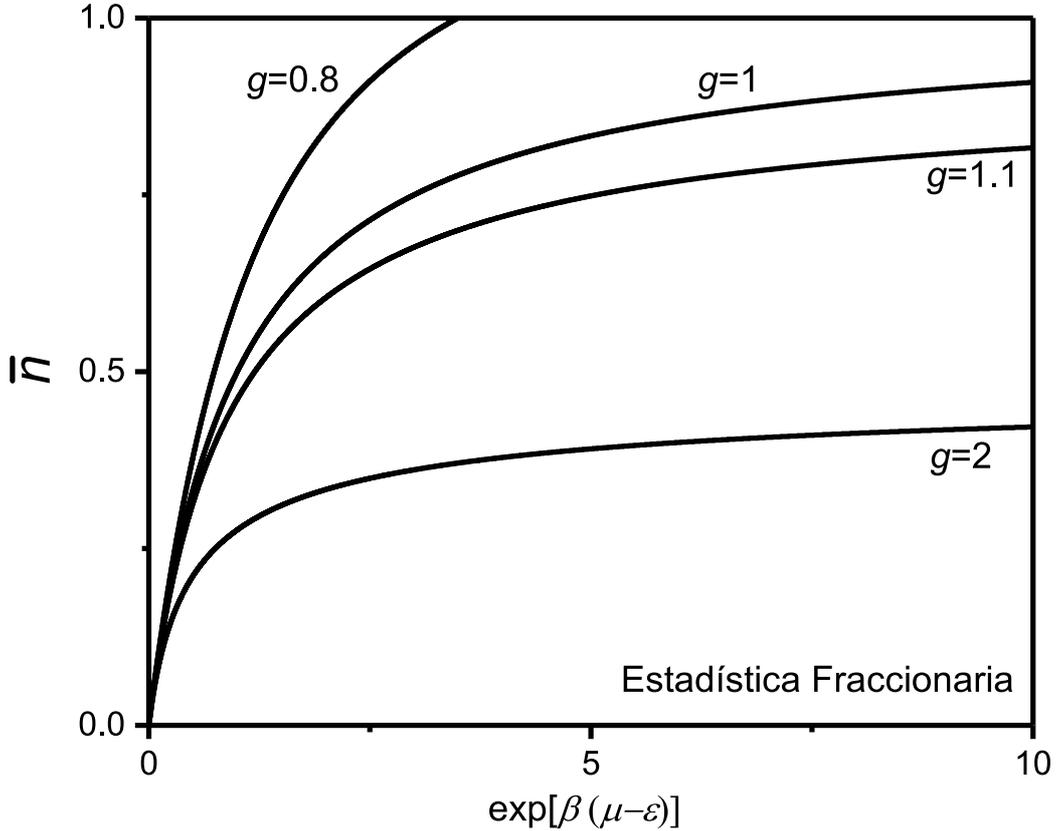}
	\caption{Estadistica Fraccionaria de Haldane. Ocupación media de estado $n$ versus $\beta(\mu-\epsilon)$ para partículas con $g>1$ (tipo superfermiones); $g=1,1.1$ y $2$ a partir de la ec. \eqref{eq.2.8.n}. Se muestra también el caso $g=$0.8 (subfermión).}
	\label{Fig.2.9.a}
\end{figure}

\noindent También representamos en la fig. \ref{Fig.2.9.a} los casos para $g>1; g=1.1 \ \textrm{y} \ g=2$ que no representan sistemas cuánticos. Sin embargo los casos $g>1$, que correspondería a una especie de \textbf{superfermiones} ya que son partículas que excluyen más de un estado, serán valiosos para interpretar el comportamiento termodinámico de gases poliatómicos clásicos como estudiaremos en el capítulo que sigue. 

\noindent Existen otras formulaciones de estadísticas cuánticas fraccionarias a las que solo nos referimos aquí sin desarrollar. La primera fue propuesta por Gentile\cite{ref34ppp} con la condición de que no puede haber mas de $p$ partículas por estado cuántico. Por otra parte Polychronakos \cite{ref34pp}  a través de una formulación basada en integrales de camino en el espacio de fases de un sistema cuántico de muchas partículas.En definitiva las formulaciones de Haldane, Polychronakos y Gentile son diferentes formas de ocupar el conjunto de estados accesibles para partículas que tienen parámetros de exclusión o interacciones estadísticas diferentes en cada caso.

\section{Extensión del formalismo de Haldane al problema de adsorción de k-meros}

En este capítulo presentamos las bases de la descripción termodinámica estadística fraccionaria de adsorción de moléculas poliatómicas o polifuncionales en un nuevo marco teórico presentado en las ref. \cite{ref34} que surge como extensión de los conceptos y formalismo de Haldane \cite{ref33}.

\noindent Esta es una descripción simple y elegante para obtener formas aproximadas de las funciones termodinámicas de un problema complejo de la mecánica estadística aplicable a un extenso conjunto de sistemas gas-sólido. Como corolario, es posible relacionar el modelo con los experimentos para obtener información físicamente significativa acerca del estado y configuración espacial de las moléculas adsorbidas y el desarrollo de fases ordenadas como consecuencia de las interacciones laterales. Esta información microscópica es algo que en la actualidad solo es accesible a partir de elaboradas técnicas espectroscópicas. 

\begin{figure}[H]
	\centering
	\includegraphics[width=1\linewidth]{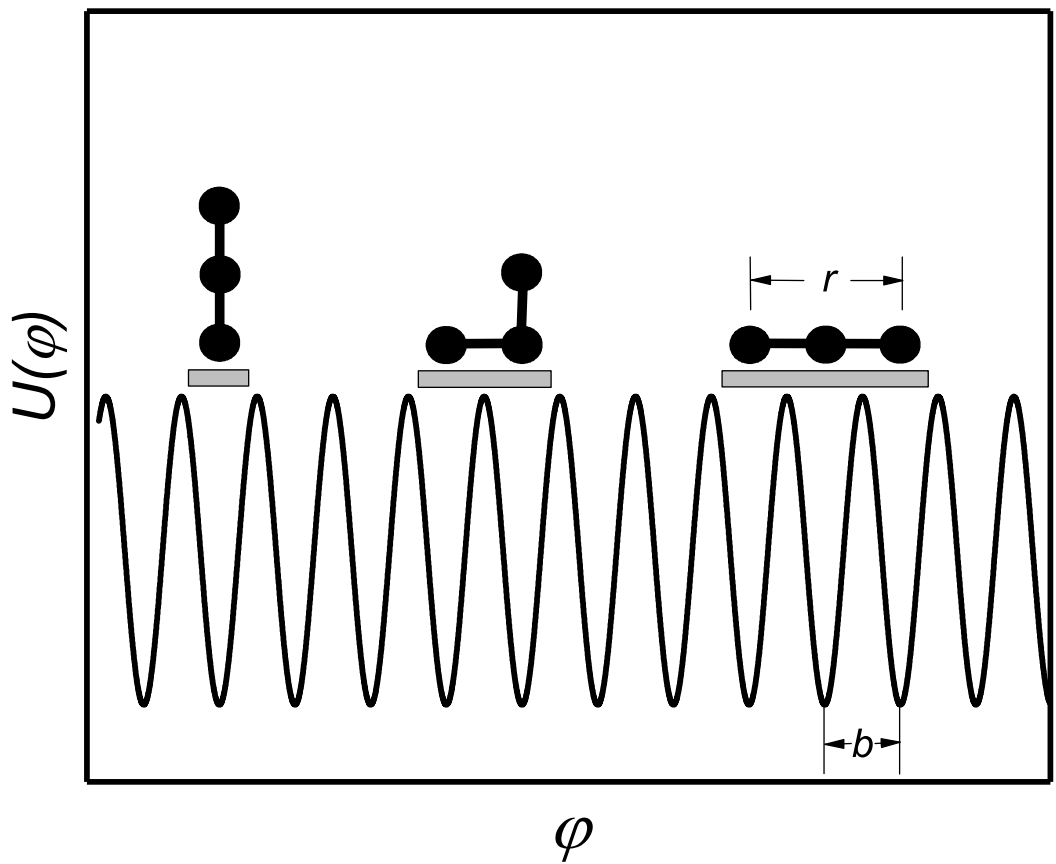}
	\caption{Representación esquemática de exclusión de estados para distintas configuraciones de trímeros adsorbidos en un potencial de adsorción $U_{o}(\varphi)$ donde $\varphi$ representa una coordenada paralela a una superficie regular. Por claridad, no se ha considerado que los minimos locales de adsorción debería tener valores diferentes para cada configuración fija de adsorción de los trimeros. Las configuraciónes de trimero vertical, horizontal y ángulo  corresponden $g=1,3 \ \textrm{y} \ 4$, respectivamente.}
	\label{Fig.potencial_de_adsorcion_exclusion}
\end{figure}

\noindent Una molécula aislada interactuando con la superficie de un material sólido regular confinado en un volúmen $V$ del espacio se puede representar por un campo de potencial de adsorción $U(\varphi)$ con un total de $G$ mínimos locales de adsorción, donde $\varphi$ representa alguna dirección paralela a la superficie adsorbente donde aparece la periodicidad del potencial $U(\varphi)$ y homogeneidad de los mínimos locales. . Cada mínimo local representa un posible estado de adsorción como se representa en la fig. \ref{Fig.potencial_de_adsorcion_exclusion}

\noindent En general, para una partícula de tamaño típico $r$ compuesta por varios átomos o grupos, la distancia separacion $b$ entre dos mínimos locales vecinos será $b<r$. Por lo tanto la ocupación de un mínimo local por una partícula puede \textbf{excluir} que otra partícula idéntica ocupe los mínimos vecinos como se representa en el esquema de la Fig. \ref{Fig.potencial_de_adsorcion_exclusion}. En general, la ocupación de un estado por una molécula \textbf{excluye} un número $g$ de estados del total $G$ que no pueden ser ocupados por otra molécula idéntica.

\noindent Análogamente al desarrollo de la sección \ref{Estadistica Fraccionaria} definimos $g$ como el número de estados excluidos por una molécula (partícula) adsorbida (o incorporada el volúmen $V$), de un total de $G$ estados. 
Denominamos a $g$ \textbf{parámetro de exclusión estadística} y es una medida de las \textbf{interacciones estadísticas} entre partículas debido a su tamaño,  estructura geométrica espacial y en general también a los potenciales de interacción entre ellas. 

\noindent Aunque hemos denominado por simplicidad a $g$ como parámetro, en esta extensión conceptual del formalismo de Haldane a sistemas moleculares clásicos, su valor dependerá en general del número de partículas presentes en el volúmen $V$, $g\equiv g(N)$ como discutiremos más adelante.  Esto no parece necesario en sistemas cuánticos pero si es crucial en el tratamiento de los fenómenos de interés en este trabajo. En contraste a los sistemas cuánticos, para los cuales $0\leq g\leq1$, en ésta formulación  $g\geq 1$ \cite{ref34}. En este caso las partículas se comportan como \textbf{superfermiones} y la estadística resultante es \textbf{suprafraccionaria}. 

\noindent Dadas $(N-1)$ partículas idénticas en un volúmen $V$ con $G$ estados accesibles a cada una de ellas en forma aislada, el número de estados accesibles para la inclusión de la $N$-ésima partícula en $V$, $d_{N}$, es 

\begin{equation}\label{eq.3.1}
d_{N}\equiv d(N)=G-\sum_{N^{'}=1}^{N-1} g(N^{'})=G-G_{o}(N)
\end{equation}

\noindent El número de configuraciones de $N$ moléculas en $G$ estados y exclusión $g(N)$ (análogo a la ec. \ref{eq.2.3}) es 

\begin{equation}\label{eq.3.2}
W(N)= \frac{\left( d_{N}+N-1\right) !}{N! \ \left( d_{N}-1\right) !} 
\end{equation}

\noindent Si $U_{o}$ es la energía de adsorción por partícula, y no hay interacciones laterales, entonces las funciones del Conjunto Canónico son

\begin{equation}\label{eq.3.3}
Z(N,T,G)= W(N)  \: e^{-\beta N U_{o}}
\end{equation}

\begin{equation}\label{eq.3.4}  
\beta F(N,T,G)= -\ln Z(N,T,G)
\end{equation}

\begin{equation}\label{eq.3.5}
\mu\equiv\mu(N,T,G)=\left( \frac{\partial F}{\partial N}\right)_{T,G}
\end{equation}

\noindent Combinado las ecs. \ref{eq.3.2} , \ref{eq.3.3} y \ref{eq.3.4} ,junto con la aproximación de Stirling, obtenemos $F$ 

\begin{equation}\label{eq.3.6}
\begin{aligned}
\beta F(N,T,G)=\beta N U_{o}&-\left(d_{N}+N-1 \right) \ln\left(d_{N}+N-1 \right)+ \left( d_{N}-1\right) \ln \left( d_{N}-1\right)\\ 
&+ N \ln N 
\end{aligned}
\end{equation}

\noindent y la entropía $S$  

\begin{equation}\label{eq.3.7}
\frac{S(N,T,G)}{k_{B}}= \left(d_{N}+N-1 \right) \ln\left(d_{N}+N-1 \right) - \left( d_{N}-1\right) \ln \left( d_{N}-1\right) - N \ln N 
\end{equation}

\noindent De la ec. \eqref{eq.3.5} obtenemos la isoterma de adsorción en la representación estadística fraccionaria.

\begin{equation}\label{eq.3.8}
K(T) \: e^{\beta \mu}=\frac{n \: [1-\tilde{G_{o}}\left(n \right) +n]^{\tilde{G_{o}}^{'}-1}}{[1-\tilde{G_{o}}(n) ]^{\tilde{G_{o}}^{'}}}
\end{equation}

\noindent donde $n= \lim_{N,G\to \infty} N/G$ es la ocupación media de estados y $K(T)= q e^{-\beta U_{o}}$. A la cantidad $n$ podemos interpretarla como una densidad relativa que se puede relacionar directamente con el cubrimiento superficial de un gas de red $\theta$ o su análogo en otros modelos. Asi $n= a\theta$, con $\theta=N/N_{m} \ \textrm{o} \ (v/v_{m})$ con $N_ {m} \ (v_{m})$ siendo el valor de $N \ (v)$ en la saturación, esto es el máximo número de partículas que pueden alojarse en los $G$ estados (monocapa completa) , o el máximo volúmen $v_{m}$ que pueden ocupar en un modelo contínuo. 

\noindent Además 

\begin{equation}\label{eq.3.8.a}
\tilde{G_{o}}\left( n \right) =\lim_{N,G\to \infty}\frac{G_{o}\left( N\right) }{G}                   
\end{equation} 

\noindent 

\begin{equation}\label{eq.3.8.b}
\tilde{G_{o}}^{'}\equiv \frac{d \tilde{G_{o}}}{d n } 
\end{equation}

\begin{equation}\label{eq.3.8.a1}
\tilde{d}\left( n \right) =\lim_{N,G\to \infty}\frac{d\left( N\right) }{G}=1-\tilde{G_{o}}\left( n \right)                   
\end{equation}

\noindent La Energía Libre $ F $ por estado, $\tilde{f}\left(n,T \right)$ se define por 

\begin{equation}\label{eq.3.8.c}
\tilde{f}\left(n,T \right)= \lim_{N,G\to \infty} \frac{F(N,T,G)}{G}
\end{equation}

\noindent Reemplazando \eqref{eq.3.1} en \eqref{eq.3.6} y tomando el $\lim_{N,G\to \infty}$ 

\begin{equation}\label{eq.3.9}
\begin{aligned}
\beta \tilde{f}(n,T)= \beta n U_{o} &- \left( 1-\tilde{G_{o}}\left(n \right) +n\right) \ln \left( 1-\tilde{G_{o}}\left(n \right) +n\right) \\ 
&+ \left( 1-\tilde{G_{o}}\left( n \right)\right)  \ln  \left( 1-\tilde{G_{o}}\left( n \right)\right) + n \ln n
\end{aligned}
\end{equation}

\noindent Análogmente , la entropía por estado $\tilde{S}(n,T)$

\begin{equation}\label{eq.3.10}
\begin{aligned}
\frac{\tilde{S}(n,t)}{k_{B}}&=  \left( 1-\tilde{G_{o}}\left(n \right) +n\right) \ln \left( 1-\tilde{G_{o}}\left(n \right) +n\right) \\ &- \left( 1-\tilde{G_{o}}\left( n \right)\right)  \ln  \left( 1-\tilde{G_{o}}\left( n \right)\right) - n \ln n
\end{aligned}
\end{equation}

Una consecuencia significativa de la función isoterma de adsorción  \eqref{eq.3.8} es que permite investigar termodinámicamente los cambios de la configuración espacial de las partículas en el estado adsorbido en función de la densidad o cubrimiento superficial. La ec. \eqref{eq.3.8} se puede escribir de la forma 

\begin{equation}\label{eq.3.10.a}
\beta \mu= \ln \left[\frac{n \: [1-\tilde{G_{o}}\left(n \right) +n]^{(\tilde{G_{o}}^{'}-1)}}{[1-\tilde{G_{o}}\left( n \right) ]^{\tilde{G_{o}}^{'}}}  \right] - \ln K(T)
\end{equation}

\noindent y también 

\begin{equation}\label{eq.3.11}
\tilde{G_{o}}^{'}(n)-\frac{ \beta(\mu-U_{o}) + \ln \left[ \frac{ 1-\tilde{G_{o}}\left(n \right) +n }{n}\right] }{\ln \left[ \frac{ 1-\tilde{G_{o}}\left(n \right) +n}{ 1-\tilde{G_{o}}\left( n \right)}\right] }=0
\end{equation}

\noindent \eqref{eq.3.11} es una ecuación diferencial no lineal de la forma $f^{'}(x) - \mathcal{F}(x,f(x))=0$ cuya solución nos permitiría obtener  $\tilde{G_{o}}$ y $\tilde{G_{o}}^{'}=g(n)$ a partir de isotermas experimentales $n(\mu,T)$

Esta consecuencia de la teoría estadística fraccionaria de adsorción no ha sido estudiada aún, como tampoco su consistencia para interpretar los posibles cambios de configuraciones del estado adsorbido de una molécula a medida 
 que el cubrimiento varía. Esta posibilidad, que se denominó \textbf{espectro termodinámico de configuraciones} \cite{ref34} será motivo de estudios posteriores. 
\section{Gas de red de k-meros: significado físico de $g$}

\begin{figure}[H]
	\centering
	\includegraphics[width=1\linewidth,clip=true,trim = 0 50 0 200]{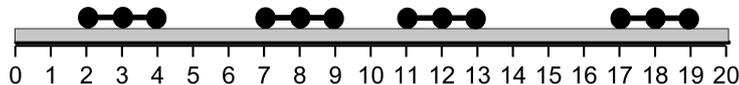}
	\caption{Representación de una configuración de 4 trímeros sobre una red unidimensional de 20 sitios. Elegimos enumerar los estados distinguibles para un k-mero aislado en \textbf{1D} por la posición del extremo izquierdo de la partícula. Los estados correspondientes a la ocupación de los sitios 5,6,15 y 16 están excluidos de ser ocupados. El estado correspondiente al sitio 14 esta  accesible a un nuevo trímero. }
	\label{Fig.3.4}
\end{figure}

Tomamos por simplicidad el caso de un gas de red de k-meros unidimensional. Cada partícula es una cadena de unidades idénticas que ocupan $k$ sitios de la red de $M$ sitios, como se representa esquemáticamente en la Fig. \ref{Fig.3.4}.

\noindent En este caso $G=M$, ya que hay solo $M$ estados o configuraciones distinguibles para cada k-mero aislado sobre la red de sitios. Cada k-mero aislado sobre la red excluye $k$ estados del total $G$ que no pueden ser ocupados por otro k-mero. Estos son, el estado que ocupa (que puede ser identificado por el sitio que ocupa la primera unidad) más $(k-1)$ estados que excluyen el resto de las unidades.  

\noindent La forma mas simple de proponer como se excluyen estados a medida que se agregan más k-meros a la red es suponer que $g$ es constante y en consecuencia $\tilde{G_{o}}$ depende linealmente de $n$ o $\theta$

\begin{equation}\label{eq.3.12}
\tilde{G_{o}}(n)=g\: n= g\: a \: \theta
\end{equation}

y 

\begin{equation}\label{eq.3.12.a}
\tilde{d}(n)=1-g\: n
\end{equation}

\noindent con $a=\textrm{const}$ y $g=k$ para k-meros en \textbf{1D}.

\noindent Como $\tilde{G_{o}}(n)=0$ y $\tilde{G_{o}}(n_{m})=1$ donde 

\begin{equation}\label{eq.3.13}
n_{m}=\frac{N_{m}}{G}=\frac{M/k}{G} 
\end{equation}		

\noindent siendo $N_{m}$ el número máximo de k-meros que se pueden acomodar en $M$ sitios $(G \: \textrm{estados})$,

\begin{equation}\label{eq.3.14}
\tilde{G_{o}}(n_{m})=g \, n_{m}= g \, a \: \textrm{,} \ \theta_{m}=1 \quad  \implies \quad a=\frac{1}{g}=\frac{1}{k}
\end{equation}

\noindent El cubrimiento de la red $\theta$ esta definido por $ \theta= k N/M $,  $ \theta_{m}=1$, y representa la fracción total de sitios de la red ocupados por k-meros. 

\noindent Finalmente, reemplazando \eqref{eq.3.12} y \eqref{eq.3.14} en \eqref{eq.3.8} obtenemos la \textbf{isoterma de adsorción} de k-meros sin interacción en\textbf{ 1D}  

\begin{equation}\label{eq.3.15}
e^{\left( \mu- U_{o}\right)}= \frac{a \theta \, \left[ 1-a\theta \left(g-1 \right) \right]^{g-1} }{\left[1-a\theta g \right]^{g} } 
\end{equation} 

\noindent o su equivalente

\begin{equation}\label{eq.3.16}
e^{\left( \mu-k u_{o}\right)}=\frac{\theta/k \, \left[ 1-\theta \frac{\left(k-1 \right)}{k} \right]^{k-1} }{\left[1-\theta \right]^{k} } 
\end{equation}

\noindent donde $U_{o}=k u_{o}$ y $u_{o}$ la energía de interacción por sitio (o por unidad que incluye el k-mero). 

\noindent La ec. \eqref{eq.3.16} es idéntica a la solución exacta del gas de red de k-meros en \textbf{ 1D } obtenida por otra vía en la ref. \cite{ref42}.

\begin{figure}[h]
	\centering
	\includegraphics[width=1\linewidth,clip=true,trim = 0 0 0 0]{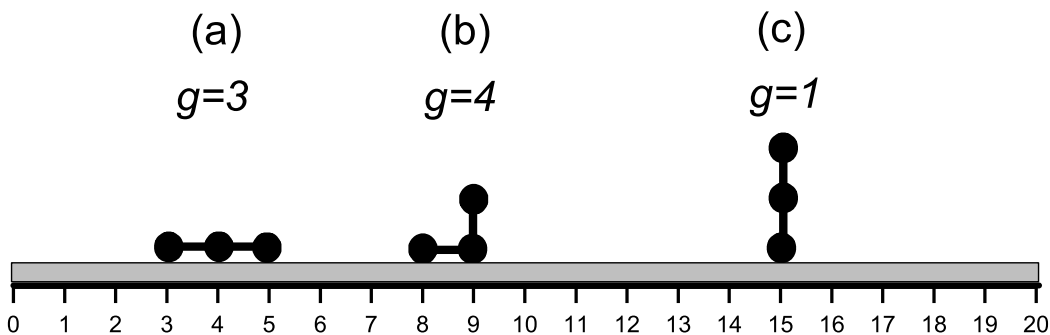}
	
	\caption{k-meros sobre una red unidimensional de sitios. Representación esquemática de diferentes configuraciones en el estado adsorbido. En cada caso se especifica el valor del parámetro de exclusión estadística $g$. La forma asimétrica del trímero en la configuración (b) corresponde a $g=4$ en vez de $g=2$ que corresponde a una partícula simétrica que ocupa 2 sitios vecinos de la red }
	\label{Fig.3.5}
\end{figure}

\begin{figure}[H]
	\centering
	\includegraphics[width=1\linewidth,clip=true,trim = 0 0 0 0]{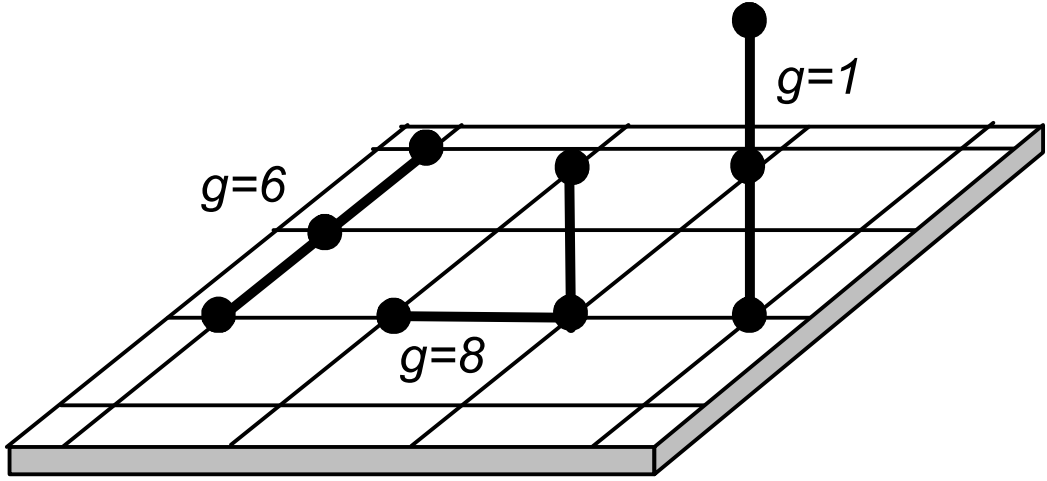}

	\caption{Similar a la Fig. \ref{Fig.3.5} para k-meros sobre una red bidimensional cuadrada.}
	\label{Fig.3.6}
\end{figure}

\noindent Este caso es muy ilustrativo porque nos muestra que el parámetro de exclusión estadística es igual al tamaño de la partícula, $g=k$, para una configuración de k-meros lineales que se adsorben con todas sus unidades en contacto con los sitios de la red en \textbf{1D}. En general $g$ dependerá de la configuración espacial con que la partícula se adsorbe , además de su tamaño y de la geometría y conectividad de la red de sitios. En las Figs. \ref{Fig.3.5}, \ref{Fig.3.6} se muestran diferentes configuraciones de trímeros en \textbf{ 1D} y \textbf{2D} y sus correspondientes valores de $g$.

\noindent El número total de  estados $G$ esta relacionado con el número de sitios de la red $M$, $G=m\: M$ donde $m$ es el número de configuraciones distinguibles de una partícula aislada por sitio, y como se observó antes, depende de la geometría de la red y de configuración espacial de la partícula en estado adsorbido. Ej., si un k-mero lineal se adsorbe con $k'$ unidades ocupando sitios de la red y $(k-k')$ alejados de la red sin ocupar sitios, entonces en \textbf{1D}, $m=1, \: g=k', \: a=1/k'$,. En cambio, en una red cuadrada 

\begin{equation}\label{key}
\begin{aligned}
&m=2, \ g=2k, \ a=\frac{1}{2k} \qquad &\mbox{para k-meros paralelos a la red} \\
&m=1, \ g=1, \ \ a=1 \qquad &\mbox{para k-meros perpendiculares a la red} \\
&m=2, \ g=2k', \ a=\frac{1}{2k'} \qquad &\mbox{para k-meros con $\mathrm{k'}$ unidades sobre la red}
\end{aligned} 
\end{equation}

En general para k-meros lineales ocupando $k$ sitios sobre una red con conectividad $\gamma$ las funciones termodinámicas de las ecs. \eqref{eq.3.8}, \eqref{eq.3.9} y \eqref{eq.3.10} toman la forma

\begin{equation}\label{eq.3.17}
\begin{aligned}
\beta \tilde{f}_{k,\gamma}(\theta,T)=\beta \theta u_{o} &+ \frac{\theta}{k} \ln\left( \frac{2\theta}{k\gamma}\right) + \frac{\gamma}{2}(1-\theta) \ln (1-\theta) \\ 
&- \frac{\gamma}{2}\left[ 1-(\frac{k\gamma -2}{k\gamma}) \theta \right] \ln \left[ 1-(\frac{k\gamma -2}{k\gamma}) \theta \right]
\end{aligned}
\end{equation}

\begin{equation}\label{eq.3.18}
\begin{aligned}
\tilde{S}_{k,\gamma}(\theta,T)=&- \frac{\theta}{k} \ln\left( \frac{2\theta}{k\gamma}\right) - \frac{\gamma}{2}(1-\theta) \ln (1-\theta) \\ 
&+ \frac{\gamma}{2}\left[ 1-(\frac{k\gamma -2}{k\gamma}) \theta \right] \ln \left[ 1-(\frac{k\gamma -2}{k\gamma}) \theta \right]
\end{aligned} 
\end{equation}

\begin{equation}\label{eq.3.19}
\begin{aligned}
e^{\left( \beta(\mu - k u_{o})\right)}= \frac{\frac{2\theta}{k\gamma} \left[ 1- \frac{2\theta}{k\gamma}\left( \frac{k\gamma}{2}-1\right)\right]^{\frac{k\gamma}{2}-1}}{\left[1-\theta \right]^{\frac{k\gamma}{2}} } 
\end{aligned}
\end{equation}

\section{Efectos de exclusión estadística $g$ sobre $\tilde{f}$, $\tilde{S}$ y $\mu$} 

En esta sección revisaremos los efectos generales que tiene la exclusión de estados sobre la forma de la Energía Libre de Helmholtz por estado $\tilde{f}$, la Entropía por estado $\tilde{S}$ y la Isoterma de Adsorción $\mu\equiv \mu(\theta,T)$. Aquí representamos la influencia que tiene el tamaño $k$ de las partículas, la geometria $\gamma$ de la red de sitios y la configuración espacial en el estado adsorbido. Esto último en la forma que solo $k'$ unidades de $k$ ocupan sitios de la red y $(k-k')$ están alejadas de la superficie sin ocupar sitios. De esta manera las funciones que desarrollemos serán válidas también para \textbf{k-meros flexibles} (denominamos de esta forma a k-meros que tienen una configuración de adsorción no recta, como el caso (b) de la Fig. \ref{Fig.3.5}).    

\noindent El formalismo de la teoría estadística fraccionaria para adsorción se puede resumir en las funciones $\tilde{f}(n,T)$ ec. \eqref{eq.3.9}, $\tilde{S}(n,T)$ ec. \eqref{eq.3.10} y $\mu(n,T)$ ec. \eqref{eq.3.8} junto con las definiciones de las ecs. \eqref{eq.3.12}, \eqref{eq.3.13} y \eqref{eq.3.14}. 

\noindent En un gas de red con $N$ k-meros con una dada geometría sobre $M$ sitios, en el cual $k'$ de las $k$ unidades ocupan sitios de la red y cada k-mero aislado tiene $m$ \textbf{configuraciones distinguibles por sitio}, $G=mM$  y

\begin{equation}\label{eq.3.23}
\tilde{f}(n,T)=\lim_{N,G\to \infty}\frac{F(N,T)}{G}=\frac{1}{m} \:\lim_{N,M\to \infty}\frac{F(N,T)}{M}=\frac{1}{m} f_{k}(\theta,T)
\end{equation}

\noindent donde $f_{k}\equiv \lim_{N,M\to \infty} F(N,T)/M$ designa la Energía Libre de Helmholtz por sitio de la red para k-meros. Con esto 

\begin{equation}\label{eq.3.24}
\beta f_{k}(\theta,T)=m \: \tilde{f}(n,T)
\end{equation}

\noindent Si además, como se definió en \eqref{eq.3.12}, \eqref{eq.3.13} y \eqref{eq.3.14}, 

\begin{equation}\label{eq.3.24.a}
n=a \: \theta, \qquad a=\frac{1}{g}, \qquad g=m \: k',\qquad \tilde{G}_{o}(n)=n \: g=\theta, \qquad U_{o}=k' \: u_{o}
\end{equation}

\noindent reemplazando en las ecs. \eqref{eq.3.9}, \eqref{eq.3.10}) y \eqref{eq.3.8}, obtenemos 

\begin{equation}\label{eq.3.25}
\begin{aligned}
f_{k}(\theta,T)=\beta \theta u_{o}&-\left[ m(1-\theta)+\frac{\theta}{k'}\right] \ln \left[ m(1-\theta)+\frac{\theta}{k'}\right] \\
&+ \left[ m (1-\theta)\right] \ln \left[ m (1-\theta)\right]+ \frac{\theta}{k'} \ln\frac{\theta}{k'} 
\end{aligned}
\end{equation}

\begin{equation}\label{eq.3.26}
\begin{aligned}
\frac{S_{k}(\theta,T)}{k_{B}}=& \left[ m(1-\theta)+\frac{\theta}{k'}\right] \ln \left[ m(1-\theta)+\frac{\theta}{k'}\right] \\
&- \left[ m (1-\theta)\right] \ln \left[ m (1-\theta)\right]- \frac{\theta}{k'} \ln\frac{\theta}{k'} 
\end{aligned}
\end{equation}

\noindent Por último, la isoterma de adsorción es

\begin{equation}\label{eq.3.27}
e^{\beta[ \mu(\theta,T)-k'u_{o}]}=\frac{\frac{\theta}{mk'}\left[ 1-\frac{\left(mk'-1 \right) }{mk'} \theta\right]^{mk'-1 }}{\left[ 1-\theta \right]^{mk'}}
\end{equation}

\noindent Para el caso $m=1$ y $k'=1$, que corresponde a monómeros, la ec. \eqref{eq.3.27} se reduce a la distribución de \textbf{FD}. Si $m=1$ y  $k'=k$ se reduce al caso de k-meros en \textbf{1D} (ec.\eqref{eq.3.16}). Para caso de k-meros lineales sobre una red con conectividad $\gamma$, $k'=k$ y $m=\gamma/2$ y recuperamos la función isoterma \eqref{eq.3.19}.  

\begin{figure}[H]
	\centering
	\includegraphics[width=1\linewidth]{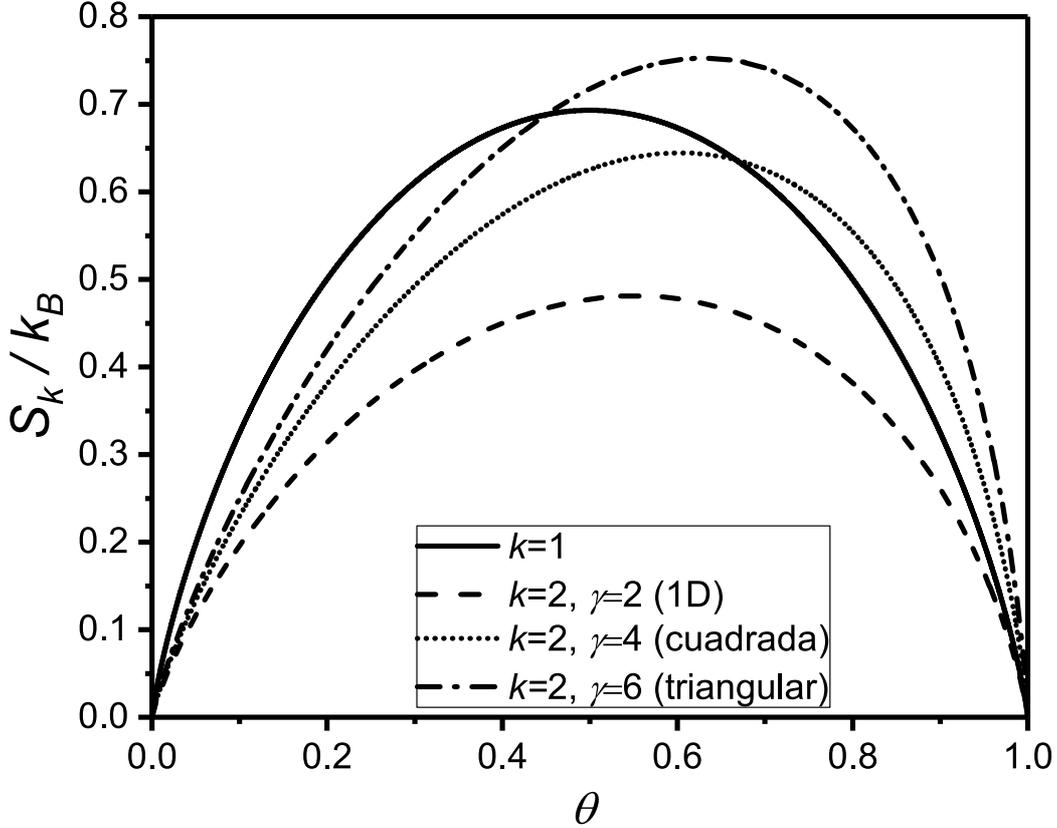}
	\caption{$S_{k}/k_{B}$ versus $\theta$ (ec. \eqref{eq.3.26}) para: $k=1$ (monómeros); $\gamma=2$, $k=k'=2$ (dimeros/1D); $\gamma=4$, $k=k'=2$ (dimeros/red cuadrada); $\gamma=6$, $k=k'=2$ (dimeros/red triangular).}
	\label{Fig.3.7}
\end{figure}

\begin{figure}[H]
	\centering
	\includegraphics[width=1\linewidth]{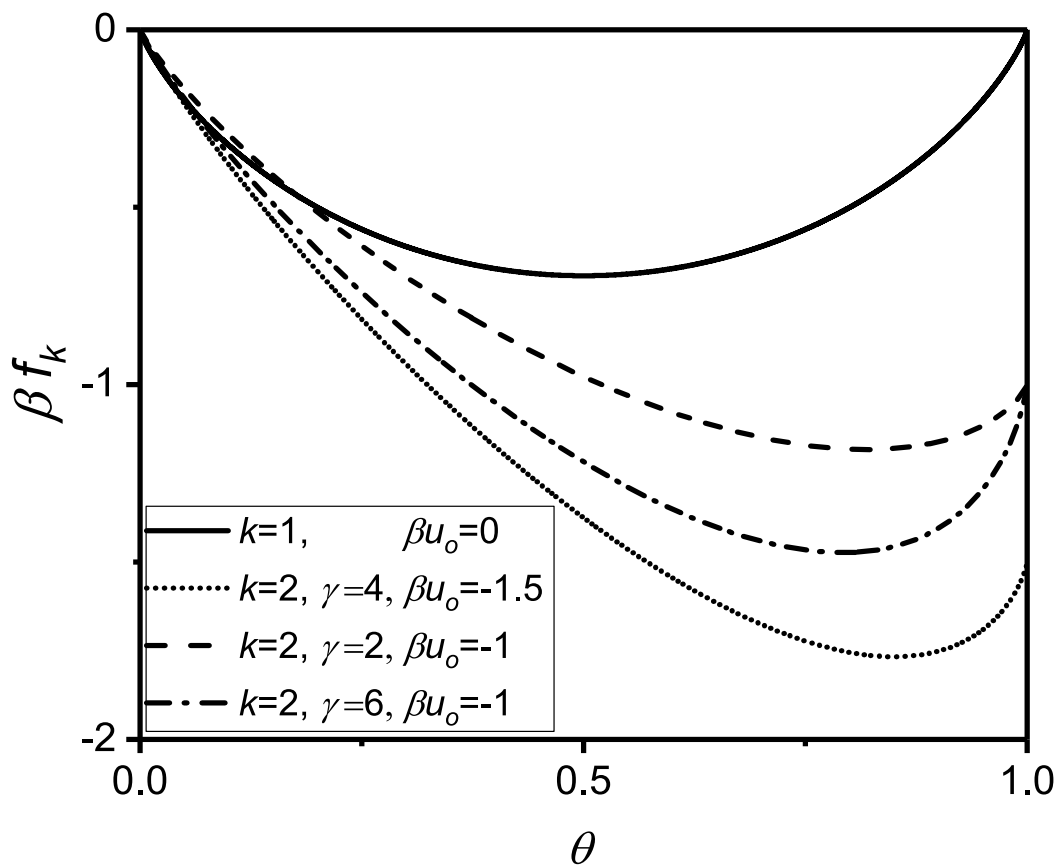}
	\caption{$\beta f_{k}$ versus $\theta$ (ec. \eqref{eq.3.25}) para k-meros con energía de adsorción por sitio $\beta u_{o}$.  $k=1$ y $\beta u_{o}=0$ (monómeros); $\gamma=2$, $k=k'=2$ y $\beta u_{o}=-1$ (dimeros/1D); $\gamma=4$, $k=k'=2$ y $\beta u_{o}=-1.5$ (dimeros/red cuadrada); $m=\gamma$, $k=k'=2$ y $\beta u_{o}=-1$ (dimeros/red triangular). Valores negativos de $\beta u_{o}$ corresponden a interacción atractiva }
	\label{Fig.3.8}
\end{figure}

\begin{figure}[H]
	\centering
	\includegraphics[width=1\linewidth]{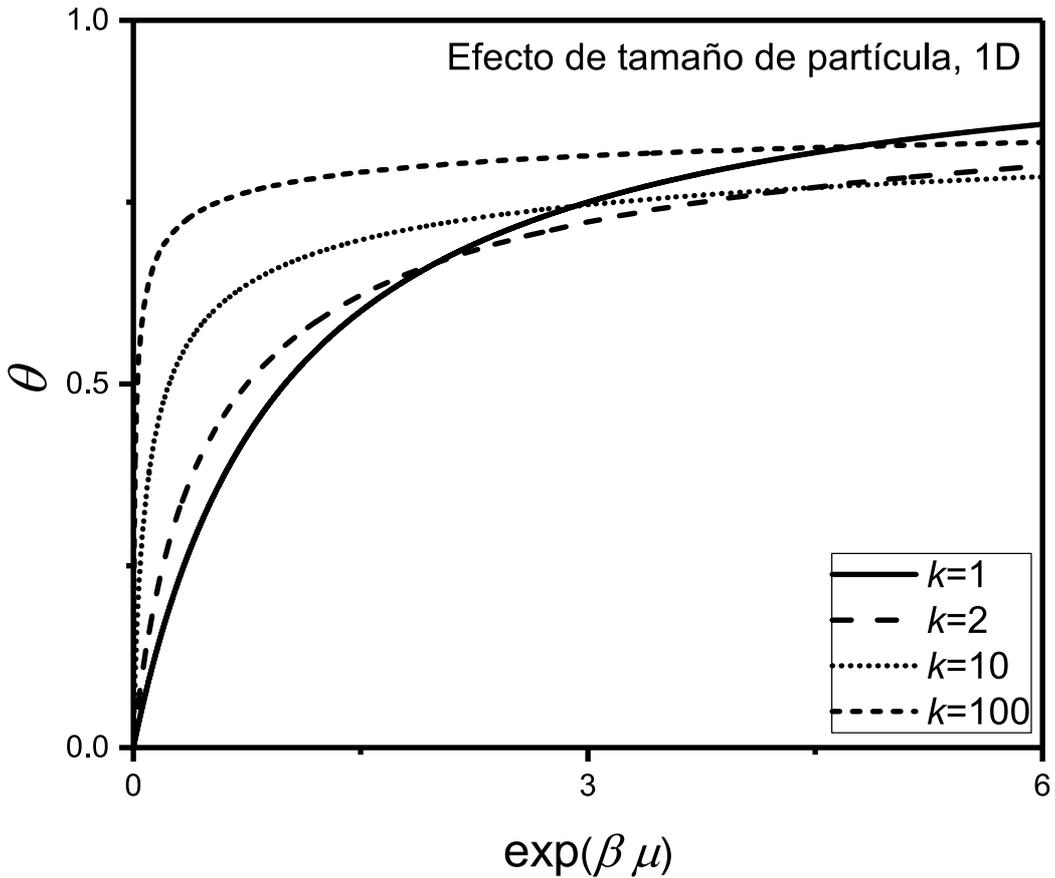}
	\caption{Efecto de tamaño de partícula sobre la isoterma de adsorción. $\theta$ versus $e^{\beta \mu}$  (ec. \eqref{eq.3.27}) para: $k=1,2,10,100$ en 1D ($\gamma=2$).}
	\label{Fig.3.8.a}
\end{figure}

\noindent En las figuras \ref{Fig.3.7}, \ref{Fig.3.8} y \ref{Fig.3.8.a} representamos la entropía por sitio $S_{k}$, la energía libre de Helmholtz por sitio $f_{k}$ y la isoterma de adsorción $\mu(\theta)$, para diferentes casos en \textbf{1D} y \textbf{2D}.

\noindent El comportamiento de $S_{k}/k_{B}$ es no trivial y muy interesante. Muestra cualitativamente y cuantitativamente la importancia de considerar adecuadamente los efectos de tamaño y forma de las partículas en su descripción termodinámica. 

\noindent Para $k=1$ (monómeros), $S_{k}$ es simétrica con un máximo en $\theta=0.5$. Los límites $S_{k}=0$ para $\theta \to 0$ y $ S_{k}=0$ para $\theta \to 1$ resultan de que solo hay una configuración posible $(\varOmega=1)$ en esos límites: red completamente vacía y red llena, respectivamente. 

\noindent El tamaño de la partícula produce que  $\forall \: k>1$ la entropía ya no es simétrica y tiene un máximo para $\theta >0.5$. Para $k>1$ en  \textbf{1D}, siempre $S_{k}<S_{1}$, excepto en los límites que son iguales. En general dada una geometría de la red, $\gamma$, $S_{k}<S_{k'}\ \forall \: k>k'$. De igual manera, dado $k$, a mayor conectividad $\gamma$ la entropía es mayor en todo el rango y el máximo se alcanza a valores mayores de $\theta$.

\noindent Es interesante ver que si comparamos el comportamiento para $k>1$ a medida que la conectividad de la red $\gamma$ aumenta, siempre existe un intervalo de $\theta$ para el 
cual $S_{k}$ es mayor para la red de mayor conectividad.

\noindent El comportamiento de $\beta f_{k}$ se muestra en la fig. \ref{Fig.3.8}, en la cual se incluye el efecto de la energía de adsorción por sitio $\beta u_{o}$ (negativa). La forma de la figura se puede entender a partir de la relación general $F=U-TS$. La energía de adsorción por sitio es $u_{o}=\textrm{const}$ y la energía total por sitio varía linealmente con $\theta$. El mínimo de $f_{k}$ se produce por la forma en que decrece la entropía para $\theta >\theta^{*}$ donde $\theta^{*}$ es el valor donde $S_{k}$ alcanza su máximo $S_{k}^{*}\equiv S_{k}(\theta^{*})$.     

\noindent Con respecto a la isoterma de adsorción, la ec. \eqref{eq.3.27} se puede escribir directamente en términos de la conectividad de la red de sitio $\gamma$, dado que $m=\gamma /2$. Con esto , 

\begin{equation}\label{eq.3.27.a}
e^{\beta\left(\mu - k' u_{o} \right) }=\frac{\frac{2 \theta}{\gamma k'} \left[ 1- (1 -\frac{2}{\gamma k'})\theta \right]^{\frac{\gamma k'}{2} -1} }{\left[1-\theta \right]^{\frac{\gamma k'}{2}} }
\end{equation}

\begin{figure}[H]
	\centering
	\includegraphics[width=1\linewidth]{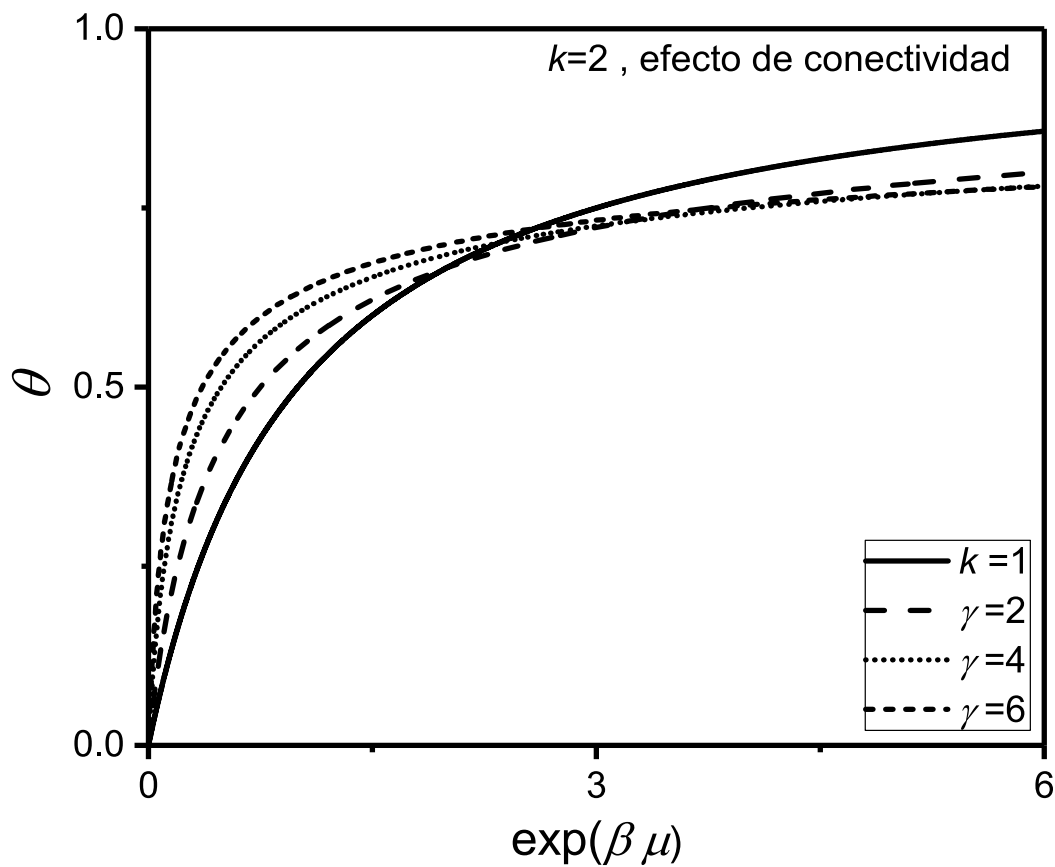}
	\caption{Efecto de conectividad de la red de sitios. $\theta$ versus $e^{\beta \mu}$  (ec. \eqref{eq.3.27}) para: $k=1$ (monómeros); $\gamma=2$, $k=k'=2$ (dimeros/1D); $\gamma=4$, $k=k'=2$ (dimeros/red cuadrada);  $\gamma=6$, $k=k'=2$ (dimeros/red triangular).} 
	\label{Fig.3.8.b}
\end{figure}

\begin{figure}{H}
	\centering
	\includegraphics[width=1\linewidth]{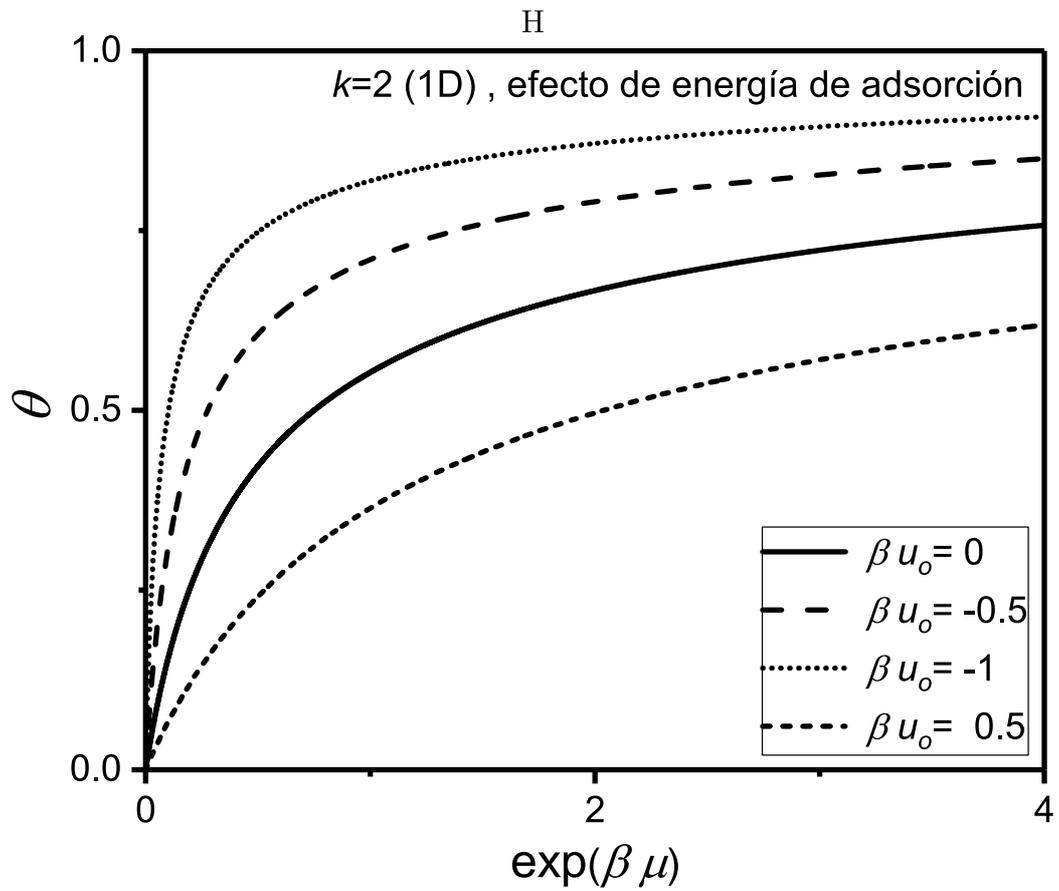}
	\caption{Efecto de interacción partícula-red para dímeros con interacción atractiva y repulsiva;  $k=2$, $\beta u_{o}=0$ $\beta u_{o}=-0.5$ $\beta u_{o}=-1$ $\beta u_{o}=0.5$}
	\label{Fig.3.8.c}
\end{figure}

Las figuras \ref{Fig.3.8.a}, \ref{Fig.3.8.b} y \ref{Fig.3.8.c}  muestran el efecto del tamaño de partículas $k$, la conectividad de la red $\gamma$ y la energía de adsorción
$u_{o}$ sobre la isoterma de adsorción que surge de la teoría estadística fraccionaria para k-meros sin interacción lateral.   

En este capítulo hemos revisado el modelos de Estadística Fraccionaria para adsorción propuesto en referencia \cite{ref34} que introduce el concepto de exclusión estadística de estados $g$ y lo relaciona con la configuración de la partícula en estado adsorbido sobre la red de sitios. El modelo muestra que la exclusión de estados tiene efecto significativo en todas las funciones termodinamicas del gas de red.   

%% file: texto/TF_Interpretacion_Cinetica_Ads_Des.tex
\chapter{Interpretación cinética del equilibrio adsorción-desorción con exclusión de estados}
	
	El estado de equilibrio de adsorción/desorción de partículas en una superficie se puede interpretar en término de una ecuación cinética y esta misma interpretación la podemos generalizar para la isoterma de adsorción o función de ocupación de estados de un sistema de partículas con exclusión estadística de estados. Esto nos sera muy útil para interpretar los resultados de simulación cuando los estados del sistema no son independientes entre sí (como es implícito la formulación de estadística fraccionaria de Haldane) estan correlacionados espacialmente como en el caso de k-meros sobre una red regular de sitios. Esta correlación es la que consideramos de manera general en la Estadística de Múltiple Exclusión. 
	
\section{Ecuación Maestra para Estadísticas de Exclusión}	
	\noindent Seguimos las mismas definiciones de los capítulos 4 y 5 y formularemos nuestro sistema de forma general para luego analizar el caso especial de un gas de red de k-meros. Si un sistema de $G$ estados de energía $U_{o}$ en un volúmen $V$ se pone en contacto, y puede intercambiar partículas caracterizadas por parámetro de exclusión $g$, con un baño térmico con potencial químico $\mu$ y temperatura $T$ entonces la evolución en el tiempo del número medio de ocupación de estados $n$ en $V$ estará dado por la ecuación
	
	\begin{equation}\label{eq.c6.1}
	\frac{dn}{dt}=P_{\circ} \: W_{in} - P_{\bullet} \: W_{out}
	\end{equation}  
	
	\noindent donde $P_{\circ}$, $P_{\bullet} $ son las probabilidades de encontrar un estado accesible vacío $(\circ)$, lleno $ (\bullet)$ en $V$, respectivamente, y  $W_{in}$, $W_{out} $ la probabilidad ingresar (adsorber), sacar (desorber) una partícula a (de) un estado en $V$. Como en la ecuación maestra de procesos estocásticos, $W_{in}$ y $W_{out} $ son en rigor velocidades de transición o probabilidades de transicion por unidad de tiempo.
	
	\noindent El equilibrio termodinámico se alcanza bajo la condición $dn/dt=0$, de donde surge
	
	\begin{equation}\label{eq.c6.2}
	\frac{W_{in}}{W_{out}}=\frac{P_{\bullet}}{P_{\circ}}=e^{\beta(\mu-U_{o})}
	\end{equation}
	\noindent y resolviendo $P_{\circ} $
	
	\begin{equation}\label{eq.c6.3}
	P_{\circ}=P_{\bullet} \ e^{-\beta\left( \mu-U_{o}\right) }
	\end{equation}
	
	\noindent Dado que  $P_{\bullet}=n$, e.d. la fracción de estados ocupados en $V$, y tomando la forma general de la isoterma de adsorción ec. \eqref{eq.3.8}
	
	\begin{equation}\label{ec.c6.4}
	P_{\circ}=\frac{\left[ 1-\tilde{G}_{o}(n)\right]^{\tilde{G}_{o}^{'}}}{\left[ 1-\tilde{G}_{o}(n)+n\right]^{\tilde{G}_{o}^{'}-1}}
	\end{equation}
	
	\noindent Esta expresión es general y útil para interpretar los resultados posteriores.
	
	\noindent Otra forma de expresarla es 
	
	\begin{equation}\label{ec.c6.5}
	P_{\circ}=\frac{\left[ \tilde{d}(n)+n\right]^{\tilde{d}^{'}(n)+1}}{\left[ \tilde{d}(n)\right]^{\tilde{d}^{'}(n)} }
	\end{equation} 

\noindent donde queda claro que $P_{\circ}$ y $\tilde{d}(n)$ son cosas distintas ya que ésta última mide el numero normalizado de estados disponibles a una partícula, dada la fracción de estados ocupados $n$. Se asume que ese número es constante para cada $n$. 

\noindent Sin embargo $P_{\circ}$ es la fracción de estados disponibles promediada sobre todas las configuraciones del Conjunto estadístico  ya que un número de estados en $V$ no están ni ocupados ni accesibles sino simplemente excluidos de ser ocupados.  Notemos que $\tilde{d}(n)+n\neq 1$, ya que el complemento de $\tilde{d}(n)$ es $\tilde{G_{o}(n)}$.	
	
	\noindent Para esto es útil observar que un estado cualquiera del conjunto $G$ puede estar solo en alguna de las tres condiciones siguientes:$\bullet\equiv $ lleno, $\circ \equiv$ vacío, ó $\varnothing\equiv$ excluido, y
	
	\begin{equation}\label{ec.c6.6}
	P_{\varnothing}=1 - P_{\bullet} -P_{\circ}=1-n-\frac{\left[ \tilde{d}(n)+n\right]^{\tilde{d}^{'}(n)+1}}{\left[ \tilde{d}(n)\right]^{\tilde{d}^{'}(n)} }
	\end{equation}

\noindent Es útil representar $P_{\varnothing}(n)$ (o  $P_{\varnothing}(n=\frac{\theta}{g})$) para interpretar la exclusión estadística en la estadística fraccionaria y en la estadística de múltiple exclusión. 

\section{Gas de red de k-meros: $\tilde{d}$, $P_{\circ} $ y $P_{\varnothing} $ vs. $\theta$}

El gas de red de k-meros tiene la siguiente dependencia de  $\tilde{d}$, $P_{\circ} $ y $P_{\varnothing} $ con $\theta$. 

\noindent En el caso de la Estadística Fraccionaria para adsorción $\tilde{d}(\theta)=1-\theta$,  $P_{\circ}^{EF} $ y $P_{\varnothing}^{EF}$ son 

\begin{equation}\label{ec.c6.10}
P_{\circ}^{EF}(\theta)=\frac{\left[ 1-\theta \right]^{g}}{\left[ 1-\frac{\left(g-1 \right) }{g} \theta\right]^{g-1 }}
\end{equation}

\begin{equation}\label{ec.c6.11}
P_{\varnothing}^{EF}(\theta)=1-\frac{\theta}{g}-\frac{\left[ 1-\theta \right]^{g}}{\left[ 1-\frac{\left(g-1 \right) }{g} \theta\right]^{g-1 }}
\end{equation}

\noindent donde usamos el superíndice $EF$ para identificar a la forma particular de las funciones en la Estadística Fraccionaria. Por simplicidad, hemos usado la notación $g$. Para unificar con notación de gas de red de los capítulos anteriores solo hay que reemplazar $g=m\:k'$ en las ecuaciones.

\begin{figure}[h]
	\centering
	\includegraphics[width=1\linewidth]{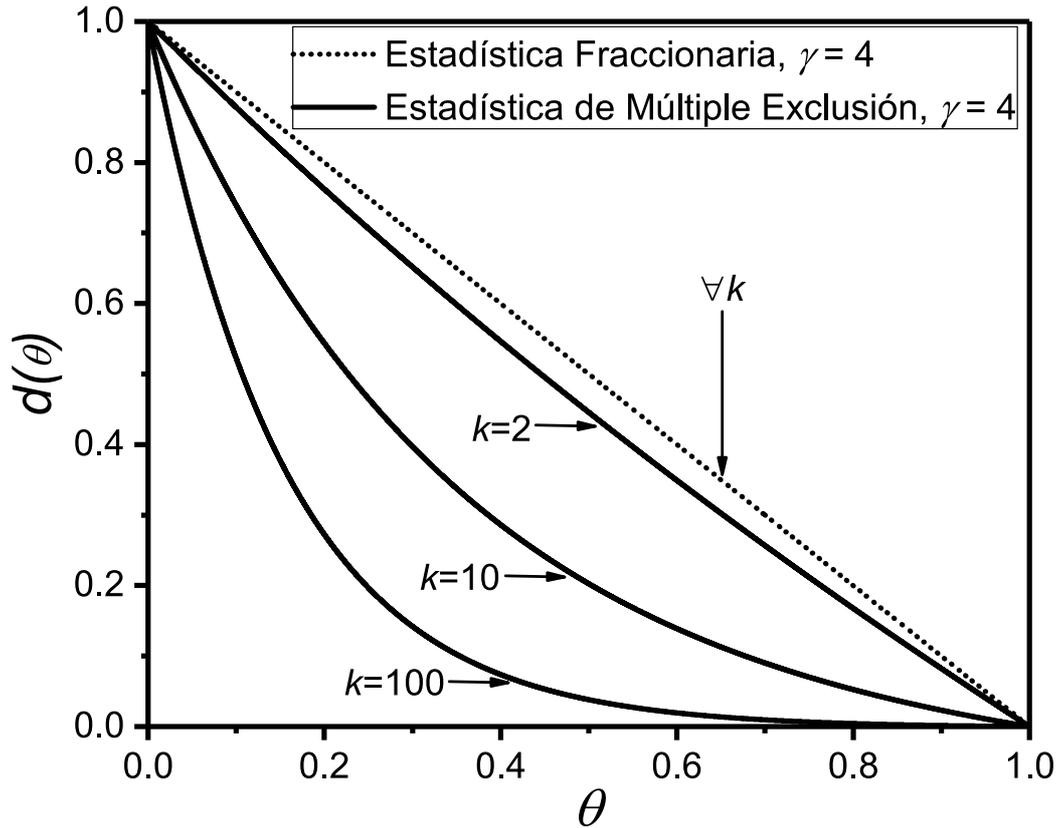}
	\caption{$\tilde{d}(n)$ versus $\theta$ para $k=2,10$ y $100$ sobre red cuadrada $(\gamma=4)$ en las estadísticas fraccionaria y de múltiple exclusión. Para $k=2$ ambas estadísticas coinciden}
	\label{fig.dn.EFyME}
\end{figure}
\begin{figure}[H]
\centering
\includegraphics[width=1\linewidth]{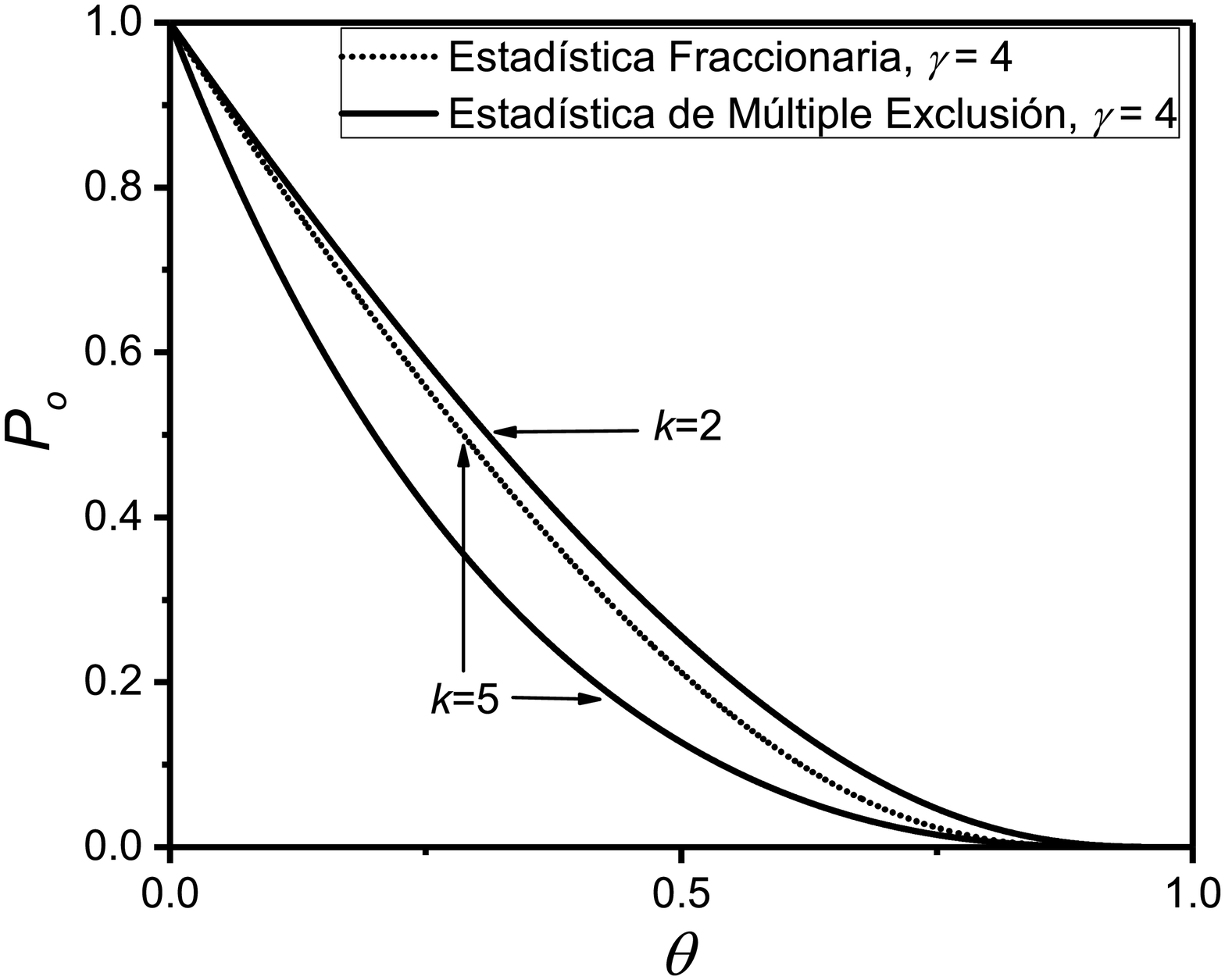}
\caption{$P_{\circ}$ versus $\theta$ para $k=2,5$ sobre red cuadrada $(\gamma=4)$ en las estadísticas fraccionaria y de múltiple exclusión. Para $k=2$ ambas estadísticas coinciden}
	\label{fig.Po.EFyME}
\end{figure}
\begin{figure}[H]
	\centering
	\includegraphics[width=1\linewidth]{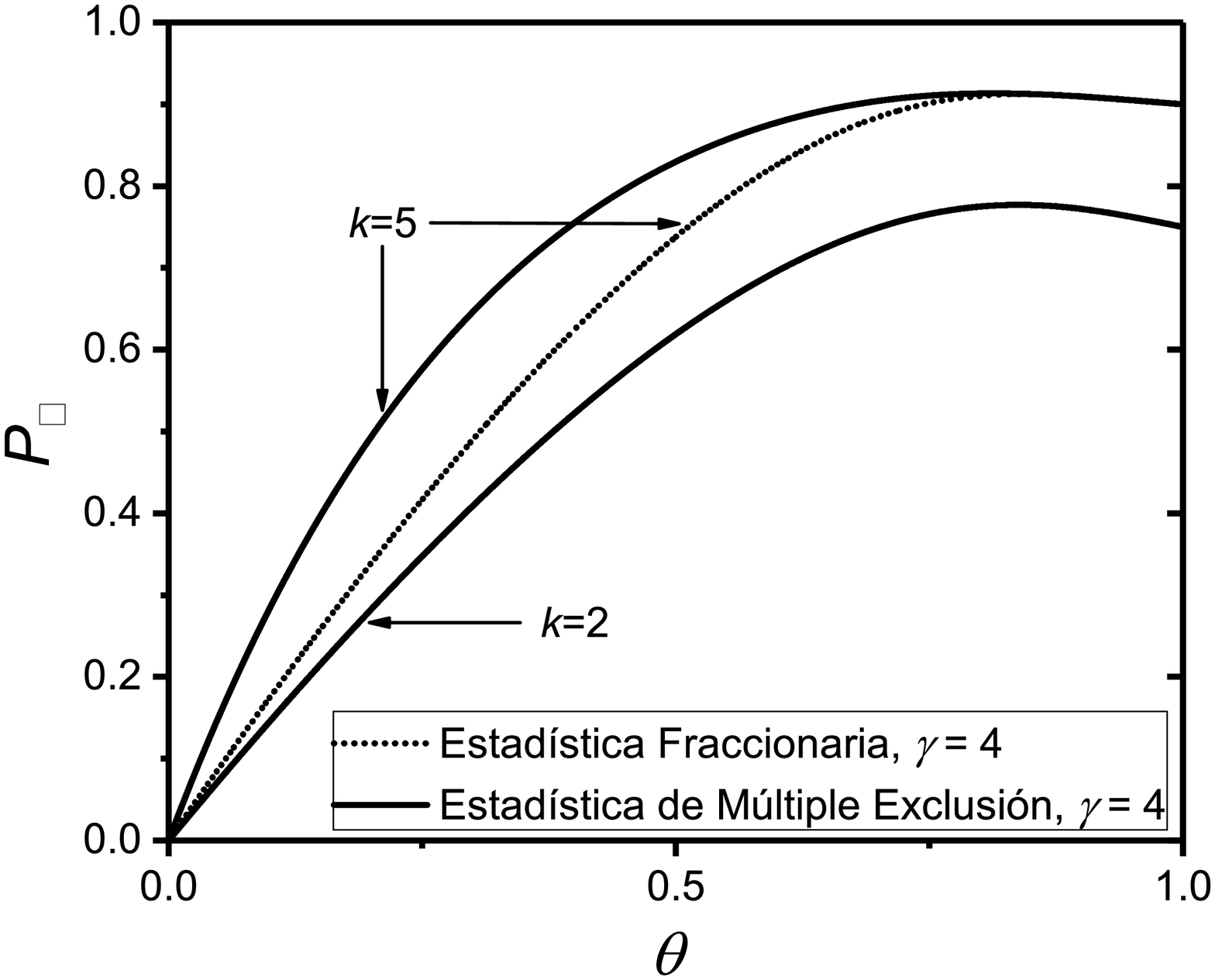}
	\caption{$P_{\varnothing}$ versus $\theta$ para $k=2,5$ sobre red cuadrada $(\gamma=4)$ en las estadísticas fraccionaria y de múltiple exclusión.Para $k=2$ ambas estadísticas coinciden}
	\label{fig.Pexcl.EFyME}
\end{figure}

\noindent En las Figs. \ref{fig.dn.EFyME}, \ref{fig.Po.EFyME} y \ref{fig.Pexcl.EFyME} se muestra la dependencia de $\tilde{d}$, $ P_{\circ}$ y $P_{\varnothing} $ para diferentes sistemas de k-meros sobre redes en \textbf{1D} y \textbf{2D}. En el capítulo \ref{Capitulo_Est_M_E} introducimos las funciones correspondientes a una nueva estadística denominada Estadística de Múltiple Exclusión, que solo se muestran en este punto para comparación.    

\section{Caso simple de k-meros en 1D. Interpretación de $P_{\circ} $}

El caso unidimensional es muy ilustrativo acerca de la diferencia entre $\tilde{d}(\theta)$ y $P_{\circ}(\theta)$. 

\noindent Consideremos $N$ k-meros sobre una red unidimensional de $M$ sitios. Como sabemos podemos resolver la función de partición en esta red original $\mathfrak{R}$ ó en una red transformada equivalente $ \mathfrak{R}'$ con $N$ monómeros sobre $M'=M-N(k-1)$ sitios. En $\mathfrak{R}'$ el cubrimiento es $\theta'=N/M'$. La ecuación de equilibrio adsorción/desorción análoga a \eqref{eq.c6.2} es 

\begin{equation}\label{ec.c6.12}
\frac{P_{\bullet}^{'}}{P_{\circ}^{'}}=\frac{\theta^{'}}{(1-\theta^{'})}=e^{\beta(\mu-U_{o})}
\end{equation}

\noindent ya que se trata de un problema de monómeros, donde los superíndices refieren a las cantidades en $ \mathfrak{R}'$. 
\noindent Relacionando $\theta^{'}$ y $\theta$ obtenemos la isoterma exacta

\begin{equation}\label{ec.c6.13}
\theta^{'}=\frac{N}{M^{'}}=\frac{N}{M-N\left( k-1\right)}=\frac{\theta/k}{\left[1-\frac{\left( k-1\right) }{k} \theta\right] }
\end{equation}

\begin{equation}\label{ec.c6.14}
\begin{aligned}
e^{\beta(\mu-U_{o})}&=\frac{\theta/k}{\left[1-\frac{\left( k-1\right) }{k} \theta\right]} \frac{\left[1-\frac{\left( k-1\right) }{k} \theta\right]^{k}}{\left[1-\theta \right]^{k} }\\
&= \frac{\theta}{k} \frac{\left[1-\frac{\left( k-1\right) }{k} \theta\right]^{k-1}}{\left[1-\theta \right]^{k} }
\end{aligned}
\end{equation} 

\noindent Por otra parte, en la red original $\mathfrak{R}$ también vale $P_{\bullet}=N/M=\theta/k$ y 

\begin{equation}\label{ec.c6.15}
\frac{P_{\bullet}}{P_{\circ}}=e^{\beta(\mu-U_{o})}
\end{equation}  

\noindent e igualando \eqref{ec.c6.14} y \eqref{ec.c6.15}, $P_{\circ}$ resulta

\begin{equation}\label{ec.c6.16}
P_{\circ}=\frac{\left[1-\theta \right]^{k}}{\left[1-\frac{\left( k-1\right) }{k} \theta\right]^{k-1}}
\end{equation} 

donde nuevamente $P_{\circ}\neq\tilde{d}(\theta)=1-\theta$ ya $P_{\circ}$ es la probabilidad de encontrar $k$ sitios vacíos consecutivos en la red y refleja la correlación topológica entre el tamaño de la partícula y la geometría que impone la red. En cambio $\tilde{d}$ es simplemente la medida de los sitios vacíos, que en la hipótesis de que los estados del sistema fuesen independientes, cada sitio vacío podría ser ocupado por el extremo de k-mero (excluyendo además $(g-1)$ sitio vacíos adicionales). Esta es la diferencia conceptual entre ambas cantidades que queda evidente en este caso.

%% file: texto/TF_Simulacion_Monte_Carlo.tex
\chapter{Simulación de Monte Carlo}
\label{SimMonCar}

En los últimos años, la simulación computacional se ha convertido en otra forma de investigar. En algunos casos, la simulación provee bases teóricas para entender los resultados experimentales. En otros casos, provee datos como si se tratase de ``experimentos numéricos'' con los cuales comparar la teoría. 

En este capítulo veremos cómo se realizan las \emph{simulaciones de Monte Carlo} (MC) para simular el proceso de adsorción-desorción de $k$-meros lineales.

\section{Simulación de Monte Carlo para la adsorción de $k$-meros}

La simulación en física estadística es una herramienta básica en la actualidad para contrastar la validez de los modelos e hipótesis. Cuanto mas complejo el sistema que se estudia mas valiosa se vuelve la posibilidad de simular el modelo y calcular los observables físicos de interés. En este capítulo reproducimos los resultados de simular la adsorción de k-meros lineales en una y dos dimensiones en el conjunto Gran Canónico mediante el procedimiento de Monte Carlo.

Para estudiar la adsorción de $k$-meros en superficies homogéneas, realizamos simulaciones de MC en la asamblea Gran Canónica (GCMC). En esta asamblea trabajamos con la temperatura $T$, el potencial químico $\mu$ y el volumen $M$ del sistema como parámetros termodinámicos fijos, siendo variable el número $N$ de moléculas de adsorbato.

Los sitios de arsorción se arreglan en una red cuadrada de $M= L \times L$ sitios (red de lado $L$) con condiciones periódicas de borde. De este modo, cada sitio posee cuatro primeros vecinos en la red. Además, usamos valores apropiados de $L$ para no perturbar las estructuras de la fase ordenada.

Introducimos la variable de ocupación $c_i$, igual a cero si el $i$-ésimo sitio de la red se encuentra vacío o a uno si se encuentra ocupado.

Bajo estas consideraciones, el Hamiltoniano $H$ está dado por 
\begin{equation}
H=  \frac{1}{2} \sum_i^M \sum_{j(i)} w \; c_i \, c_j - w \; N \; (k-1) + (U_0 - \mu) \sum_i c_i
\label{Hamiltonianok}
\end{equation}
donde $w$ es la energía de interacción lateral a primeros vecinos, que puede ser repulsiva (positiva) o atractiva (negativa), y $j(i)$ representa a todos los primeros vecinos de $i$. La energía de interacción entre una unidad del adsorbato y el substrato, $U_0$, se toma igual a cero por simplicidad y sin pérdida de generalidad, al considerar la superficie homogénea.

La probabilidad $P(N,\mathbf{x})$ \cite{binder1,binder2} de tener $N$ $k$-meros adsorbidos en un dado estado de ocupación $\mathbf{x}$ definido por el conjunto de valores $\left\lbrace c_i \right\rbrace$ en la asamblea gran canónica puede escribirse como
\begin{equation}
P(N,\mathbf{x})=\frac{e^{-\frac{1}{k_B T} \left( H(\mathbf{x}) - \mu \sum_i c_i \right) }}{\Xi(\mu,T,M)}
\label{ecuacion22'}
\end{equation}
donde el Hamiltoniano $H(\mathbf{x})$  está dado por la Ec. (\ref{Hamiltonianok}) y $\Xi(\mu,T,M)$ es la gran función de partición.

En el esquema de Metropolis \cite{metropolis,metropolis2}, la probabilidad de transición desde un estado inicial $\mathbf{x}_{i}$ a otro final $\mathbf{x}_{f}$, $W(\mathbf{x}_{i} \rightarrow \mathbf{x}_{f})$, está dada por
\begin{equation}
W( \mathbf{x}_{i} \rightarrow \mathbf{x}_{f})= \min \left\lbrace 1, \frac{P(\mathbf{x}_{i})}{P(\mathbf{x}_{f})}\right\rbrace.
\label{ecuacion23'}
\end{equation}
La probabilidad de adsorción $W_{ads}$ y la de desorción $W_{des}$ son las probabilidades de transición de un estado con $N$ $k$-meros adsorbidas a un nuevo estado con $N+1$ y $N-1$ $k$-meros respectivamente.

La Ec. (\ref{ecuacion23'}) satisface el principio de reversibilidad microscópica

\begin{equation}\label{eq.c7.2}
P(\mathbf{x}_{i}) W(\mathbf{x}_{i}\rightarrow\mathbf{x}_{f})=P(\mathbf{x}_{f}) W(\mathbf{x}_{f}\rightarrow\mathbf{x}_{i})
\end{equation} 

\noindent y es condición suficiente para obtener una cadena de estados con una distribución estacionaria dada por la Ec. (\ref{ecuacion22'}).

Una vez definida la probabilidad de transición, el equilibrio de adsorción-desorción en una simulación de MC estándar será alcanzado mediante un algoritmo del tipo \emph{spin flip} (dinámica de Glauber) que podemos describir en la forma siguiente para un paso elemental de Monte Carlo (un MCS) \cite{ramirez1995}:

\begin{enumerate}
\item Se fija un valor de $\mu$ y de $T$.
\item Se fija un estado de ocupación inicial $\mathbf{x}_{0}$ sobre la red, colocando $N$ $k$-meros sobre la red.
\item Se elige al azar un sitio $i$ sobre la red. 
\item Se selecciona una dirección al azar. Si la dirección es vertical, se considera la $k$-upla de sitios $i$, $i+L$, $i+2L$, ... , $i+(k-1)L$. Si la dirección es horizontal, se considera la $k$-upla $i$, $i+1$, $i+2$, ... , $i+(k-1)L$. En ambos casos se tendrán en cuenta las condiciones de contorno periódicas al momento de identificar la $k$-upla. 
\item Se genera un número aleatorio $\xi \in [0,1]$.
\begin{enumerate}
\item Si la $k$-upla de sitios elegida está completamente vacía, se adsorbe una partícula si $\xi \leq W_{a}$.
\item Si la $k$-upla de sitios elegida está completamente ocupada por una misma partícula, se desorbe la misma si $\xi \leq W_{d}$.
\item Si la $k$-upla de sitios no está ni llena ni vacía, el intento es rechazado.
\end{enumerate}
\item Adicionalmente se realiza un paso de difusión o "relajación" de partículas, que en nuestro caso se tratan de $k$-meros lineales. Este proceso consiste en una desorción seguida de una adsorcion hacia posiciones vecinas, tanto a través de desplazamientos a lo largo del eje del $k$-mero, como también a través de reptación o rotación del $k$-mero en torno a alguna de sus unidades. Este paso es esencial para alcanzar el equilibrio en un tiempo razonable, principalmente para valores crecientes de $k$.
\end{enumerate}

Cuando se realizan estudios de simulación de sistemas que presentan transiciones de fase, con frecuencia se presenta el problema de \textit{dinámica lenta}. Este efecto aparece en una amplia variedad de sistemas en donde los ingredientes presentes generalmente son desorden, frustración o competencia entre diferentes interacciones (atractivas o repulsivas). En estos sistemas, las fases de baja temperatura poseen mínimos locales difíciles de sortear. En una simulación de MC estándar como la descrita anteriormente, el tiempo para escapar de uno de estos mínimos aumenta rápidamente a medida que la temperatura decrece, haciendo que no sea posible acceder a un conjunto representativo de regiones del espacio de fases. Esto es particularmente evidente cuando se trabaja a altos cubrimientos.

Una primera aproximación para superar este inconveniente, consiste en incluir en la dinámica de simulación un paso difusivo de "relajación", tal como el descripto anteriormente en el paso 6. Esto, en nuestro caso, resulta ser una solución efectiva al problema, al menos para tamaños pequeños de $k$ (hasta valores de $k=6$). 

Es importante mencionar que existen estrategias mas sofisticadas para superar este problema, las cuales pueden ser clasificadas en dos grandes categorías. La primera son los algoritmos de clusters (o de actualización no local), tales como el de Swendsen y Wang \cite{swendsen}, el de Wolff \cite{wolff}, el algoritmo de pocket  \cite{pocket} y mas recientemente el algoritmo presentado por Kundu et al \cite{kundu}, el cual puede ser implementado en forma paralela. 

La segunda categoría son los denominados métodos de asamblea extendida, entre los cuales podemos mencionar: el método multi-canónico \cite{berg1991,berg1992}, el templado simulado \cite{marinari} y la versión paralelizada de este último llamada templado paralelo \cite{hukushima,earl,yan}.
Este último consiste en generar un sistema compuesto por $R$ replicas no interactuantes del sistema bajo estudio, donde cada réplica está asociada con un potencial químico $\mu_n$ y una temperatura $T$ diferentes. El proceso de templado paralelo consiste en combinar la simulación simultánea e independiente de cada réplica, junto con un proceso de intercambio de configuraciones, $\mathbf{x}_{i}$ y $\mathbf{x}_{i}$, entre dos réplicas a potenciales químicos y/o temperaturas diferentes, el cual es aceptado con cierta probabilidad.

\bigskip

Los seis pasos enumerados anteriormente representarán el paso elemental (MCs) en el proceso de simulación de MC. Típicamente, el estado de equilibrio puede ser reproducido después de descartar los primeros $r'=4 \times 10^6$ MCs. Luego, los próximos $r=4 \times 10^6$ MCs se usan para computar promedios. 

Las diferentes cantidades de interés, cubrimiento y energía configuracional promedio, se obtienen de la siguiente manera:
\begin{equation}
\left\langle \theta \right\rangle = \frac{1}{M} \sum_{i} \left\langle c_i \right\rangle
\end{equation}
\begin{equation}
\langle U \rangle = \left\langle H \right\rangle -\mu \left\langle N \right\rangle
\end{equation}
donde $\left\langle ... \right\rangle$ representa el promedio temporal sobre $r$ pasos de simulación de MC. 
El calor diferencial de adsorción $q_d$ puede ser obtenido de la siguiente manera. A partir de la gran funcion de partición $\Xi(\mu,T,M)$, se puede derivar la bien conocida relación termodinámica \cite{bakaev}: $(\partial \ln z / \partial \beta)_\langle N \rangle = \partial \langle U \rangle / \partial \langle N \rangle$ donde $\beta=1/k_BT$ y $z=\exp(\mu/k_BT)/h^3(2\pi m k_BT)^{3/2}$. Para un gas ideal, $z\rightarrow k_BTP$, siendo $P$ la presión, luego,
\begin{equation}
q_d = RT^2\left( \frac{\partial \ln P}{\partial T}  \right)_{\langle N \rangle}-RT = -\frac{\partial \langle U \rangle}{\partial \langle N \rangle} = -\frac{\langle UN \rangle - \langle U \rangle\langle N \rangle}{\langle N^2 \rangle - \langle N \rangle^2},
\end{equation}
donde $R$ es la constantes de los gases ($R/k_BT =$ número de Avogadro); $\langle UN \rangle$, $\langle N \rangle$ y $\langle N^2 \rangle$ pueden ser obtenidos a partir de la simulación.

\bigskip

Finalmente, otra cantidad de interés será el número de $k$-uplas disponibles en el sistema, $\eta$ ($k$-uplas vacías que pueden ser ocupadas por un $k$-mero),
\begin{equation}
\langle \eta \rangle = \frac{1}{r} \sum_{t=1}^r \eta( \mathbf{x}_{t} )
\end{equation}
donde la función $\eta( \mathbf{x}_{t} )$ representa el número instantáneo de $k$-uplas disponibles en la configuración $\chi^t$ al tiempo $t$ de simulación. 

\bigskip

Los cálculos computacionales se realizan en el Cluster BACO el cual está constituido por 30 PCs con CPU tipo Intel Core 2 QUAD Q9550 y más de 40 PCs con CPU Intel i7-3370 / 2600. Hay un total de 500 cores actualmente operativos en el CLUSTER. Cuenta con un servidor con CPU tipo Intel Core 2 QUAD Q9550 con disco redundante de 1TB. Todas las unidades están bajo Sistema Operativo Scientific Linux 5.0 y con sistema de gestión de colas de procesos (mecanismo de checkpoint y prioridad) llamado Condor. Tiene dos GPUs NVIDIA modelos GTX 480 y GTX 580 con 1.5GB y 2GB, respectivamente, conectadas en red para procesamiento de algoritmos en paralelo con lenguaje CUDA. 

Estudiando la transición nemática en un modelos de red cuadrada, en la ref. \cite{kundu} se propuso un  esquema de probabilidades de transición  alternativo que prueba ser ergódico y que cumple con el principio de reversibilidad microscópica también. Allí los k-meros son adsorbidos y desorbidos con probabilidades $p$ y $1-p$, respectivamente, donde $p$ esta relacionado con la actividad química $\lambda=e^{\beta(\mu-U_{o})}$ de la forma 

\begin{equation}\label{eq.c7.4}
\lambda=\frac{\theta p}{2k(1-p)}
\end{equation}     

En el futuro ampliaremos nuestro estudio sobre k-meros  con interacción con éste y otros esquemas que optimicen las simulaciones.   
  
\section{Resultados de Simulación para k-meros con interacción repulsiva}

Una de las motivaciones para analizar el comportamiento de el gas de red de k-meros es la rica variedad de transiciones de fase que desarrollan cuando interactúan entre si, especialmente para interacciones repulsivas que pueden producir fases ordenadas \cite{dimeros_int_repulsivas}. Aún en el caso de k-meros sin interacción existe evidencia de transición nemática para $k>6$ en redes cuadradas. En definitiva existe una amplia variedad de fenómenos críticos asociados a un gas de red de partículas lineales y es nuestra motivación interpretarlos en función de los conceptos de exclusión estadística y estadísticas fraccionarias.

Mostramos y analizamos a continuación las isotermas de adsorción $\theta$ versus $\beta \mu$ y la exclusión de estados a partir de la probabilidad $P_{o}$ en función de $\theta$.          

\begin{figure}[H]
	\centering
	\includegraphics[width=1\linewidth]{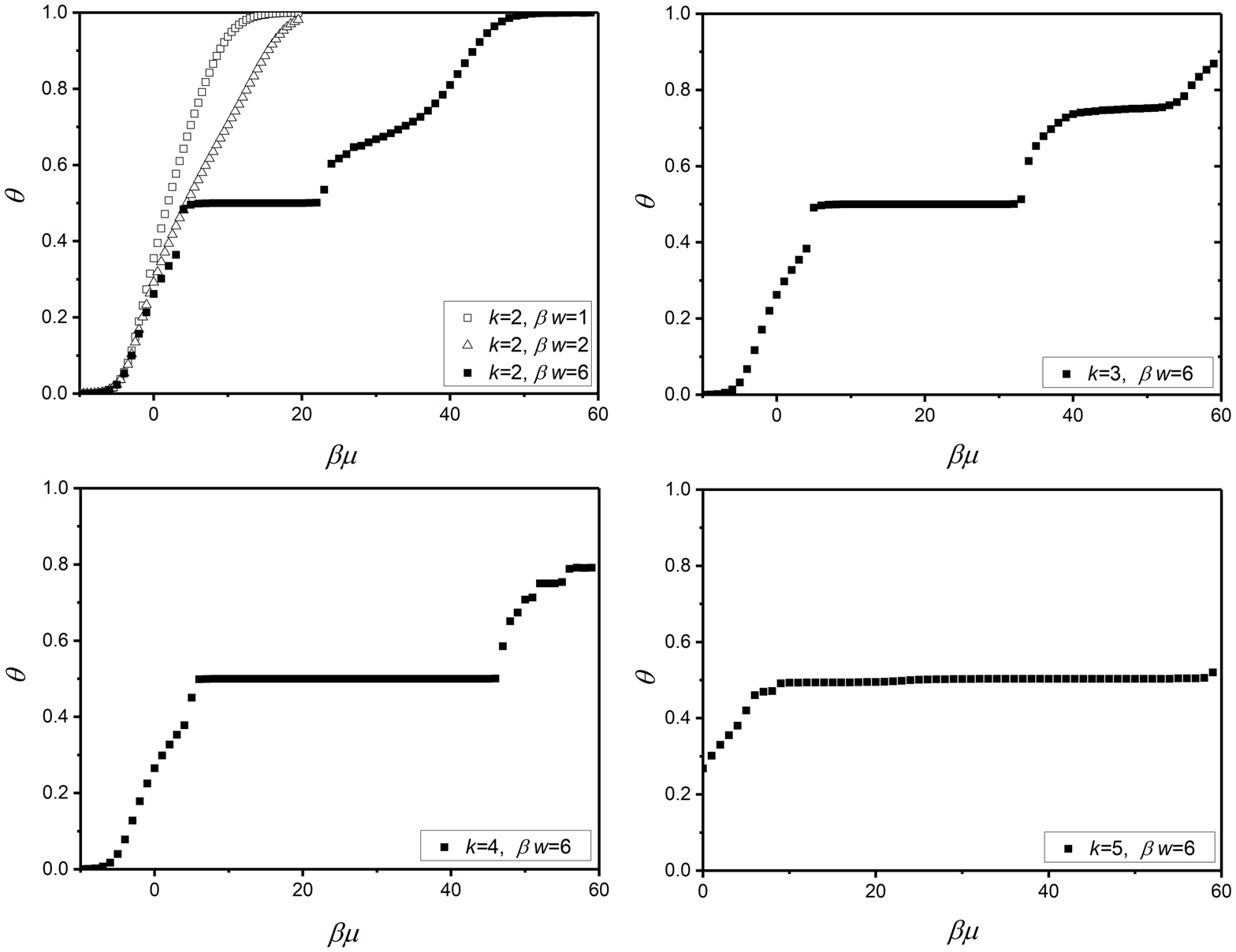}
	\caption{Isotermas de Adsorcion de k-meros sobre red cuadrada con interacciones repulsivas}
	\label{Fig.isotermas_k_2_3_4_5_repulsiva_merged}
\end{figure}

\begin{figure}[h]
	\centering
	\includegraphics[width=1\linewidth]{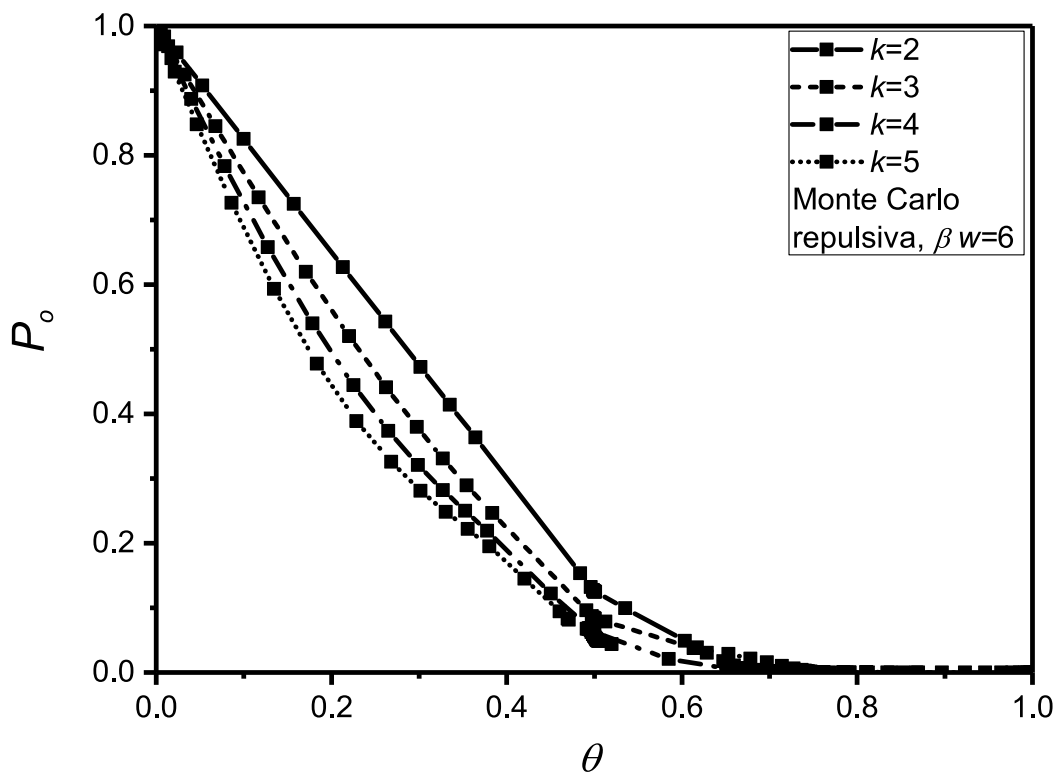}
	\caption{$P_{o}$ versus $\theta$ para k-meros con interacción repulsiva sobre red cuadrada. Los símbolos conectados con líneas representan los resultados de Monte Carlo para $k=2,3,4$ y $k=5$  }
	\label{fig:pokmerosconinteraccionrepulsiva}
\end{figure}

Las isotermas para el caso repulsivo $\beta W=6$ muestran una meseta bien definida a $\theta=0.5$ para todos los valores de $k$. Sin embargo, la forma de la función $P_{o}$ es diferentes para los distintos $k$, con segmentos claramente lineales de pendientes diferentes para $k=2$ que induce a pensar en rangos de cubrimiento con un orden espacial diferente de las partículas sobre la red en cada uno de ellos. 

\noindent En el otro extremo, $k=5$ muestra un decaimiento con pendiente variable hasta  $\theta=0.5$ y aún cuando la meseta de la isoterma podría ser indicio de orden en el gas de red el resultado no es concluyente. 

\noindent Como veremos en adelante la cantidad $P_{o}$ será valiosa para interpretar el comportamiento termodinámico del gas de k-meros a la luz del concepto de exclusión estadística de estados.            

\section{K-meros sin interacción. Comparación con Estadística Fraccionaria}

En esta sección presentamos el caso de k-meros sin interacción y comparamos con las predicciones de la estadística fraccionaria para adsorción. De los resultados surge que esta estadística tiene limitaciones para describir los sistemas con $k>2$ donde los efectos de exclusión múltiple de estados son más significativos.
\begin{figure}[h]
	\centering
	\includegraphics[width=1\linewidth]{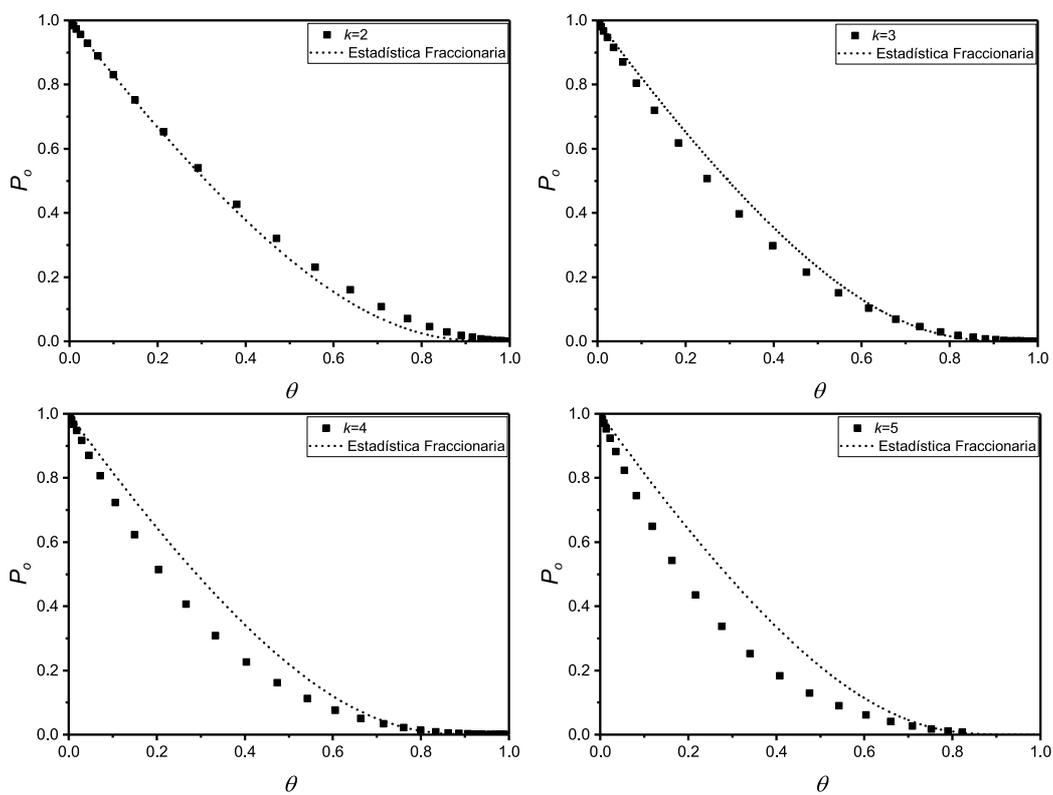}
	\caption{$P_{o}$ versus $\theta$ para k-meros sin interacción. Comparación con resultados de la estadística fraccionaria (linea punteada).}
	\label{fig:Po_kmeros_sin_interaccion}
\end{figure}

Estos resultados de $P_{o}$ muestran que la Estadística Fraccionaria reproduce los resultados aproximadamente sólo para $k=2$, sin embargo difiere apreciablemente para $k>2$ y el desacuerdo es mayor a medida que aumenta $k$.  
 
\begin{figure}[h]
	\centering
	\includegraphics[width=1\linewidth]{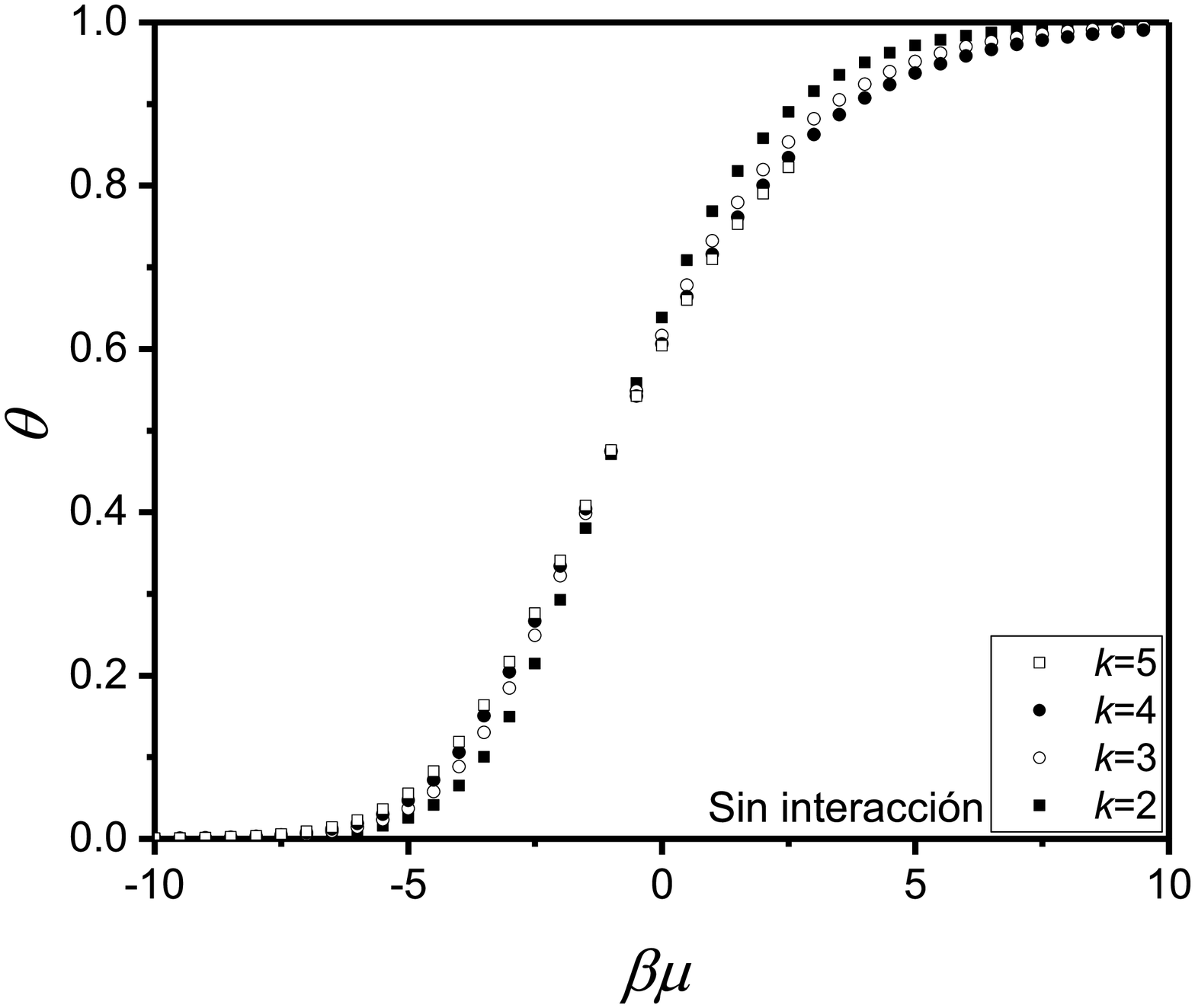}
	\caption{Isotermas de adsorcion de k-meros sin interacción; $k=2,3,4$ y $5$.}
	\label{fig:Isotermas_kmeros_sin_interaccion}
\end{figure}

%% file: texto/TF_Estadistica_de_Multiple_Exclusion_notacion_mas_reducida.tex
\chapter{Estadística de Múltiple Exclusión}{\label{Capitulo_Est_M_E}}

\section{Una nueva estadística para la adsorción de k-meros}
En este capítulo desarrollamos una nueva estadística, denominada \textbf{Estadística de Múltiple Exclusión} para partículas poliatómicas que resuelve las limitaciones de la estadística cuántica fraccionaria de Haldane y su extensión al fenómeno de adsorción. 

\noindent  Las soluciones de gas de red de k-meros obtenidas en el capítulo 3 están basadas en las formas mas simples del fenómeno de exclusión de estados, y en consecuencia para $G_{o}(N)$ y $ d_{N}$

\begin{equation}\label{eq.3.27.d}
\begin{aligned}
d_{N}\equiv d(N)&=G-G_{o}(N)=G-\sum_{N^{'}=1}^{N-1} g(N^{'})=G-\sum_{N^{'}=1}^{N-1} g \,N^{'} \\
&=G- g\,(N-1)
\end{aligned}
\end{equation}

\noindent donde la exclusión por partícula $g$ es constante e independiente del número de partículas presentes en en sistema. Esto es equivalente a considerar que los $G$ estados accesibles son \textbf{independientes} entre si, y que cada los $g$ estados excluidos por una dada partícula \textbf{no pueden ser excluidos por ninguna otra}. en otros términos, si designamos $\mathbf{E_{i}^{l}}$ y $\mathbf{E_{j}^{l}}$ a los  conjuntos de estados excluidos por las partículas $i$ y $j$ en la configuración $l$ del sistema, entonces $ \mathbf{E_{i}^{l}} \bigcap \mathbf{E_{j}^{l}}=\emptyset\ \  \forall \ \mathbf{l},\ \forall \ \mathbf{i,j}  \in \:N $, es decir, tienen intersección vacía.

\section{Conjetura de múltiple exclusión de estados }

\begin{figure}[h]
	\centering
	\includegraphics[width=1\linewidth]{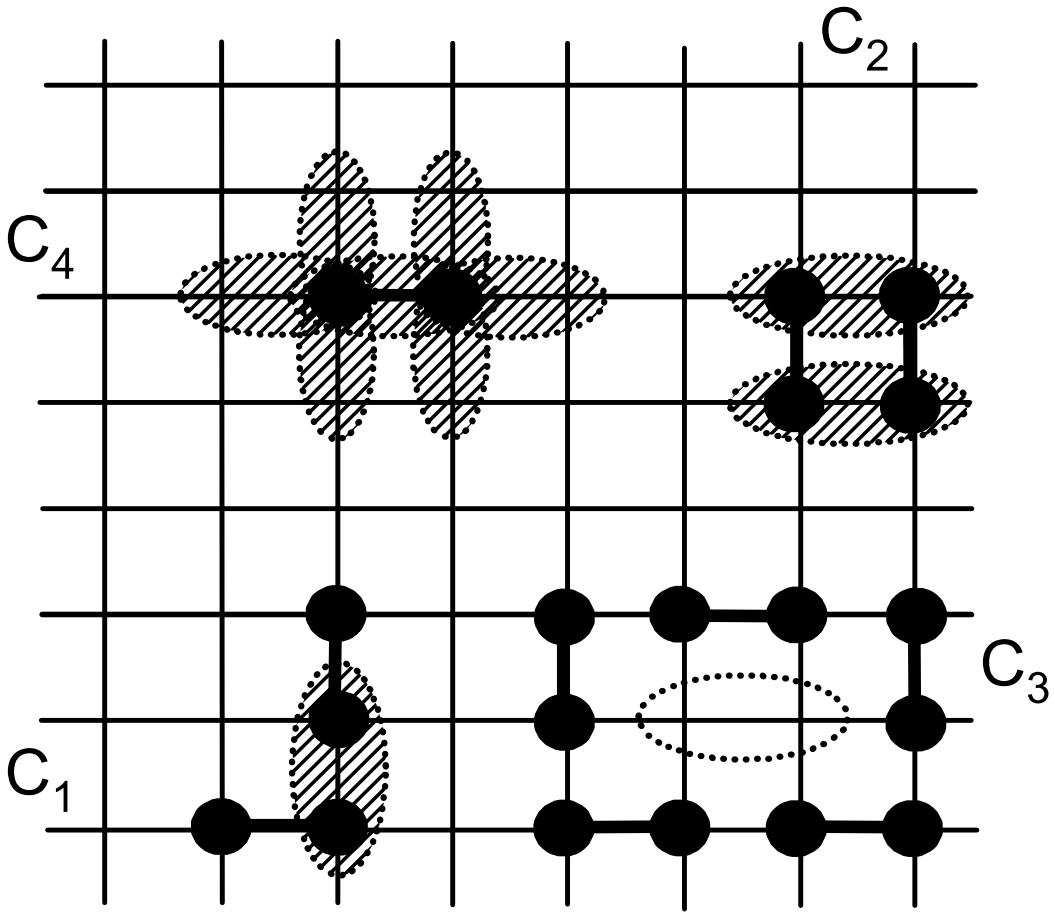}
	\caption{Representación esquemática de configuraciones de dimeros sobre red cuadrada donde se muestran en areas sombreadas los estados mutuamente excluidos por partículas vecinas.}
	\label{Fig.3.9}
\end{figure}

\noindent Es claro que cuando tratamos con partículas sobre una red de sitios con geometría dada esta condición no se cumple porque existe \textbf{múltiple exclusión de estados}. La \ref{Fig.3.9} muestra esta fenómeno para el caso simple de dímeros sobre una red cuadrada. 

\noindent De los estados excluidos por ambos dímeros en la configuración $C_{1}$ hay uno que es excluido por ambos. Para la configuración $C_{2}$ hay dos estados mutuamente excluidos. Por otra parte es claro que el dímero que ocupe el estado accesible de la configuración $C_{3}$ solo excluye un estado (el que ocupa) efectivamente ya que los demás estados los excluye mutuamente con las partículas vecinas. También es evidente de la configuracion $C_{4}$ que un dimero aislado $\theta \to 0$ excluye $7$ estados (en general $k^2+2k-1$ para un k-mero). Se concluye que en general la exclusión estadística no es constante y tampoco depende solo de $N$ sino que depende de la configuración en que se encuentre el sistema. Como veremos, en el gas de red el parámetro $g$ representa la exclusión estadística por partícula  a $\theta \to 1$.     

\noindent Esta nueva estadística que desarrollamos se basa en considerar de manera general el hecho de que un sistema de muchas partículas con estructura geométrica arbitraria sobre una red (o en el contínuo) excluyen mutuamente estados del conjunto total que deben ser adecuadamente cuantificados para calcular la entropía correctamente. 

\noindent Sea $d_{N}$ el número de estados disponibles del total $G$ para la $N$-ésima partícula que se agrega al volúmen $V$ y $g$ el número de estados que excluye cada partícula aislada a $\theta \to 1$. Así

\begin{equation}\label{eq.3.28}
d_{1}=G
\end{equation}

\noindent A partir de \eqref{eq.3.28}, podemos escribir la siguiente relación de recurrencia para las demás partículas $2,3,\dots,N$ que se incorporen en el volumen $V$

\begin{equation}\label{eq.3.29}
\begin{aligned}
d_{1}&=G \\
d_{2}&=d_{1} - \mathcal{N}_{1} \\
.\\
.\\
.\\
d_{N}&=d_{N-1} - \mathcal{N}_{N-1} 
\end{aligned}                         
\end{equation}  

\noindent donde $\mathcal{N}_{j}$ es el número de estados ``efectivamente'' excluidos por la partícula $j$. Debido a que las partículas excluyen estados mutuamente, es decir hay estados excluidos por dos partículas vecinas como se observa en la Fig. \ref{Fig.3.9}, entonces en general el número de estados efectivamente excluidos  \textbf{por partícula} sera $1+g_{e}$, distinto de $g$ que como dijimos es el valor de exclusión a cubrimiento máximo $\theta \to 1$. Notar que $1+g_{e}$ cuenta el estado ocupado por la partícula, solo uno (que no puede ser excluido por ninguna otra), más $g_{e}$ estados ``excluidos'' por la partícula (sin contar el ocupado por ella). 

\noindent Nuestra \textbf{conjetura de múltiple exclusión} es la siguiente 

\begin{equation}\label{eq.3.30}
\mathcal{N}_{j}=1+g_{e} \:\mathcal{P}_{j}
\end{equation} 
   
\noindent donde $\mathcal{P}_{j}\equiv$ probabilidad de que un estado este disponible para ser excluido por la partícula $j$ y la definimos como 

\begin{equation}\label{eq.3.31}
\mathcal{P}_{j}= \frac{d_{j}}{G}
\end{equation}

\noindent e.d, el cociente del numero de estados accesibles a $j$ sobre el número total de estados $G$. Es implícito en la definición \eqref{eq.3.31} que se asume homogeneidad estadística en el conjunto de los estados accesibles a la partícula $j$, $d_{j}$, en el sentido que todos ellos son considerados equivalentes. La ec. \eqref{eq.3.30} es condición básica que tiene en cuenta que el número de estados excluidos por partícula es una función de $N$. $\mathcal{P}_{j}$ depende obviamente de la relación topológica entre el tamaño y estructura geométrica de las partículas y la geometría de la red de sitios que ocupan. 

\noindent La interpretación de la ec. \eqref{eq.3.30} es que $\mathcal{N}_{j}$ es igual a un (1) estado ocupado por la partícula mas una fracción de los $g_{e}$ estados adicionales que puede excluir en presencia de las demás ,  $g_{e} \:\mathcal{P}_{j}$, donde el factor  $\mathcal{P}_{j}$ es la fracción de estados vacíos para la partícula $j$ y por lo tanto es la probabilidad de que un estado este vacío para ser excluido. 

\begin{equation}\label{eq.3.32}
\mathcal{N}_{j}=1+g_{e} \: \frac{d_{j}}{G}
\end{equation}

\noindent Reemplazando \eqref{eq.3.32} en el sistema de ecuaciones recurrentes \eqref{eq.3.29}, obtenemos

\begin{equation}\label{eq.3.33}
\begin{aligned}
d_{1}&=G \\
d_{2}&=d_{1} -\left[ 1+ g_{e} \frac{d_{1}}{G} \right]=d_{1}\left[1-\frac{g_{e} }{G} \right]-1=G \left[1-\frac{g_{e} }{G} \right]-1  \\
d_{3}&=d_{2} -\left[ 1+ g_{e} \frac{d_{2}}{G} \right]=d_{2}\left[1-\frac{g_{e} }{G} \right]-1  \\
&=\left[G\left[1-\frac{g_{e} }{G}\right]-1 \right] \:\left[ 1-\frac{g_{e} }{G}\right]-1   \\
&=G\left[1-\frac{g_{e} }{G}\right]^{2}-\left[1-\frac{g_{e} }{G}\right]-1 \\
.\\
.\\
.\\
d_{N}&=d_{N-1} -\left[ 1+ g_{e} \frac{d_{N-1}}{G} \right]\\
&=G\left[1-\frac{g_{e} }{G}\right]^{N-1} - \left[1-\frac{g_{e} }{G}\right]^{N-2}-\dots-\left[1-\frac{g_{e} }{G}\right]-1
\end{aligned}
\end{equation}

\noindent Ahora de la ecs.\eqref{eq.3.1} y \eqref{eq.3.33} obtenemos $G_{o}(N)$ y $d_{N}=d(N)$ que son las funciónes básicas para escribir las funciones termodinámicas (ver ecs. \eqref{eq.3.8}, \eqref{eq.3.9} y \eqref{eq.3.10})

\begin{equation}\label{eq.3.34}
G_{o}(N)= G-d_{N}
\end{equation}

\noindent y tomamos el  límite

\begin{equation}\label{eq.3.34.a}
\begin{aligned}
\tilde{G}_{o}(n)\equiv \lim_{N,G \to \infty} \frac{G_{o}(N)}{G}&=\lim_{N,G \to \infty}\frac{1}{G}\left(G-d_{N}\right)\\
&=1-\lim_{N,G \to \infty}\frac{d_{N}}{G}=1-\tilde{d}(n)                    
\end{aligned}
\end{equation}

 con $n=N/G$.
 
\noindent Para resolver este límite usamos la propiedad $\lim (f(x)+g(x))= \lim f(x)+\lim g(x)$, entonces

\begin{equation}\label{eq.3.34.b}
\begin{aligned}
\tilde{d}(n)&=\lim_{N,G \to \infty}\frac{d_{N}}{G}\\
&=\lim_{N,G \to \infty} \left\lbrace \left[1-\frac{g_{e} }{G}\right]^{N-1}-\frac{1}{G}\left[\sum_{j=1}^{N-2} \left[1-\frac{g_{e} }{G}\right]^{j} +1\right] \right\rbrace \\
&=\lim_{N,G \to \infty} \left[1-\frac{g_{e} }{G}\right]^{N-1} - \lim_{N,G \to \infty}\frac{1}{G}\left[\sum_{j=1}^{N-2} \left[1-\frac{g_{e} }{G}\right]^{j} +1\right]
\end{aligned}
\end{equation}


\noindent En el límite de la sumatoria los términos que incluyen el factor $\frac{g_{e}-1}{G}$ no sobreviven porque tienden a zero y solo los $(N-2)$ términos constantes contribuyen, con los cual 

\begin{equation}\label{eq.3.34.c}
\begin{aligned}
\lim_{N,G \to \infty}\frac{1}{G}\left[\sum_{j=1}^{N-2} \left[1-\frac{g_{e} }{G}\right]^{j} +1\right]&=\lim_{N,G \to \infty}\frac{(N-2)+\mathcal{O}\left(\nicefrac{1}{G}\right)+1}{G} \\
&=\lim_{N,G \to \infty} \frac{N-1}{G}=n
\end{aligned}
\end{equation}

\noindent Por otra parte, para obtener el $\lim_{N,G \to \infty} \left[1-\frac{g_{e} }{G}\right]^{N-1}$ definimos la función

\begin{equation}\label{eq.3.37}
h(N)=\left( 1-\frac{g_{e} }{G}\right) ^{N-1}=\left( 1-\frac{n g_{e} }{N}\right)^{N-1}
\end{equation} 

\noindent y usamos la propiedad $\ln\left[ \lim_{N \to \infty} h(N)\right] = \lim_{N \to \infty} \ln h(N)$

\begin{equation}\label{eq.3.38}
\lim_{N \to \infty} \ln h(N)=\lim_{N \to \infty}\left[ (N-1) \ln \left( 1-\frac{n g_{e} }{N}\right)  \right] 
\end{equation}

\noindent haciendo el cambio de variables $\xi=n g_{e} /N$ en \eqref{eq.3.38} y desarrollando alrededor de $\xi=0$, $\ln(1-\xi)=-\xi -\xi^{2}/2 - \xi^{3}/3- \dots$ 

\begin{equation}\label{eq.3.39}
\begin{aligned}
\lim_{N \to \infty} \ln h(N)
&= \lim_{\xi \to 0} \left( \frac{n g_{e} }{\xi}-1\right)  \ln \left( 1-\xi\right)\\
&=  \lim_{\xi \to 0} \left(-n g_{e}  -\left( \frac{ng_{e} }{2}-1\right) \xi  + \mathcal{O}(\xi^{2})  \right)\\
&=-ng_{e}   
\end{aligned}
\end{equation}

\noindent luego 
\begin{equation}\label{eq.3.39.a}
\lim_{N \to \infty} h(N)=e^{-n g_{e} } 
\end{equation}

Finalmente, de \eqref{eq.3.34.a}, \eqref{eq.3.34.b}, \eqref{eq.3.34.c} y \eqref{eq.3.39.a} resulta

\begin{equation}\label{eq.3.45}
\tilde{d}(n)= e^{-n g_{e} }-n 
\end{equation}
 
\noindent

\begin{equation}\label{eq.3.42}
\tilde{G}_{o}(n)=1-\left[ e^{-n g_{e} }-n\right] 
\end{equation}

\noindent Este resultado es muy interesante porque muestra que $\tilde{d}(n)$ tiene decaimiento exponencial típico y que el parámetro de múltiple exclusión estadística por partícula $g_{e}$ es la constante de decaimiento. La ec. \eqref{eq.3.45}, o su análoga \eqref{eq.3.42},  es la relación fundamental de la \textbf{Estadística de Múltiple Exclusión } de la cual se derivan toda su propiedades termodinámicas. Es interesante también que la derivada negativa de $\tilde{d}$ a $n\to 0$ es $\tilde{G}_{o}^{'}(0)=-\tilde{d}'(0)=g_{e}$, es decir, es el valor de exclusión estadística efectiva por partícula en presencia de múltiple exclusión. 

\noindent $\tilde{d}(n)$ en \eqref{eq.3.45} esta definida excepto  constantes ya que debe cumplir con los límites

\begin{equation}\label{eq.3.43}
\tilde{d}(0)=1  \quad y \quad \tilde{d}(n_{m})=0 
\end{equation}

\noindent Haciendo $\tilde{d}(n)=C_{1} e^{-n g_{e} }-C_{2} n$  y determinando $C_{1}$ y $C_{2}$  de la condición \eqref{eq.3.43} con $n_{m}=1/g$, $C_{1}=1$ y  

\begin{equation}\label{eq.3.43.b}
C_{2}=g \: e^{-\frac{g_{e}}{g}}
\end{equation}

\begin{equation}\label{eq.3.43.c}
\tilde{d}(n)= e^{  -n g_{e}  }-g\: e^{ -\frac{g_{e} }{g}  } n
\end{equation}

\begin{equation}\label{key}
\tilde{d}^{'}(n)\equiv \frac{\tilde{d}(n)}{dn}=-g_{e} e^{-ng_{e} } - g e^{-\frac{g_{e} }{g}} 
\end{equation}

\begin{equation}\label{eq.3.44}
\tilde{G}_{o}(n)=1-\left[ e^{ -n g_{e}  }-g\: e^{- \frac{g_{e} }{g} } n \right]  
\end{equation}

\noindent El valor de $g_{e}$ no es arbitrario sino que esta completamente determinado por el valor de la exclusión estadística para la partícula completamente aislada sobre la red $g_{o}=<g>_{(n=0)}$. Como mostraremos detalladamente en el Capítulo 7,  

\begin{equation}\label{eq.3.44.a}
<g>(n)= \frac{(1-P_{\circ})}{n}=\frac{1}{n}-  \left[  \frac{\left[ \tilde{d}(n)+n\right]^{\tilde{d}^{'}(n)+1}}{n \:\left[ \tilde{d}(n)\right]^{\tilde{d}^{'}(n)} }\right]   
\end{equation}	 

\noindent y de la condición $\lim_{n \to 0} <g(n)>=g_{o}$ se determina $g_{e}$. Recordemos que $g_{o}=2k-1=2g-1$ para k-meros en 1D y $g_{o}=k^2+2k-1=(g^2/4)+g-1$ para k-meros sobre una red cuadrada por ejemplo (ver Fig. \ref{Fig.3.9}). 

\noindent Resolviendo el límite,  finalmente la condición para determinar $g_{e}$ es la ecuación  

\begin{equation}\label{eq.3.44.b}
\begin{aligned}
\lim_{n \to 0} <g(n)&=\lim_{\theta \to 0} <g(\theta)> \\ &=\lim_{\theta \to 0} \frac{g (1-P_{\circ})}{\theta}=2e^{\frac{-g_{e}}{g}}+2g_{e}-1=\frac{g^2}{4}+g-1=g_{o}
\end{aligned}
\end{equation}.
\noindent cuya solución para la red cuadrada es\footnote{La solución de la ec. \eqref{eq.3.44.b} es $g_{e}=\frac{g^{2}}{8}+\frac{g}{2}+g \mathcal{L}(z)$ para $g\geq 4$, donde $\mathcal{L}(z)$ es la solución (positiva) de $z= \mathcal{W}(z) \: e^{\mathcal{W}(z)}$, y $\mathcal{W}(z)$ es la denominada Función de Lambert, que resulta ser la función inversa de $f(x)=x \: e^{x}$, $x=\mathcal{W}(x \: e^{x})$.} : para $k=2 (g=4)$, $g_{e}=0$; para  $k=3 (g=6)$, $g_{e}=4.80731$; para $k=4 (g=8)$, $g_{e}=9.55863$ y para $k=5 (g=10)$, $g_{e}=15.3442$. 
   
\noindent La función $\tilde{d}(n)$ (ec. \ref{eq.3.43.c}) ha quedado completamente determinada luego de cumplir las condiciones de borde  $\tilde{d}(0)=1$, $\tilde{d}(n_{m})=0$ y $<g>(0)=g_{o}$. Una consecuencia importante de esta formulación es que la Estadística Fraccionaria de Haldane se obtiene como caso límite particular de la Estadística de Múltiple Exclusión\footnote{Para k-meros en 1D, $g_{o}=2g-1$ y la solución de la ec. \eqref{eq.3.44.b} es $g_{e}=0$ $\forall g$ ($\forall k$).} para $g\leq4$, ya que $\tilde{d}(n)=1-n \:g$ para $g\leq4$ ($g_{e}=0$) y recuperamos la forma \eqref{eq.3.12}
 
\noindent Además, de la ec. \eqref{eq.3.44}   

\begin{equation}\label{eq.3.46}
\tilde{G}_{o}^{'}(0)=g_{e}+ g \:e^{-\frac{g_{e}}{g}}
\end{equation} 


\noindent Usando la relación $\tilde{d}(n)=1-\tilde{G}_{0}(n)$  vemos que $\tilde{f}$, $\tilde{S}$ y $\mu$ toman las siguiente formas simples y muy manejables. 

\begin{equation}\label{eq.3.47.a}
\beta \tilde{f}(n)= \beta n U_{o}- \left[  \tilde{d}(n)+n\right]  \ln \left[  \tilde{d}(n)+n\right] + \tilde{d}(n) \ln \tilde{d}(n) +n \ln n
\end{equation} 

\noindent y análogamente para la entropía y la isoterma de adsorción,  

\begin{equation}\label{eq.3.48.a}
\frac{\tilde{S}(n)}{k_{B}}=  \left[  \tilde{d}(n)+n\right]  \ln \left[  \tilde{d}(n)+n\right] - \tilde{d}(n) \ln \tilde{d}(n) - n \ln n
\end{equation}

\begin{equation}\label{eq.3.49.a}
e^{\beta \left( \mu-U_{o}\right) }=\frac{n \,\left[ \tilde{d}(n)+n\right]^{-\left( \tilde{d}'+1\right) } }{\left[ \tilde{d}(n)\right]^{-\tilde{d}'}}
\end{equation}

\noindent Reemplazando la ec. \eqref{eq.3.43.c} en \eqref{eq.3.47.a}, \eqref{eq.3.48.a} y \eqref{eq.3.49.a} obtenemos las funciones termodinámicas por estado de la nueva \textbf{Estadística de Múltiple Exclusión} en su forma normalizada

\begin{equation}\label{eq.3.47}
\begin{aligned}
\beta \tilde{f}(n)= \beta n U_{o}&- \left[ e^{  -n g_{e}  }-g\: e^{ -\frac{g_{e} }{g}  } n+n\right]  \ln \left[e^{  -n g_{e}  }-g\: e^{ -\frac{g_{e} }{g}  } n +n\right]\\
&+ \left[e^{  -n g_{e}  }-g\: e^{ -\frac{g_{e} }{g}  } n\right] ln \left[e^{  -n g_{e}  }-g\: e^{ -\frac{g_{e} }{g}}  n\right]+ n \ln n
\end{aligned}
\end{equation} 

\begin{equation}\label{eq.3.48}
\begin{aligned}
\frac{\tilde{S}(n)}{k_{B}}=&  \left[ e^{  -n g_{e}  }-g\: e^{ -\frac{g_{e} }{g}  } n +n\right]  \ln \left[e^{  -n g_{e}  }-g\: e^{ -\frac{g_{e} }{g}  } n  +n\right] \\
&- \left[e^{  -n g_{e}  }-g\: e^{ -\frac{g_{e} }{g}  } n \right] \ln \left[ e^{  -n g_{e}  }-g\: e^{ -\frac{g_{e} }{g}  } n\right]  - n \ln n
\end{aligned}
\end{equation}

\begin{equation}\label{eq.3.49}
\begin{aligned}
e^{\beta \left( \mu-U_{o}\right) }=\frac{n \,\left[e^{  -n g_{e}  }-g\: e^{ -\frac{g_{e} }{g}  } n +n\right]^{\left[ g_{e} e^{-ng_{e} } + g e^{-\frac{g_{e} }{g}} -1\right]  } }{\left[e^{  -n g_{e}  }-g\: e^{ -\frac{g_{e} }{g}  } n \right]^{\left[ g_{e} e^{-ng_{e} } + g e^{-\frac{g_{e} }{g} }\right] }}
\end{aligned}
\end{equation}

\begin{equation}\label{ec.c6.8}
P_{\circ}(\theta)=\frac{\left[e^{  -n g_{e}  }-g\: e^{ -\frac{g_{e} }{g}  } n +n\right]^{-\left[ g_{e} e^{-ng_{e} } + g e^{-\frac{g_{e} }{g}} -1\right]  } }{\left[e^{  -n g_{e}  }-g\: e^{ -\frac{g_{e} }{g}  } n \right]^{-\left[ g_{e} e^{-ng_{e} } + g e^{-\frac{g_{e} }{g} }\right] }}
\end{equation}

\begin{equation}\label{ec.c6.9}
\begin{aligned}
P_{\varnothing}(\theta)=1&-\frac{\theta}{g} \\
&-\frac{\left[e^{  -n g_{e}  }-g\: e^{ -\frac{g_{e} }{g}  } n +n\right]^{-\left[ g_{e} e^{-ng_{e} } + g e^{-\frac{g_{e} }{g}} -1\right]  } }{\left[e^{  -n g_{e}  }-g\: e^{ -\frac{g_{e} }{g}  } n \right]^{-\left[ g_{e} e^{-ng_{e} } + g e^{-\frac{g_{e} }{g} }\right] }}
\end{aligned}
\end{equation}

\section{Formas de $\tilde{f}$, $\tilde{S}$ y $\mu$ en nomenclatura de gas de red}

Las funciones \eqref{eq.3.47}, \eqref{eq.3.48} y \eqref{eq.3.49} las podemos escribir en nomenclatura de gas de red, con las definiciones usadas anteriormente: $M$ sitios, $n_{m}=N_{m}/G=N_{m}/mM= N_{m}k'/(mk'M)=\theta_{m}/mk'=1/g$ y $n=a\theta=\theta /g \ (ec.\eqref{eq.3.14})$.

\noindent Del cambio de variables $n=\theta/g$ en la ec.\eqref{eq.3.43.c} se obtiene 

\begin{equation}\label{eq.3.49.x}
\begin{aligned}
\tilde{d}(n=\nicefrac{\theta}{g})=\varTheta(\theta)=  e^{  -\theta \frac{g_{e}}{g}  }- e^{ -\frac{g_{e} }{g}  } \: \theta
\end{aligned}
\end{equation}

\noindent Notar que $\varTheta(\theta)=\tilde{d}(n=\nicefrac{\theta}{g})$ y que $\varTheta^{'}\equiv d\varTheta(\theta)/d\theta=\frac{1}{g} d \left( \tilde{d}(n)\right) /dn \mid_{n=\theta/g}= \frac{1}{g}\tilde{d}^{'}(\theta/g)$. Además, $f_{k}(\theta)=m\tilde{f}(n)$, y operando resulta

\begin{equation}\label{eq.3.48.x}
\begin{aligned}
f_{k}(\theta)=\beta \frac{\theta}{k'} U_{o}&- \left[m\, \varTheta(\theta)  +\frac{\theta}{k'}\right]  \ln \left[m\, \varTheta(\theta)  +\frac{\theta}{k'}  \right] + m\, \varTheta(\theta) \ln\left[  m\, \varTheta(\theta)\right]  \\ 
&+ \frac{\theta}{k'} \ln \frac{\theta}{k'}
\end{aligned}
\end{equation}

\noindent

\begin{equation}\label{eq.3.50}
\frac{S_{k}(\theta)}{k_{B}}= \left[m\, \varTheta(\theta)  +\frac{\theta}{k'}\right]  \ln \left[m\, \varTheta(\theta)  +\frac{\theta}{k'}  \right] - m\, \varTheta(\theta) \ln\left[  m\, \varTheta(\theta)\right]  - \frac{\theta}{k'} \ln \frac{\theta}{k'}
\end{equation}
\noindent y 
\begin{equation}\label{eq.3.52}
\begin{aligned}
e^{\beta \left( \mu-U_{o}\right) }&=\frac{\frac{\theta}{g} \,\left[ \varTheta\left( \theta\right) +\frac{\theta}{g}\right]^{-\left( g \varTheta'+1\right) } }{\left[ \varTheta\left( \theta\right) \right]^{-g\varTheta'}} \\  \\
&= \frac{\theta \,\left[g\, \varTheta\left( \theta\right) +\theta\right]^{-\left( g \varTheta'+1\right)}}{\left[ g\, \varTheta\left( \theta\right) \right]^{-g\varTheta'}}
\end{aligned}
\end{equation}

\noindent con lo que queda completo el conjunto de funciones termodinámicas analíticas de la \textbf{Estadística de Múltiple Exclusión}. 

\noindent En las siguientes secciones realizamos en forma preliminar la comparación de estos resultados con simulaciones de Monte Carlo . 

\section{Comparación entre Estadísticas}
\subsection{Isotermas de adsorción}

En esta sección realizamos la comparación de las isotermas de adsorción de k-meros obtenidas por simulación en el conjunto gran canónico y las isotermas analíticas obtenidas en el marco de la Estadística de Múltiple Exclusión para varios casos. También incluimos la comparación con las isotermas analíticas de la Estadística Fraccionaria para mostrar cuál es  la solución mas aproximada. 

\begin{figure}[H]
	\centering
	\includegraphics[width=1\linewidth]{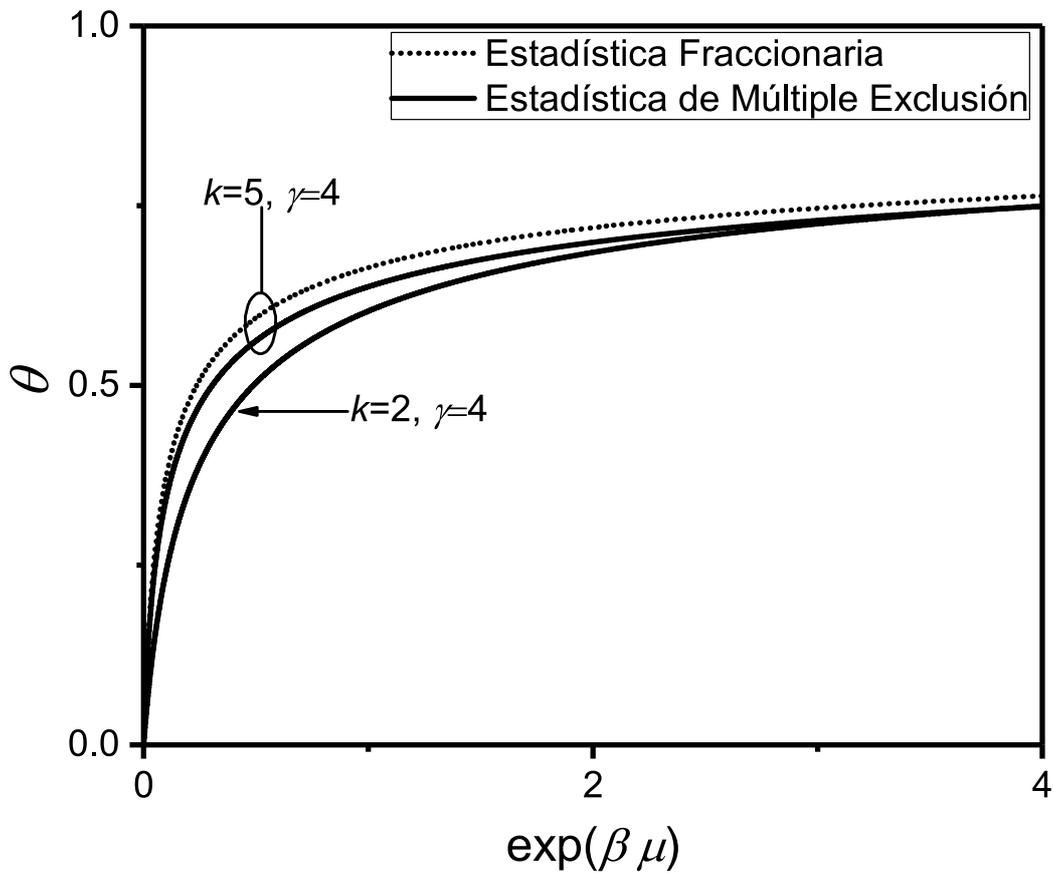}
	\caption{Comparación de isotermas de adsorción de las Estadística de Múltiple Exclusión (ec. \eqref{eq.3.52}) y Estadística Fraccionaria (ec. \eqref{eq.3.49}), para $k=2$ y $k=5$ en red cuadrada $(\gamma=4)$. Para $k=2$ ambas estadísticas coinciden}
	\label{fig.IsotCompMEyFE}
\end{figure}

\begin{figure}[H]
	\centering
	\includegraphics[width=1\linewidth]{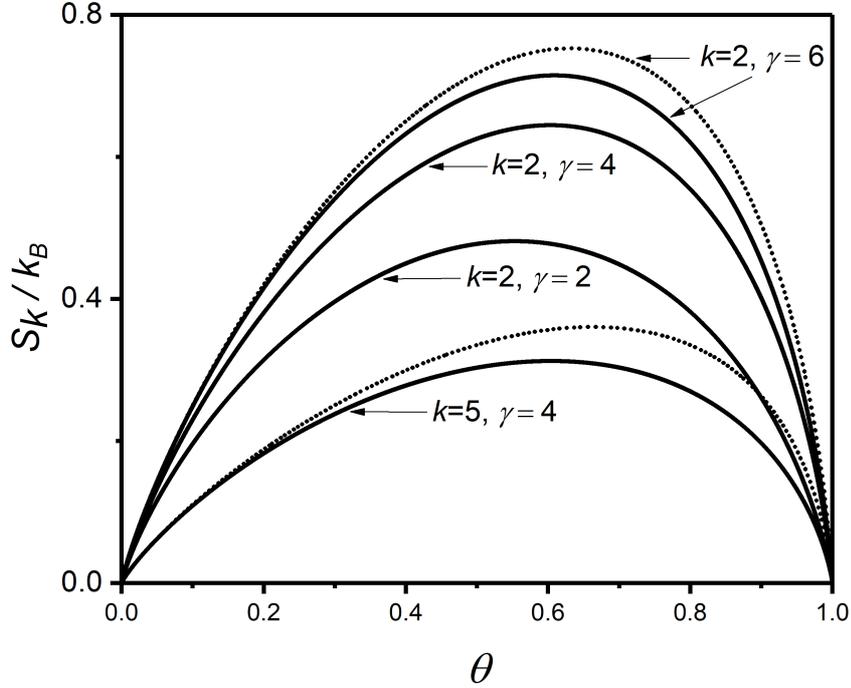}
	\caption{Comparación de la entropía por sitio de las Estadística de Múltiple Exclusión (ec. \eqref{eq.3.50}) y Estadística Fraccionaria (ec. \eqref{eq.3.48}), para $k=2$ en 1D ($\gamma=2$),  $g=2$; red cuadrada ($\gamma=4$), $g=4$; red triangular $(\gamma=6)$, $g=6$; y $k=5$ ($\gamma=4$), $g=10$.}
	\label{fig.EntropCompMEyFE}
\end{figure}

\noindent En la Figs. \ref{fig.IsotCompMEyFE} y \ref{fig.EntropCompMEyFE} se muestra la comparación de los resultados de la isoterma de adsorción y entropía por sitio de las Estadísticas de Múltiple Exclusión y Fraccionaria.

\subsection{Entropía}

En esta sección realizamos la comparación de la entropía por sitio $S_{k}$ de k-meros obtenida de la Estadística de Múltiple Exclusión para varios casos e incluimos la comparación con las funciones correspondientes de la Estadística Fraccionaria.  

\noindent La diferencias en los valores de la entropía de ambos no son despreciables en el mas simple de los casos $k=2$ sobre la red triangular. Cabe notar que la Estadística de Múltiple Exclusión converge a la Estadística Fraccionaria en el caso límite $g\leq4$ por lo que los resultados coinciden para esos valores de $g$.  

\subsection{$P_{\circ}$: comparación con simulaciones de MC}

Una de las motivaciones para desarrollar una nueva estadística fue observar las limitaciones que muestra la Estadística Fraccionaria (EF) para reproducir la función $P_{o}(\theta)$ que da cuenta en detalle de la forma en que se ocupan los estados del sistema a medida que aumenta la densidad o cubrimiento y por lo tanto es muy sensible al tamaño de la partícula , la conectividad de la red de sitios y a las correlaciones que surgen de ambas. En la Fig.  \ref{fig:Po_kmeros_sin_interaccion} se observo que la EF solo reproducía relativamente bien la función $P_{o}$ para $k=2$ y fallaba crecientemente para $k>2$. 

En esta sección mostramos la comparación de los resultados de simulación de MC con los que obtenemos de la Estadística de Múltiple Exclusión (EME) propuesta, para todos los valores de $k$ estudiados. 

\begin{figure}[h]
	\centering
	\includegraphics[width=1\linewidth]{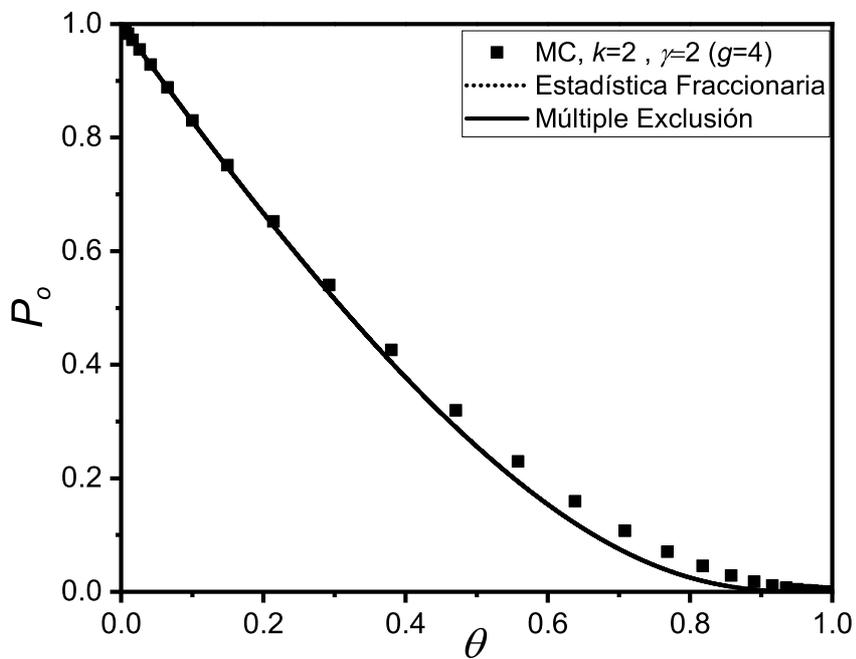}
	\caption{$P_{o}$ en función de $\theta$ para dímeros ($k=2$) sobre red cuadrada ($\gamma=4$). Los símbolos representan los resultados de simulación de MC, la línea sólida la Estadística de Múltiple Exclusión y la línea de puntos la Estadística Fraccionaria (no visible en este caso porque ambas estadísticas coinciden para $g=4$)} 
	\label{Fig.Po_Simulacion_cuadrada_k=2}
\end{figure}

\begin{figure}[H] \vspace{2cm}
	\centering
	\includegraphics[trim = 0mm 10mm 0mm 12mm,width=1\linewidth]{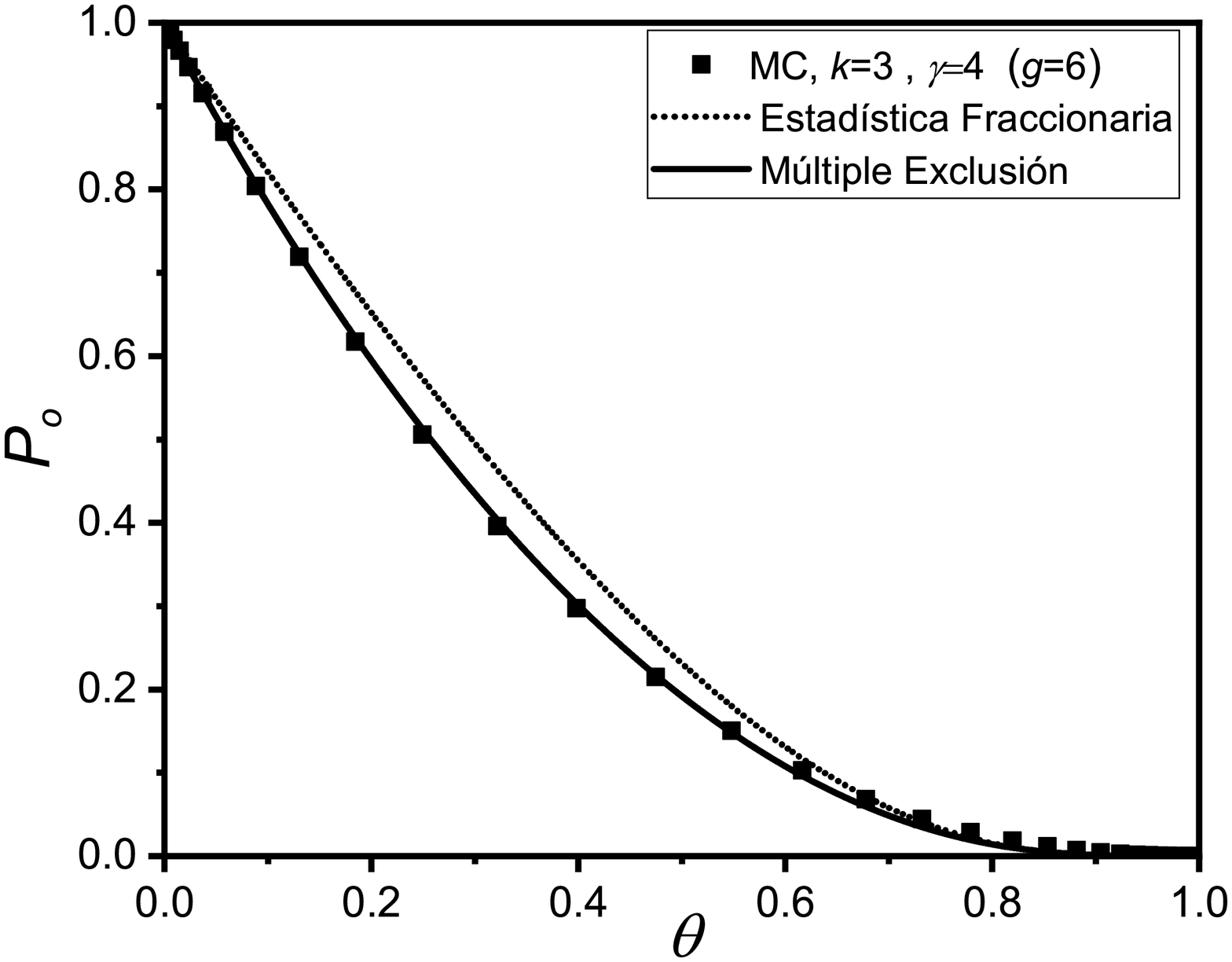}
	\caption{Similar a la Fig. \ref{Fig.Po_Simulacion_cuadrada_k=2} para $k=3$}
	\label{Fig.Po_Simulacion_cuadrada_k=3}
\end{figure}

\begin{figure}[H] \vspace{2cm}
	\centering
	\includegraphics[trim = 0mm 10mm 0mm 25mm,width=1\linewidth]{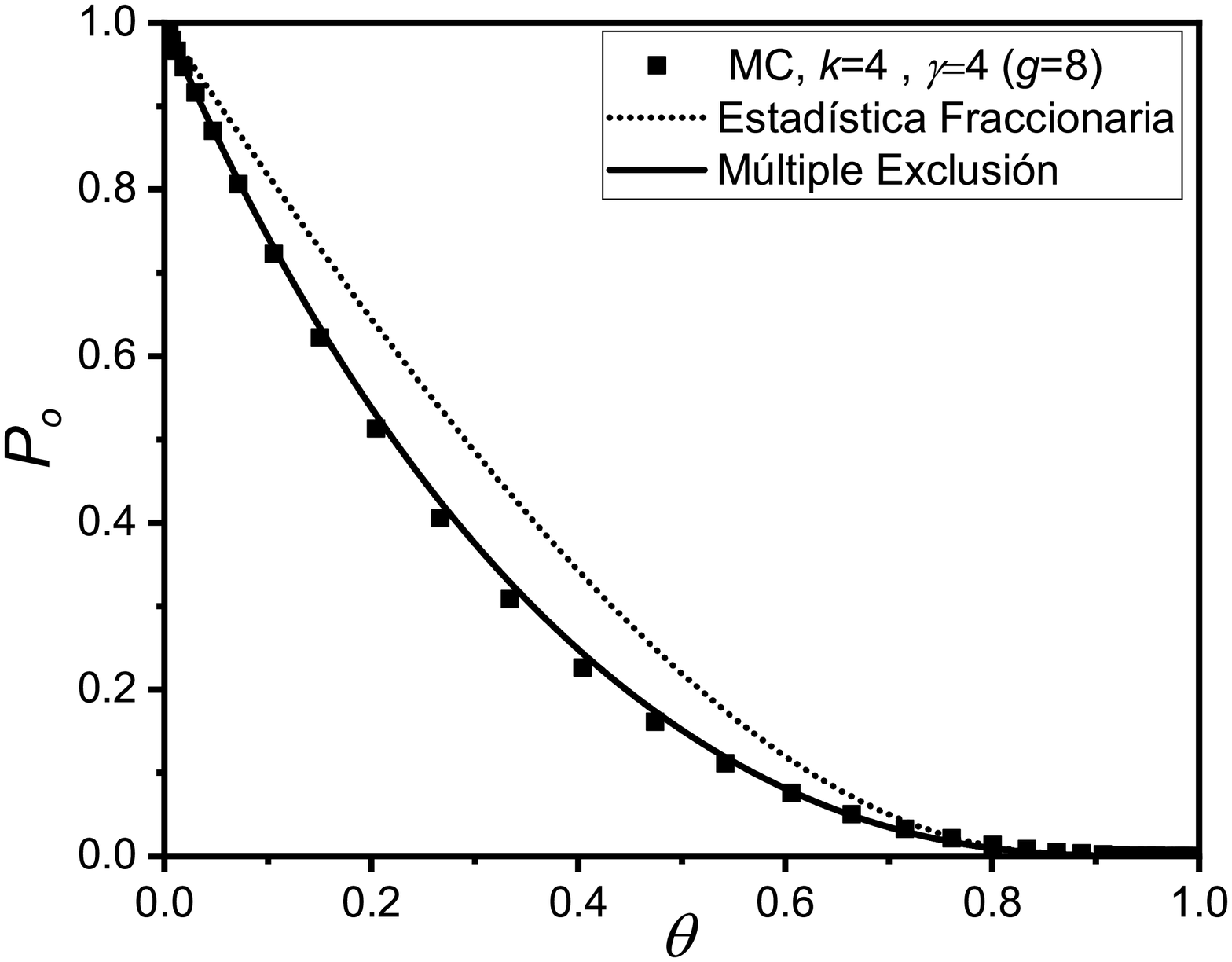}
	\caption{Similar a la Fig. \ref{Fig.Po_Simulacion_cuadrada_k=2} para $k=4$}
	\label{Fig.Po_Simulacion_cuadrada_k=4}
\end{figure}

\begin{figure}[H] \vspace{2cm}
	\centering
	\includegraphics[trim = 0mm 10mm 0mm 12mm,width=1\linewidth]{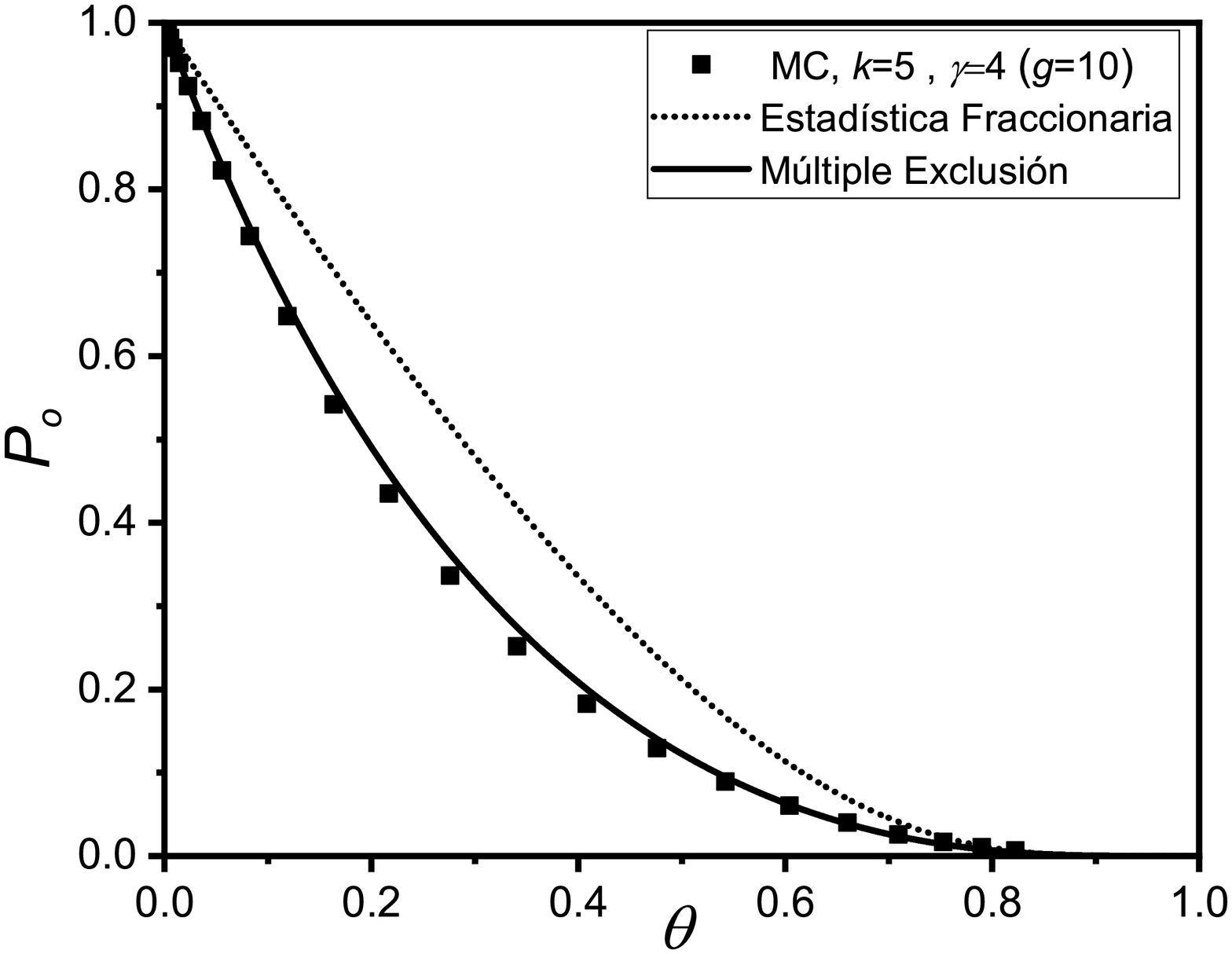}
	\caption{Similar a la Fig. \ref{Fig.Po_Simulacion_cuadrada_k=2} para $k=5$}
	\label{Fig.Po_Simulacion_cuadrada_k=5}
\end{figure}

La EME reproduce todos los resultados de simulación con fidelidad remarcable en los tamaños de partículas estudiados y en el rango completo de cubrimiento como se muestra en las Figs. \ref{Fig.Po_Simulacion_cuadrada_k=2}, \ref{Fig.Po_Simulacion_cuadrada_k=3}, \ref{Fig.Po_Simulacion_cuadrada_k=4} y \ref{Fig.Po_Simulacion_cuadrada_k=5}.    

\begin{figure}[H] \vspace{2cm}
	\centering
	\includegraphics[trim = 0mm 10mm 0mm 12mm,width=1\linewidth]{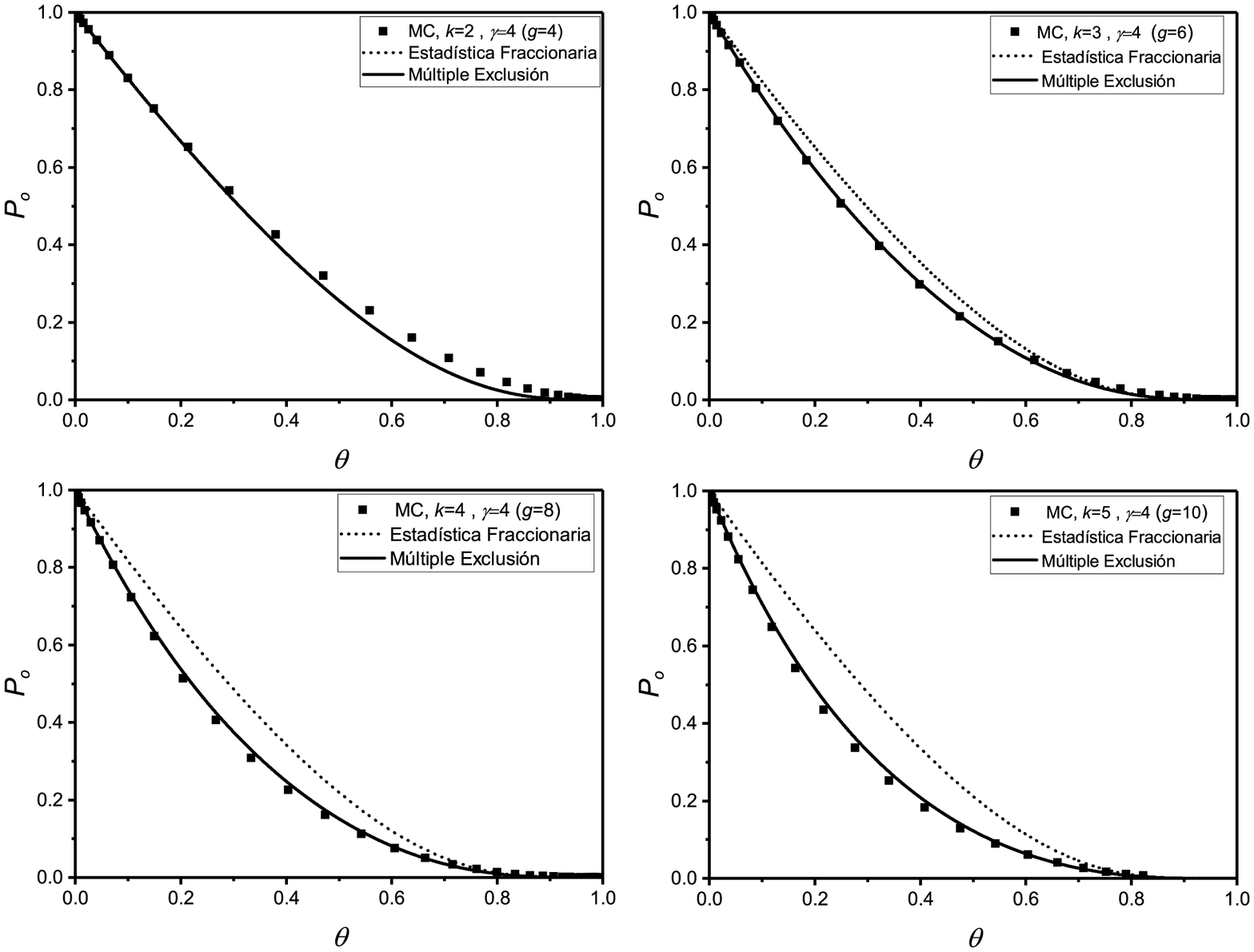}
	\caption{Composición de los resultados de $P_{o}$ para todos los valores de $k$}
	\label{fig:posimulacionefmecuadradak2345capitulo4merged}
\end{figure}

\vspace{1cm}

\begin{figure}[H] 
	\centering
	\includegraphics[trim = 0mm 70mm 0mm 0mm, clip,width=1\linewidth]{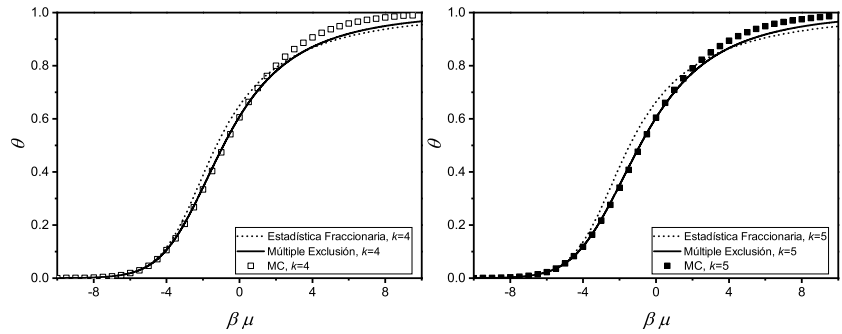}
	\caption{Isotermas de adsorción para $k=4$ y $k=5$. Los símbolos representan los datos de MC, la línea llena el Modelo de Estadística de Múltiple Exclusión y la línea de puntos la Estadística Fraccionaria }
	\label{fig:isotermak4k5mcmeefmerged1}
\end{figure}

\noindent De la comparación de los resultados para el caso sin interacción concluimos que la Estadística de Múltiple Exclusión reproduce con notable aproximación los resultados de simulación para distintos tamaños de k-meros, hasta los mas grandes estudiados en este trabajo.

%% file: texto/TF_Determinacion_de_g.tex
\chapter{Determinación de la exclusión estadística $g$ }

\section{Una nueva caracterización del fenómeno de adsorción}
En esta sección presentamos el desarrollo una simple y a la vez robusta metodología para la interpretación experimental de la termodinámica de adsorción de partículas poliatómicas en los cuales son relevantes los efectos entrópicos que surgen del tamaño y estructura del adsorbato y sus interacciones. Particularmente, desarrolamos una forma simple y directa para determinar el parámetro de exclusión estadística $g$ y eventualmente su dependencia con el cubrimiento $\theta$ a partir de las isotermas de adsorción. Este resultado, en nuestro mejor conocimiento; es la primera propuesta general que se conoce para caracterizar la configuración espacial de las moléculas en estado adsorbido a partir de isotermas de adsorción y válido para moléculas de tamaño y forma arbitraria. 

\noindent Si tenemos en cuenta que el parámetro de exclusión estadística $g$ es la suma de la cantidad de estados del espectro de estados accesibles que una partícula ocupa (1) mas los estados que excluye de ser ocupados por otras partículas idénticas ($g-1$), entonces podemos escribir que el promedio estadístico 

\begin{equation}\label{ec.c6.17}
\overline{g}=<g>=<\frac{1}{N}\sum_{i=1}^{N} g_{i}>
\end{equation} 

\noindent donde $g_{i}$ es la cantidad de estados que excluye la partícula $i$, y el promedio $<...>$ se realiza sobre todo el conjunto estadístico.

\noindent  la ec. \eqref{ec.c6.17} se puede reescribir como

\begin{equation}\label{ec.c6.18}
\overline{g}=<\frac{1}{N} \frac{G}{G}\sum_{i=1}^{N} g_{i}>=<\frac{G}{N} \frac{1}{G}\sum_{i=1}^{N} g_{i}>
\end{equation} 

\noindent Sin embargo el término $<\frac{1}{G}\sum_{i=1}^{N} g_{i}>$ es total de estados ocupados mas los estados excluidos (podríamos decir 'el total de estados excluidos' si no hacemos la diferencia entre estado excluido ocupado y estado excluido no ocupado)  promediado sobre todo el conjunto estadístico y dividido por $G$, y esto es justamente la probabilidad de encontrar un estado u ocupado o excluido, es decir el complemento de $P_{\circ}$ ya que éste último es la probabilidad de que un estado este vacío (disponible). Simbólicamente, a partir de la ec. \eqref{ec.c6.6}, $1-P_{\circ}= P_{\bullet}+P_{\varnothing}$. Con esto,

\begin{equation}\label{ec.c6.19}
\overline{g}=<\frac{G}{N} \frac{1}{G}\sum_{i=1}^{N} g_{i}>= \frac{1}{n} \left(1-P_{\circ} \right)=\frac{1-P_{\circ}}{n} 
\end{equation} 

\noindent Además de la ec. \eqref{eq.c6.3}
$P_{\circ}=P_{\bullet} \ e^{-\beta\left( \mu-U_{o}\right) }= n \ e^{-\beta\left( \mu-U_{o}\right) }$, y finalmente 

\begin{equation}\label{ec.c6.20}
\overline{g}=\frac{1-P_{\circ}}{n}= \frac{1}{n}- e^{-\beta\left( \mu-U_{o}\right)}
\end{equation}

\begin{equation}\label{ec.c6.21}
\overline{g}=\frac{1-P_{\circ}}{n}= \frac{1}{n}- \frac{1}{e^{\beta\left( \mu-U_{o}\right)}}
\end{equation}

\noindent este resultado es extraordinariamente útil ya que la ec. \eqref{ec.c6.21} implica que de la isoterma de adsorción experimental, $n$ versus $ e^{\beta \mu}$, en forma muy elemental se puede determinar la exclusión estadística $g$, dado que el valor de $U_{o}$ se obtiene experimentalmente de la región lineal de la isoterna a baja densidad $n\approx e^{-\beta U_{o}} \ e^{\beta \mu}=K(T) \ e^{\beta \mu}$. 

\noindent Definimos entonces operacionalmente la función Espectro Configuracional de Adsorción $\mathcal{G}(n)$

\begin{equation}\label{ec.c6.22}
\mathcal{G}(n)=  \frac{1}{n}- \frac{1}{e^{\beta\left( \mu-U_{o}\right)}}
\end{equation}

\noindent o

\begin{equation}\label{ec.c6.23}
\mathcal{G}(\theta)=g \ \frac{\left( 1-P_{\circ}\right) }{\theta}=   \frac{g}{\theta}- \frac{1}{e^{\beta\left( \mu-U_{o}\right)}}
\end{equation}

\noindent donde $\lim_{\theta \to 0}\mathcal{G}(\theta)=g_{o}$ representa el números estados excluidos por una partícula aislada, ej. $g_{o}=7$ para $k=2$ en la red cuadrada como surge de los resultados de simulación representados en la forma $<g>=\mathcal{G}(\theta)$. Para $\theta=1$ en cambio $<g>=\mathcal{G}(\theta)=g$ y queda claro que en el gas de red este número representa el valor de la exclusión estadística por partícula cuando todos los estados posibles han sido ocupados, ej. $g=4$ en el caso mencionado arriba. Este resultado nos permite visualizar por primera vez e interpretar de manera clara la variación de la exclusión estadística en función del cubrimiento.      

\noindent $\mathcal{G}(\theta)$ nos proporciona toda la información configuracional de las moléculas en el estado adsorbido en función de la densidad o el cubrimiento a partir de datos termodinámicos dentro del formalismo de las estadísticas de exclusión. Esto representa un avance cualitativo en la descripción de los fenómenos de adsorción.       

\ A modo de ejemplo vemos que en el caso de monómeros, $n=\theta$, $e^{\beta\left( \mu-U_{o}\right)}=\theta/(1-\theta)$ y $\mathcal{G}(\theta)=1$ $\forall \theta$. En las figs. \ref{Fig.c6.espectro.g.4}, \ref{Fig.c6.espectro.g.6}, \ref{Fig.c6.espectro.g.8} y \ref{Fig.c6.espectro.g.10} podemos ver el comportamiento de $\mathcal{G}(\theta)$ para $k=2,3,4$ y $5$ sobre una red cuadrada, que nos muestra efectivamente los efectos de la exclusión múltiple de estados y nos permite interpretar que la exclusión es una función del cubrimiento y que contiene información significativa sobre la configuración y correlación espacial  de las partículas adsorbidas.

\section{Espectro termodinámico de configuraciones $<g>$. Comparación con MC}

De los resultados de simulación representados en la forma ec. \eqref{ec.c6.23}  en  todos los casos surge que $<g>$ varía continuamente desde su valor máximo $<g>=g_{o}$ para $\theta=0$ a su mínimo $<g>=g$ a $\theta=1$. El valor a $g_{o}$ a $\theta=0$ que se obtiene  de esta representación para cada $k$ coincide exactamente con el valor esperado de estados que excluyen las partículas completamente aisladas, esto es, $g_{o}=7$ para $k=2$; $g_{o}=14$ para $k=3$, $g_{o}=23$ para $k=4$ y   $g_{o}=34$ para $k=5$. 

\begin{figure}[H]
	\centering
	\includegraphics[width=1\linewidth]{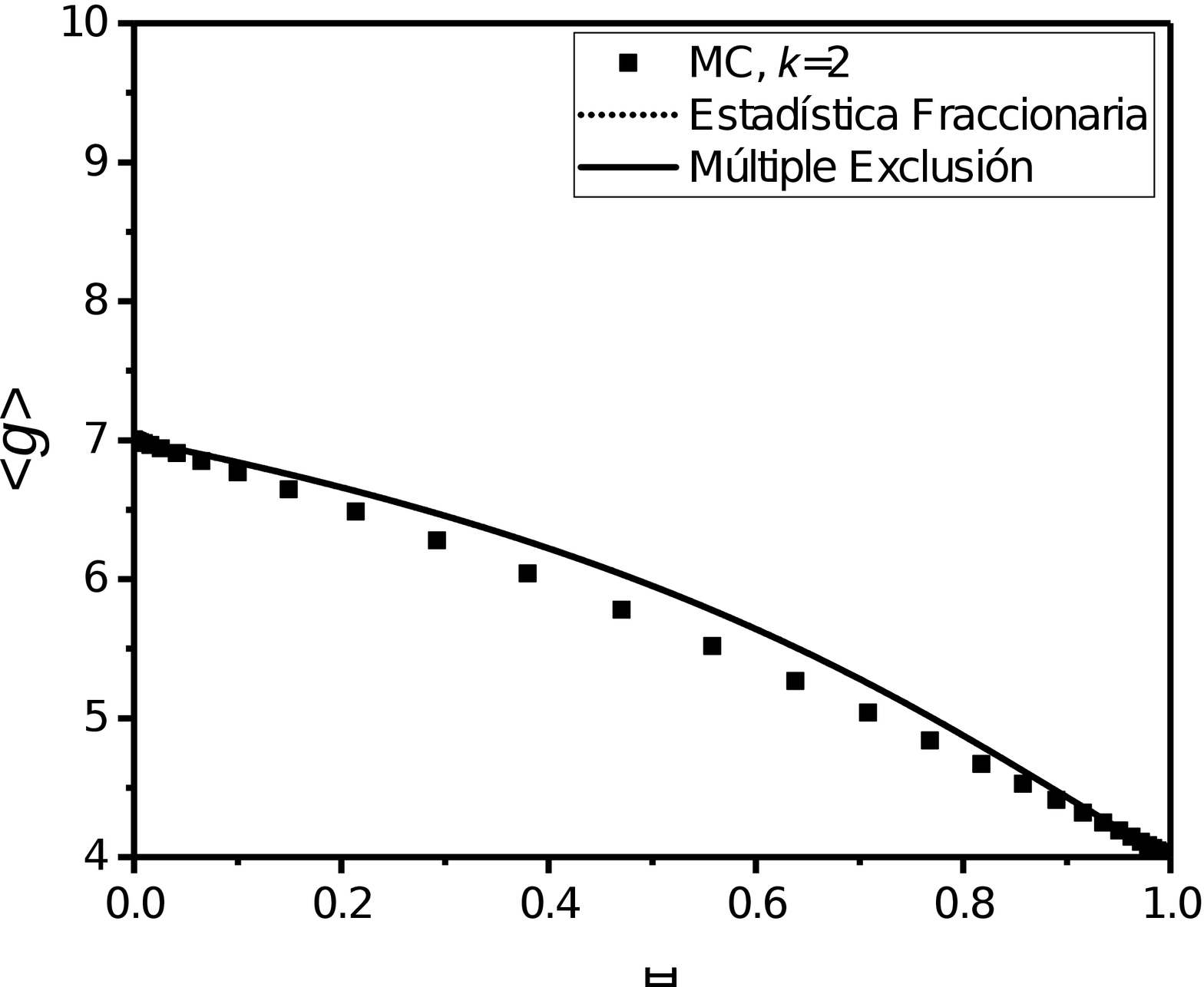}
	\caption{Exclusión estadística media, $\mathcal{G}(\theta)=<g>$ a partir de la ec. \ref{ec.c6.23}, versus $\theta$. Los símbolos representan resultados de MC para $k=2$ sin interacción lateral sobre una red cuadrada de 120x120 sitios. La linea sólida representan los resultados del modelo de Estadística Fraccionaria (ecs. \ref{eq.3.44.a} y \ref{eq.3.12.a}) y de la Estadística de Múltiple Exclusión (ecs. \ref{eq.3.44.a} y \ref{eq.3.43.c}). $g_{e}=0$ de la ec. \ref{eq.3.44.b}. Ambas estadísticas coinciden en este caso}
	\label{Fig.c6.espectro.g.4}
\end{figure}

\begin{figure}[H] \vspace{2cm}
	\centering
	\includegraphics[width=1\linewidth]{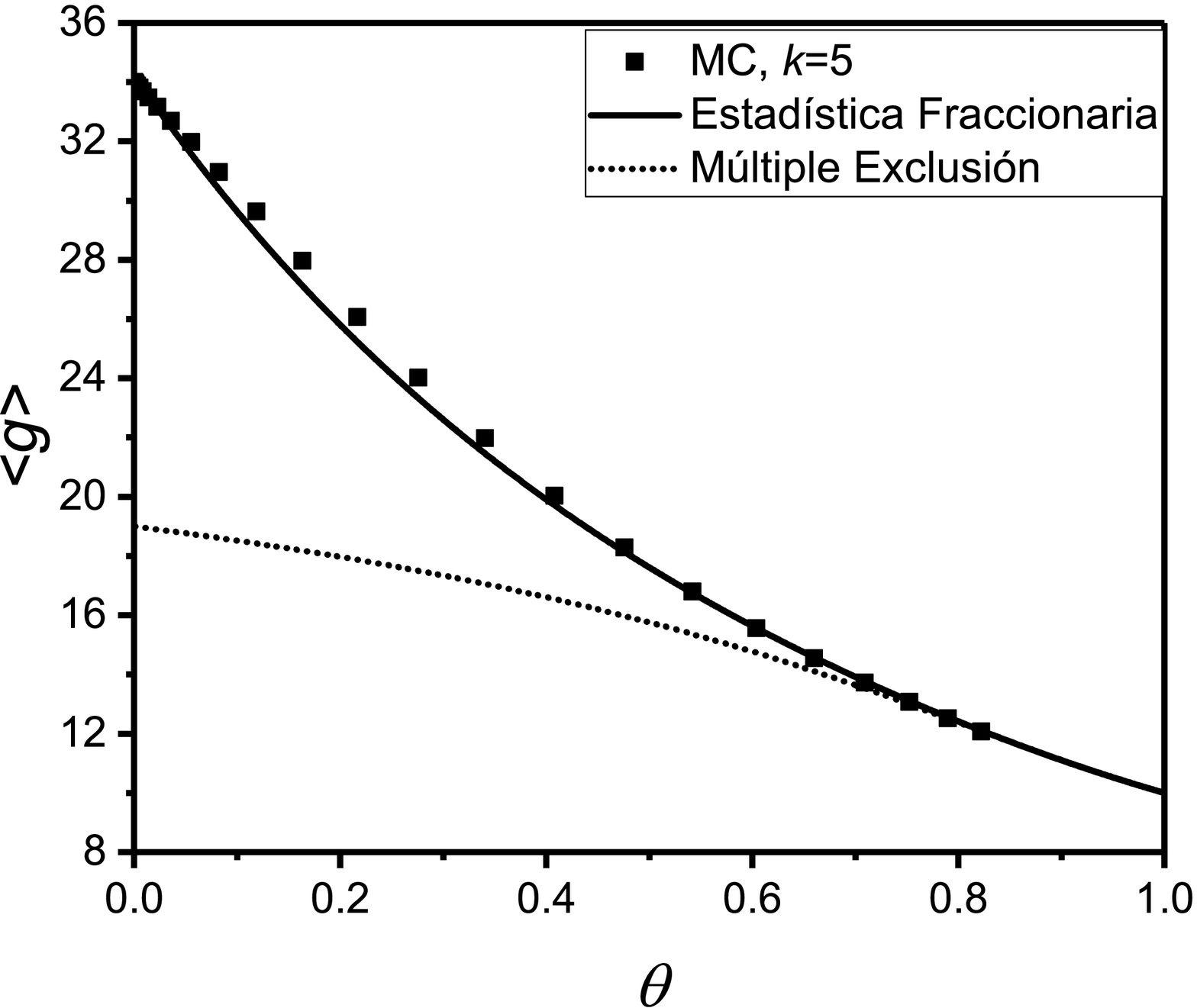}
	\caption{Similar a la Fig. \ref{Fig.c6.espectro.g.4}. Los símbolos representan MC para $k=5$ sobre una red cuadrada de 400x400 sitios. $g_{e}=15.35$ de la ec. \ref{eq.3.44.b}}
	\label{Fig.c6.espectro.g.10}
\end{figure}

\noindent También observamos en las Figs. \ref{Fig.c6.espectro.g.4} a \ref{Fig.c6.espectro.g.10} que la predicción de la Estadística Fraccionaria es muy aceptable para $k=2$. Sin embargo, difiere significativamente para $k>2$. 

\noindent Por otra parte la Estadística de Múltiple Exclusión reproduce en forma notable los resultados de simulación para todos los valores de $k$ estudiados y en todo el rango de cubrimiento $\theta$. Esta Estadística presentada preliminarmente nos da un marco formal muy potente para interpretar gases de red de partículas lineales y probablemente para formas mas complejas.

\begin{figure}[H] \vspace{2cm}
	\centering
	\includegraphics[trim = 0mm 10mm 0mm 12mm,width=1\linewidth]{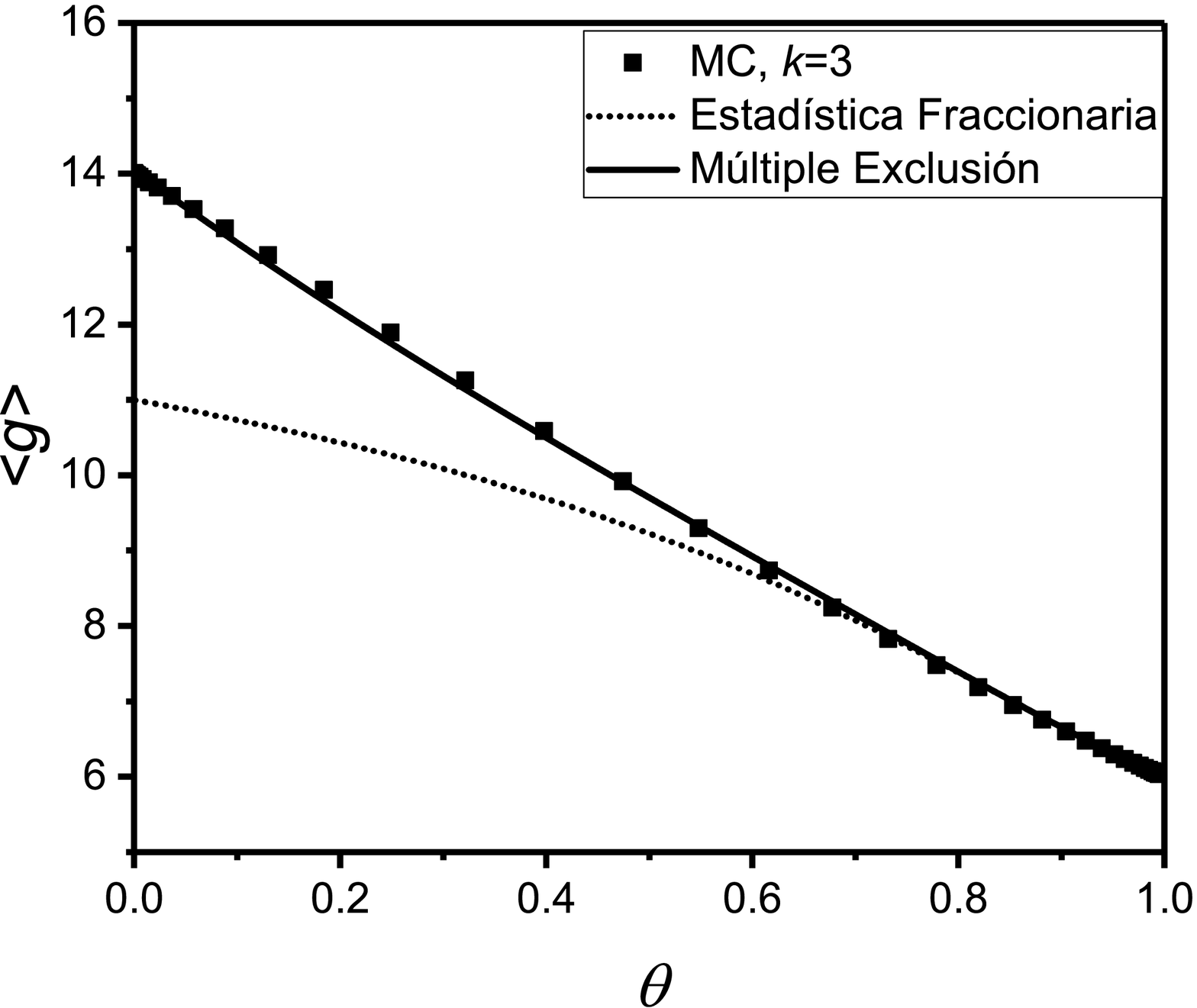}
	\caption{Similar a la Fig. \ref{Fig.c6.espectro.g.4} para $k=3$.}
	\label{Fig.c6.espectro.g.6}
\end{figure}

\begin{figure}[H] \vspace{2cm}
	\centering
	\includegraphics[trim = 0mm 10mm 0mm 24mm,width=1\linewidth]{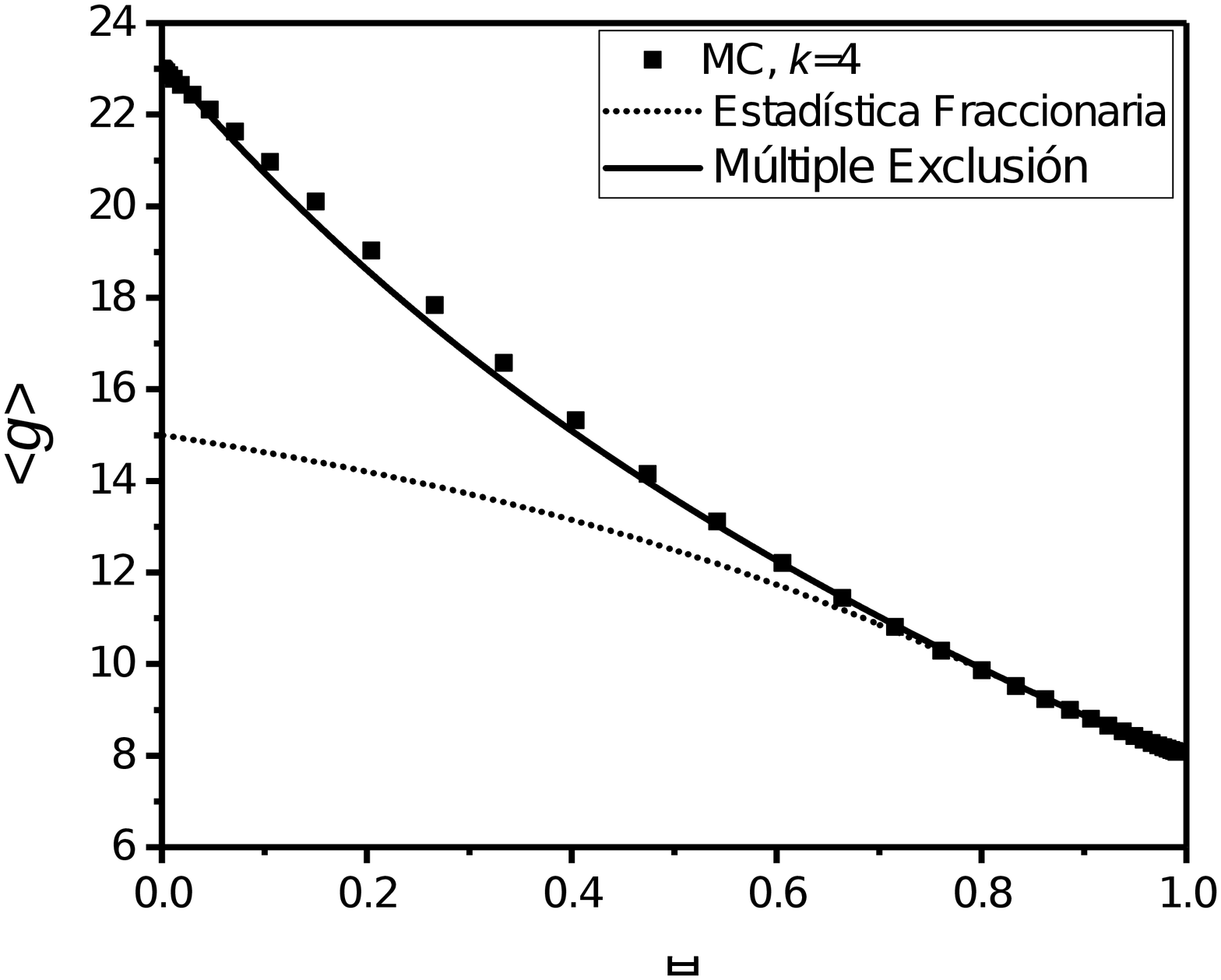} 
	\caption{Similar a la Fig. \ref{Fig.c6.espectro.g.4} para $k=4$.}
	\label{Fig.c6.espectro.g.8}
\end{figure}


\noindent Es evidente de las \ref{Fig.c6.espectro.g.4}, \ref{Fig.c6.espectro.g.6} y \ref{Fig.c6.espectro.g.8} las diferencias entre los valores máximos y míminos de $<g>$ es fuertemente dependiente del tamaño de la partícula (y también lo será de la geometría de la red de sitios). En el caso de interpretar isotermas experimentales bajo la forma de $<g>=\mathcal{G}(\theta)$, ya que a priori no conocemos los valores de $g_{o}$ y $g$ que caracterizan a las partículas adsorbidas sino que son en definitiva los parámetro a determinar, la representación 

\begin{equation}\label{ec.c6.24}
\frac{\mathcal{G}(\theta)}{g}= \frac{\left( 1-P_{\circ}\right) }{\theta}
\end{equation} 

\noindent nos permite estimar $g$ a partir de la amplitud de $<g>$ para una dada red de sitios.

\begin{figure}[H] \vspace{1cm}
	\centering
	\includegraphics[width=1\linewidth]{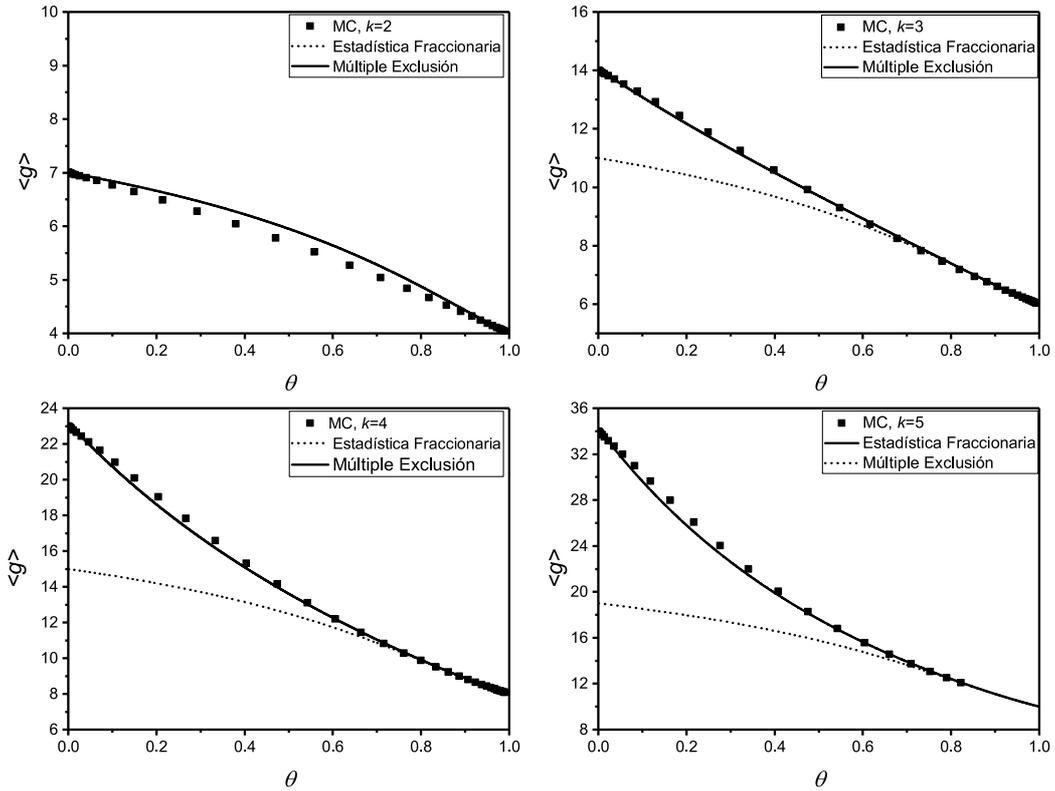}
	\caption{$<g>$ en función de $\theta$ para los casos$k=2,3,4$ y $5$ sobre una red cuadrada. Los simbolos representan los resultados de MC y las lineas llenas y punteada a las Estadísticas de Múltiple Exclusión y Estadística Fraccionaria, respectivamente.}
	\label{fig:g_medio_MC_y_EF_y_ME_cuadrada_merged}
\end{figure}

\begin{figure}[h]
	\centering
	\includegraphics[width=1\linewidth]{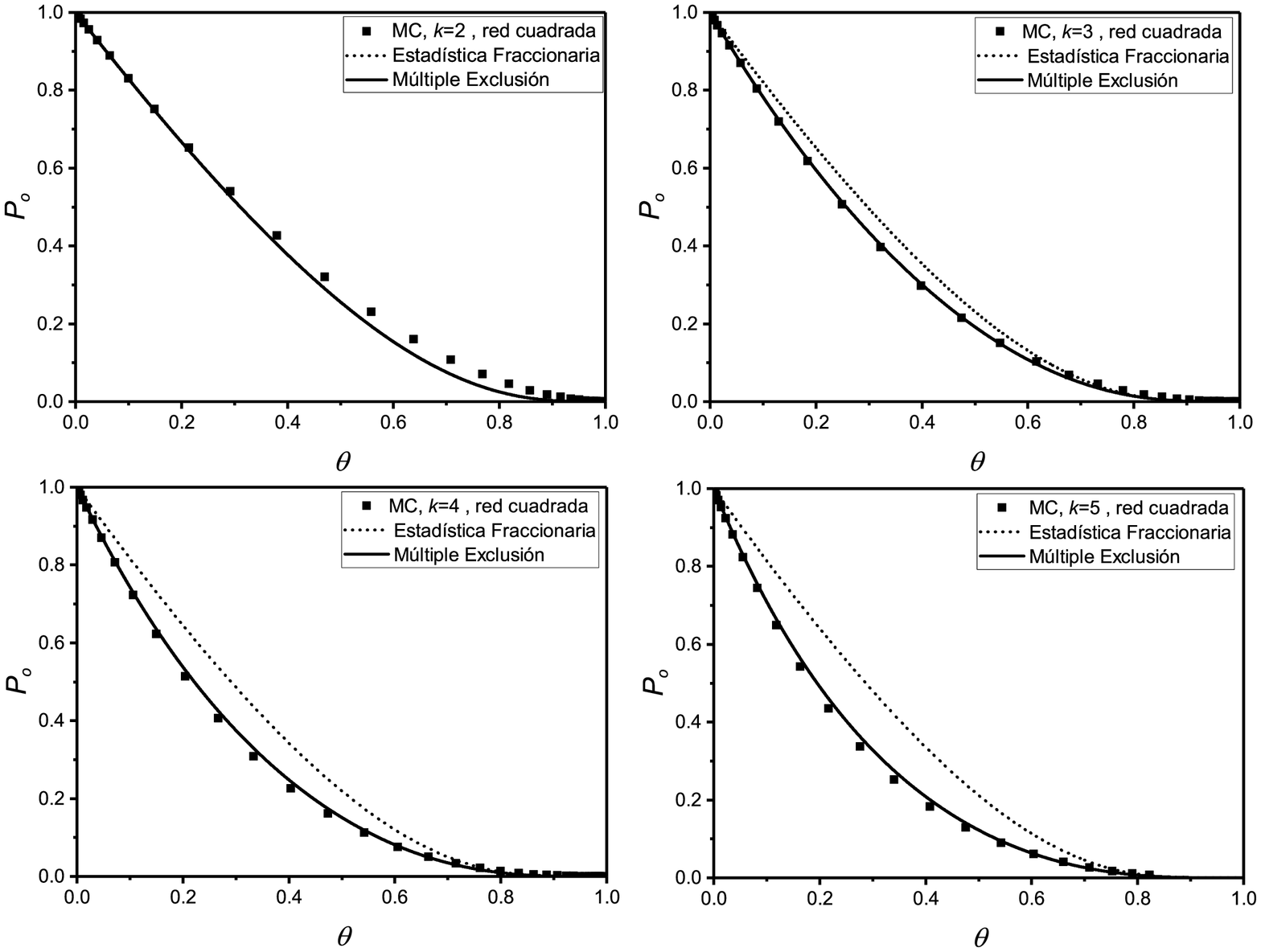}
	\caption{$P_{o}$ en función de $\theta$ para los casos$k=2,3,4$ y $5$ sobre una red cuadrada. Los simbolos representan los resultados de MC y las lineas llenas y punteada a las Estadísticas de Múltiple Exclusión y Estadística Fraccionaria, respectivamente.}
	\label{fig:Po_simulacion_y_EF_y_ME_cuadrada_merged}
\end{figure}

\noindent En todos los casos de las Figs. \ref{fig:g_medio_MC_y_EF_y_ME_cuadrada_merged}
y \ref{fig:Po_simulacion_y_EF_y_ME_cuadrada_merged} la Estadística de Múltiple Exclusión muestra una notable reproducibilidad de los resultados para todos los tamaños de partículas y cubrimientos, inclusive el cambio de curvatura de $<g>$ de cóncava a convexa a medida que $k$ crece. La Estadística Fraccionaria, es un caso límite inferior de la anterior, y solo parece adecuada para describir $k=2$ en 2D y es exacta en 1D para todo $k$.

\begin{figure}[H]
	\centering
	\includegraphics[width=1\linewidth]{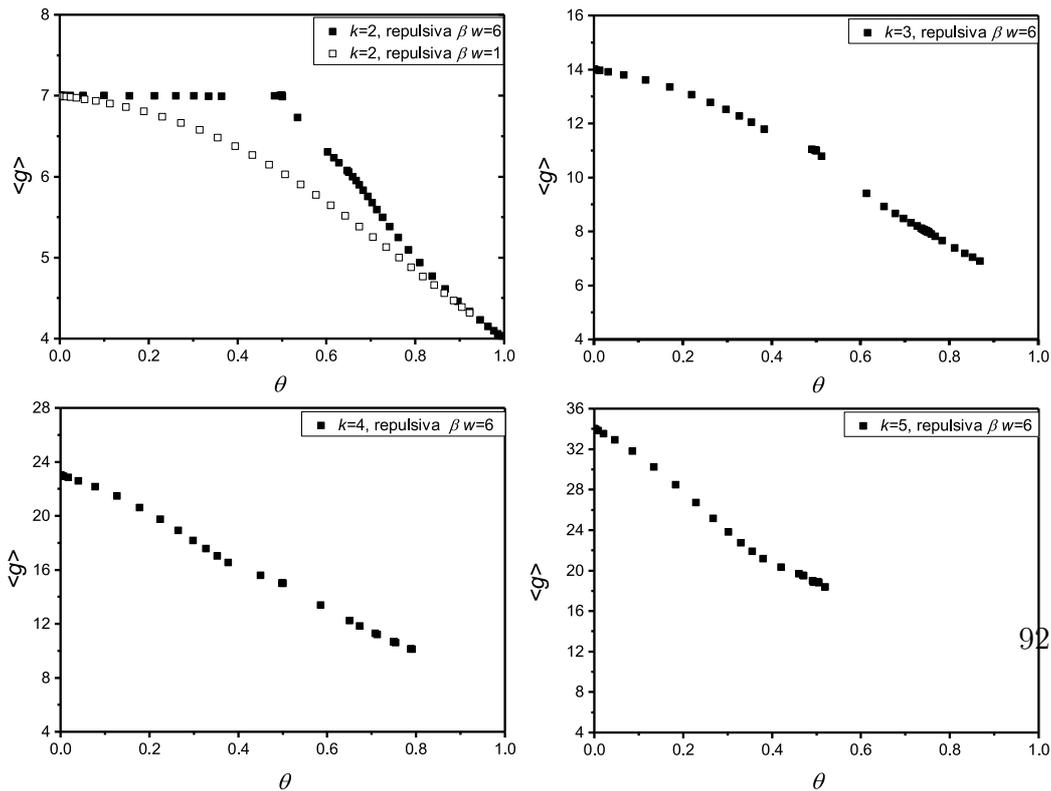}
	\caption{$<g>$ versus $\theta$ para k-meros con interacción repulsiva}
	\label{fig:gmedio_kmeros_con_interaccion_repulsiva}
\end{figure}

\noindent El caso de $<g>$ para k-meros con interacción repulsiva es particulamente interesante porque nos permite determinar la correlación espacial entre partículas vecinas y visualizar la formación de fases ordenadas. 

\noindent Como se observar en el caso $k=2$ de la Fig.\ref{fig:gmedio_kmeros_con_interaccion_repulsiva}, $<g>$ para dímeros con repulsión por debajo de la temperatura crítica se mantiene constante en el valor $<g>=7$ hasta $\theta=0.5$. Esto indudablemente corresponde a una configuración espacial de la fase adsorbida donde los dímeros se ubican aislados unos de otros sin tener unidades en sitios primeros vecinos desarrollando una estructura tipo tablero de ajedrez en estructura regular centrada C(4x2) que se completa a $\theta=0.5$. 

\noindent A partir de allí $<g>$ decrece casi linealmente a medida que los nuevos dímeros tienen necesariamente que tener al menos un sitio primer vecino lleno hasta tanto se completa una nueva fase ordenada a $\theta=2/3$ donde $<g>=6$ y cada dímero tiene dos primeros vecinos llenos formando una estructura tipo bandas diagonales de dímeros sobre la red. En este caso cada dímero excluye mutuamente dos (2) estados con los dos (2) primeros vecinos y por lo tanto el números de estados excluidos por partícula es solo uno (1); luego 7-1=6. Estas estructuras ordenadas son evidentes en los cambios de pendiente de $<g>$ a $\theta=0.5$ y $2/3$ y esta en relación con las regiones de decaimiento lineal de $P_{o}$ en la Fig. \ref{fig:pokmerosconinteraccionrepulsiva}. En contraste para $T>T_{c}$ $<g>$ decrece continuamente con $\theta$ mostrando la ausencia de fases ordenadas.

\begin{figure}[H]
	\centering
	\includegraphics[width=1\linewidth]{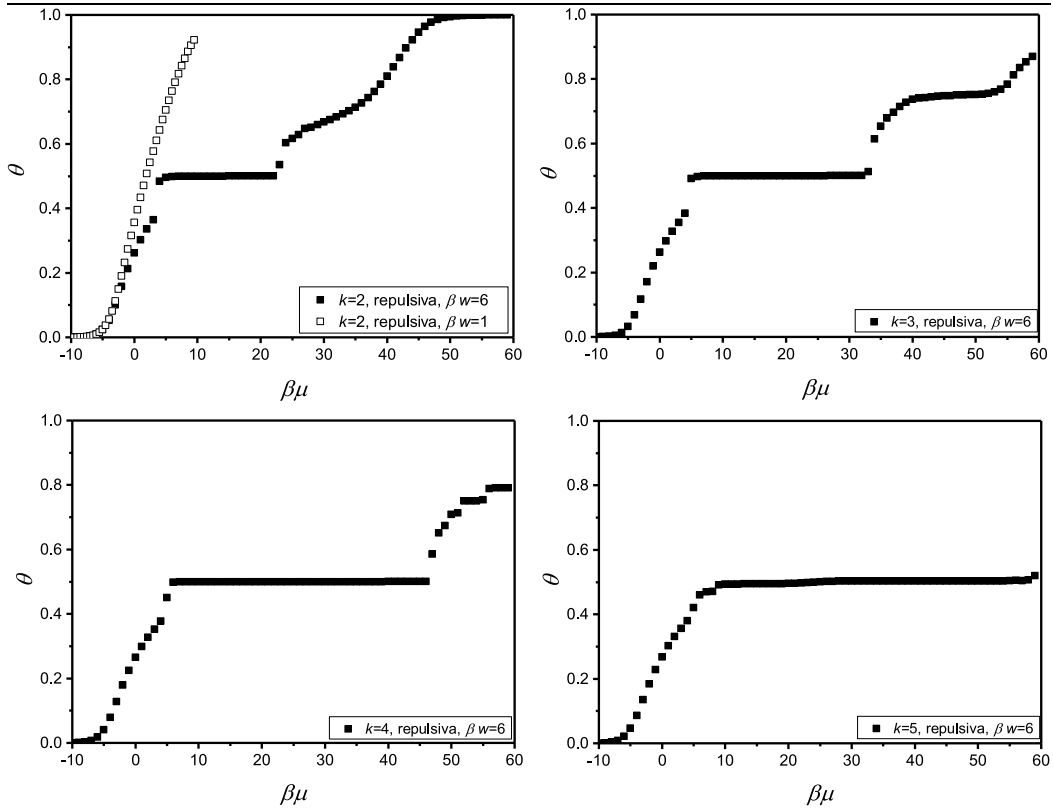}
	\caption{Isotermas de Adsorción de k-meros sobre red cuadrada con interacciones repulsivas}
	\label{Fig.isotermas_k_2_3_4_5_repulsiva_merged_auxiliar}
\end{figure}

\noindent Las Figs.\ref{fig:gmedio_kmeros_con_interaccion_repulsiva} se pueden complementar en su análisis también con las isotermas que le dan origen (Figs. \ref{Fig.isotermas_k_2_3_4_5_repulsiva_merged_auxiliar}).  

\noindent Es sorprendente que para $k=3$, aun cuando la isoterma presenta una meseta muy similar a la de $k=2$, el comportamiento de $<g>$ para este caso es muy diferente al anterior mostrando un decaimiento consistente con la falta de orden en el adsorbato. Sin embargo consideramos que es necesario realizar nuevas simulaciones para reproducir $<g>$ con mas detalle en todo el rango de cubrimientos para interpretar mejor este comportamiento.

%% file: texto/TF_Conclusiones.tex
\chapter{Conclusiones y perspectivas futuras}

La motivación inicial de este trabajo fue estudiar gases de red de k-meros con interacción lateral repulsiva debido a los fenómenos de orden-desorden y las transiciones de fase que pueden desarrollar, e interpretarlos desde el punto de vista de la Estadística Fraccionaria desarrollada en la ref. \cite{ref34}. 

Intentamos entender la configuración de los k-meros a medida que varía el cubrimiento en términos del concepto de exclusión estadística de estados. Para ello realizamos simulaciones de k-meros $(k=2,3,4,5)$ en 1D y 2D y  calculamos las isotermas de adsorción y la probabilidad de encontrar  estados vacíos en función del cubrimiento. 

Teniendo en cuenta que el modelo de Estadística Fraccionaria es exacto para 1D, resulta que reproduce relativamente bien los resultados para partículas pequeñas ($k=2$) pero no así para $k>2$. Esto es una consecuencia de que se asume en el modelo de Estadística Fraccionaria (como en su análoga la Estadística Cuántica Fraccionaria de Haldane) que los estados que ocupan más los que excluyen las partículas son independientes unos de otros. Sin embargo, en un gas de red de k-meros, y en general para todo tipo de partículas excepto el caso especial de monómeros, esta condición no se cumple porque los estados que ocupan los k-meros están correlacionados espacialmente y por esto la Estadística Fraccionaria esta limitada en este sentido para describir los gases de red de k-meros grandes ($k>2$) en 2D. 

Esto nos motivó a proponer una nueva estadística que tenga en cuenta que los estados en un gas de red de k-meros no son independientes unos de otros para ($k\geq2$)y que se produce múltiple exclusión de estados en las configuraciones del gas de red. Obtenemos así una nueva formulación estadística (Estadística de Múltiple Exclusión) que reproduce completamente en principio los resultados de simulación  para la probabilidad de estados vacíos $P_{\circ}$ en función del cubrimiento para todos los tamaños de partículas estudiados. La comparación de esta cantidad es fundamental porque en definitiva significa una prueba de la conjetura básica sobre la forma en que se ocupa el conjunto de estados del sistema, $d(n)$, propuesta para desarrollar esta nueva estadística. Las demás funciones termodinámicas se pueden expresar en términos de esta función y su derivada. La Estadística de Múltiple Exclusión de estados además contiene como caso límite a la Estadística Fraccionaria.

Por último proponemos y desarrollamos un método simple y robusto para determinar el parámetro de exclusión estadística $<g>$ a partir de datos de simulación o experimentales y su dependencia en todo el rango de cubrimiento. Esto nos permite entender con claridad el sentido físico de la exclusión estadística $g$, su relación con la configuración y correlación espacial de las partículas en estado adsorbido, e interpretar el desarrollo de fases ordenadas para k-meros con interacción repulsiva, que fue nuestra motivación inicial. 

Todos los desarrollos presentados son preliminares, y aparecen con la posibilidad de describir el comportamiento termodinámico de una amplia variedad de gases de red con partículas de forma y tamaño arbitrario. 

En el futuro, ampliaremos nuestras simulaciones a partículas de mayor tamaño y diferentes geometrías de red, completaremos el desarrollo de la Estadística de Múltiple Exclusión y analizaremos con  detalle su aplicación a la interpretación de gases de red de partículas poliatómicas de formas no lineales y a sistemas experimentales desarrollando la metodología de espectro termodinámico propuesta.